\documentclass[aps,prc,twocolumn,superscriptaddress,floatfix, nofootinbib,preprintnumbers]{revtex4-2}
\usepackage{graphicx}  
\usepackage{dcolumn}   
\usepackage{bm,relsize}        
\usepackage{amssymb, amsmath}
\usepackage{textcomp}
\usepackage{wasysym}
\usepackage{slashed}
\usepackage{braket}
\usepackage{ulem}
\usepackage{caption, subcaption}
\usepackage{multirow}

\usepackage{lipsum, color}
\usepackage[usenames,dvipsnames,svgnames]{xcolor}
\usepackage{hyperref}
\hypersetup{colorlinks, citecolor=blue, linkcolor=blue, urlcolor=blue}
\usepackage{booktabs}
\usepackage[utf8]{inputenc}

\def\beq{\begin{equation}}
\def\eeq{\end{equation}}
 \def\be{\begin{equation}} \def\ee{\end{equation}}
\def\bea{\begin{eqnarray}} \def\eea{\end{eqnarray}}

\usepackage{caption}
\usepackage{subcaption}
\usepackage{multirow}
\allowdisplaybreaks

\begin{document}

\title{Nuclear-level effective theory of $\mu\rightarrow e$ conversion: Inelastic process}

\author{W. C. Haxton}
\affiliation{Department of Physics, University of California, Berkeley, CA 94720, USA}
\affiliation{Lawrence Berkeley National Laboratory, Berkeley, CA 94720, USA}
\author{Evan Rule}
\affiliation{Department of Physics, University of California, Berkeley, CA 94720, USA}
\affiliation{Theoretical Division, Los Alamos National Laboratory, Los Alamos, NM 87545, USA}

\date{\today}

\preprint{N3AS-24-022, LA-UR-24-28651}

\begin{abstract}
Mu2e and COMET will search for electrons produced via the neutrinoless conversion of stopped muons bound in 1s atomic orbits of $^{27}$Al, improving existing limits on charged lepton flavor violation (CLFV) by roughly four orders of magnitude. Conventionally, $\mu\rightarrow e$ conversion experiments are optimized to detect electrons originating from transitions where the nucleus remains in the ground state, thereby maximizing the energy of the outgoing electron. Clearly, detection of a positive signal in forthcoming experiments would stimulate additional work --- including subsequent conversion experiments using complementary nuclear targets --- to further constrain the new physics responsible for CLFV.  Here we argue that additional information can be extracted without the need for additional experiments, by considering inelastic conversion in $^{27}$Al. Transitions to low-lying nuclear excited states can modify the near-endpoint spectrum of conversion electrons, with the ratio of the elastic and inelastic responses being sensitive to the underlying CLFV operator. We extend the nuclear effective theory of $\mu\rightarrow e$ conversion to the inelastic case, which adds five new response functions to the six that arise for the elastic process. We evaluate these nuclear response functions in $^{27}$Al and calculate the resulting conversion-electron signal, taking into account the resolution anticipated in Mu2e/COMET. We find that $^{27}$Al is an excellent target choice from the perspective of the new information that can be obtained from inelastic $\mu \rightarrow e$ conversion.
\end{abstract}


\maketitle

\section{Introduction}
Unlike their neutral counterparts, which famously oscillate between flavor eigenstates as they propagate, standard-model processes involving charged leptons are expected to conserve flavor. The level of charged lepton flavor violation (CLFV) induced by neutrino oscillations, although technically non-zero, is unobservably small. Thus, any signal of CLFV would be definitive evidence of new physics.  Existing upper limits on CLFV branching ratios provide strong constraints on proposed extensions of the standard model \cite{BARBIERI1994212,Bernstein:2013hba,Calibbi:2017uvl}. Among the most sensitive tests of CLFV are those employing stopped muons, including $\mu^+\rightarrow e^++\gamma$, $\mu^+\rightarrow e^+ e^- e^+$, and $\mu^-+(A,Z)\rightarrow e^-+(A,Z)$. 

The last of these processes, known as $\mu\rightarrow e$ conversion, involves the capture of a negative muon by the Coulomb field of an atomic nucleus with $A$ nucleons and $Z$ protons. Essentially, the nucleus is used as a laboratory to study flavor violation, with proper interpretation of conversion experiments requiring a detailed understanding of the relevant nuclear physics. If treated properly, the nuclear physics can be leveraged in order to maximize the information about CLFV that one can extract from experiment.

The quantity reported by $\mu\rightarrow e$ conversion experiments is the branching ratio 
\begin{equation}
    R_{\mu e}=\frac{\Gamma\left[\mu^-+(A,Z)\rightarrow e^-+(A,Z)\right]}{\Gamma\left[\mu^-+(A,Z)\rightarrow \nu_\mu + (A,Z-1)\right]},
\end{equation}
where the numerator is the rate for the CLFV process and the denominator is the rate for standard muon capture, which is well measured in many different nuclear targets \cite{Suzuki:1987jf}. Currently, the best limit is $R_{\mu e}< 7.0\times 10^{-13}$ at 90\% confidence level, which was obtained by SINDRUM II using a gold target \cite{SINDRUMII:2006dvw}.

By the end of the current decade, a new generation of experiments, Mu2e \cite{Bernstein_2019} at Fermilab and the COherent Muon-to-Electron Transition (COMET) experiment \cite{10.3389/fphy.2018.00133} at Japan Proton Accelerator Research Complex (J-PARC), will improve existing limits by up to four orders of magnitude, reaching a single-event sensitivity of $R_{\mu e}\lesssim 10^{-17}$. Both experiments have selected $^{27}$Al as the target nucleus.

As the captured muon quickly de-excites to the 1$s$ orbital of the nuclear Coulomb potential, the muon energy is known precisely. The momentum of the conversion electron (CE) is then determined by kinematics
\begin{equation}
\vec{q}^{\,2}=\frac{M_T}{m_\mu + M_T}\left[\left(m_{\mu}-E^\mathrm{bind}_{\mu}-\Delta E_\mathrm{nuc}\right)^2-m_e^2\right], 
\label{eq:inelastic_energy}
\end{equation}
where $M_T$ is the mass of the target nucleus, $\Delta E_\mathrm{nuc}=E_f-E_i$ is the energy gap between the final and initial nuclear states, and $E_\mu^\mathrm{bind}$ is the muon binding energy, defined here to be a positive quantity. The CLFV signal is a mono-energetic electron with energy $E_\mathrm{CE}= \sqrt{m_e^2 + |\vec{q}\,|^2}$.  For elastic conversion, $\Delta E_\mathrm{nuc}=0$ and $E_\mathrm{CE} \rightarrow E^\mathrm{elastic}_\mathrm{CE} \approx m_\mu$. The primary background is due to bound muons decaying via the standard-model process $\mu\rightarrow e+2\nu$. Near the endpoint, the spectrum of electrons produced by decay-in-orbit (DIO) is suppressed by $(E-E_\mathrm{end})^5$, where $E_\mathrm{end}=E^\mathrm{elastic}_\mathrm{CE}$. While the DIO background for elastic $\mu \rightarrow e$ conversion is nonzero only because of an experiment's finite energy resolution, it becomes progressively more troublesome 
for inelastic transitions characterized by increasing $\Delta E_\mathrm{nuc}$.

If a positive CLFV signal is observed in Mu2e and/or COMET, some modification of the standard model would be indicated. There would be intense interest in further constraining the source of the new physics. In contrast to the positive muons employed in $\mu\rightarrow e\gamma$ and $\mu\rightarrow 3e$ searches, the captured negative muons in $\mu\rightarrow e$ conversion provide sensitivity to a wider range of CLFV mechanisms, which can be explored by performing additional conversion experiments with different nuclear targets. Frequently, this task is considered within the limited context of coherent conversion (see, e.g., \cite{Cirigliano:2009bz,DAVIDSON2019380,Cirigliano:2022ekw,HEECK2022115833}), where the underlying operator is assumed to couple to nuclear charge and consequently the response is coherently enhanced. Under these assumptions, one can attempt to disentangle the relative coupling strength to protons vs neutrons and to determine whether $\mu\rightarrow e$ conversion is mediated by a scalar or vector coupling to quarks or a dipole coupling to photons. To probe the relative couplings to protons vs neutrons, one can compare rates in targets with different neutron excesses. The difference between scalar and vector mediators amounts to the sign of the relativistic correction from the muon's lower component. In both cases, the sharpest contrast can be made between light and heavy nuclear targets, for example $^{27}$Al and $^{197}$Au.

 However, there is no guarantee that CLFV, if realized in nature at an observable level, will be mediated by a coherent operator. In Refs. \cite{Rule:2021oxe,Haxton:2022piv}, the most general description of the elastic $\mu\rightarrow e$ conversion process was derived in the context of non-relativistic effective theory (NRET). General symmetry arguments dictate that transitions to the nuclear ground state can be mediated by six independent nuclear responses, only one of which corresponds to the much-studied coherent response. (If distinct couplings to protons and neutrons are considered,
 there would be 12 and 2 such responses, respectively.)  The additional responses are nuclear spin- and/or velocity-dependent and can be probed
 by selecting nuclear targets having ground states with the requisite sensitivity to spin, convection current, or spin-orbit correlations \cite{Haxton:2022piv}. 
 
 In subsequent work, the nucleon-level NRET description was matched to relativistic, quark-level weak effective theory (WET) \cite{Haxton:2024lyc}. When combined with existing standard-model effective field theory (SMEFT) and WET formulations \cite{Aebischer:2018bkb,Fuentes-Martin:2020zaz}, this yields a
 tower of effective theories connecting the low-energy nuclear scale where  elastic $\mu\rightarrow e$ conversion experiments are performed with the UV scale where theories of CLFV are formulated. This framework can be utilized to perform a top-down reduction of a specific UV theory to determine the elastic $\mu\rightarrow e$ conversion rate it will predict, or alternatively, to port a given experimental limit from the bottom upward, to constrain entire classes of UV theories.

 The completeness of the NRET operator basis guarantees, in a consistent top-down reduction, that the low-energy consequences of a given UV theory will be
 properly captured.  This happens even though each step of the reduction will, in general, filter out degrees of freedom relevant at higher 
 energies.
 
 Conversely, given that the source of CLFV is unknown, the bottom-up approach should be executed in a manner that takes into account all candidate nuclear-scale operators.  These are the operators generating the six independent elastic $\mu\rightarrow e$ conversion nuclear responses described above: the coefficient of each response is, in principle, a measurable quantity. Determining the precise values of these coefficients and their consequences for UV theories
 will be a challenging task requiring observation of CLFV in a large number of conversion experiments, performed with nuclear targets carefully
 selected for their ground-state properties. 

Experimental considerations place various constraints on the choice of nuclear target. The impressive sensitivities expected at Mu2e and COMET are due in part to the use of a pulsed muon beam, which allows electrons to be observed during a delayed time window largely free of backgrounds from radiative pion capture (RPC). However, the population of captured muons can be depleted during the time that one waits for the background to subside. In heavier nuclei, bound muons decay primarily by ordinary muon capture, with rates scaling approximately as $Z^{4}$ \cite{Primakoff59,Suzuki:1987jf}. For a sufficiently heavy target nucleus, the muon lifetime becomes shorter than the RPC timescale, so that most of the captured muons are depleted before the electron observation window opens. To avoid such losses, the muon capture lifetime must be $\gtrsim 250$ ns \cite{Mu2e-II:2022blh}, restricting target nuclei to those with $Z\lesssim 25$. Thus, while the pulsed-beam technique promises to greatly extend experimental sensitivities,
it does somewhat limit the nuclear targets that can be used.\footnote{Proposed muon storage-ring experiments \cite{KUNO2005376,BARLOW201144, CGroup:2022tli} could potentially provide competitive sensitivity without being restricted to light target nuclei.}

While we can learn more about CLFV by performing experiments with different targets, this strategy also has an obvious drawback: the
additional time each new experiment will require.  Consequently, it is important to consider other strategies for extracting additional information about CLFV. Here and in the accompanying Letter \cite{Haxton:2024ecp}, we show that observation of the \textit{inelastic} contribution to $\mu\rightarrow e$ conversion,
together with the elastic contribution, can provide such information, without the need for an additional experiment.  This strategy is also compatible with new techniques like pulsed muon beams, because the lighter nuclear targets employed in such experiments are more likely to have the necessary attributes, namely low-lying states sufficiently separated to be
distinguishable, given expected experimental resolutions. With increasing nuclear mass, level densities increase, transition strengths become increasingly fractionated, and, for many operators, stronger collectivity removes strength from low-lying states --- all trends that make finding states with the necessary properties less likely.  Thus lighter targets, like $^{27}$Al, that are suitable for the Mu2e and COMET pulsed-beam technique also tend to be better for inelastic $\mu \rightarrow e$ studies.  In fact,
we find $^{27}$Al to be a particularly good choice for this purpose.

The new information available from inelastic $\mu \rightarrow e$ conversion comes in two forms.  First, there exist CLFV operators that can only be probed in the inelastic process, as selection rules prevent them from contributing to the ground-state process. The axial charge operator is one example. Second, even if the CLFV interaction generates elastic conversion, contributions from excited states can modify the spectrum of conversion electrons in distinctive ways. In the case of a positive signal at Mu2e and/or COMET, the precise shape of the CE spectrum can be used to distinguish between possible conversion mechanisms, potentially influencing the design and target selection of future experiments. In effect, instead of changing nuclear targets in order to probe new aspects of CLFV, inelastic $\mu \rightarrow e$ conversion allows
one to accomplish similar goals by changing the nuclear final state.

The physics of inelastic $\mu\rightarrow e$ conversion is essentially governed by two quantities: the nuclear excitation energy $\Delta E_\mathrm{nuc}$ and the relative branching ratio 
\begin{equation}
    \frac{R_{\mu e}(gs\rightarrow f)}{R_{\mu e}(gs\rightarrow gs)}=\frac{\Gamma_{\mu e}(gs\rightarrow f)}{\Gamma_{\mu e}(gs\rightarrow gs)},
    \label{eq:inelastic_branching}
\end{equation}
for transitions to an excited final nuclear state $f$. The quantity in the denominator is the rate for the elastic conversion process, which can be readily computed from the general form described in Ref. \cite{Haxton:2022piv}. In the present work, we derive --- within the context of nuclear effective theory --- the most general expression for the inelastic $\mu\rightarrow e$ conversion rate $\Gamma_{\mu e}(gs\rightarrow f)$, where $f$ is any final nuclear state. 

The nuclear effective theory of inelastic $\mu\rightarrow e$ conversion shares many similarities with the elastic analog. Most importantly, both processes are governed by the same single-nucleon effective theory. Significant differences arise when this single-nucleon description is embedded into a nuclear system. In the elastic process, the approximate parity (P) and time-reversal (T) symmetries of the nuclear ground state restrict the operators that can contribute \cite{Rule:2021oxe,Haxton:2022piv}. In principle, two nuclear charges and three nuclear currents can give rise to 11 independent nuclear response functions (each charge has one multipole projection, each current can be decomposed into longitudinal, transverse-electric, and transverse-magnetic responses). P and T symmetries permit only 6 of these responses to contribute in the elastic case. On the other hand, if the nucleus transitions to an excited final state, the constraint of time-reversal no longer applies, and additional operators can contribute.

In their seminal work on $\mu\rightarrow e$ conversion \cite{PhysRevLett.3.111}, Weinberg and Feinberg considered both the elastic and inelastic responses that arise when the conversion process is mediated by a virtual photon. The authors estimated the total response from all nuclear excited states, concluding that the elastic response, which is coherently enhanced (over the protons), is roughly six times stronger than the total inelastic response for a nucleus like Cu. Subsequent investigations by other authors, who frequently use the term \textit{incoherent} to refer to the inelastic process, are similarly focused on the total response from all excited states \cite{KOSMAS1990641,CHIANG1993526,KOSMAS1994637,PhysRevC.62.035502,KOSMAS2001443,PhysRevC.99.065504}. These previous studies focused on specific CLFV operators --- charge or spin --- and not on the current question of interest, how the simultaneous observation of elastic and inelastic $\mu \rightarrow e$ conversion might be exploited as a 
diagnostic of the source of CLFV.  Moreover, the inclusive inelastic response is not measurable, as it will be buried beneath DIO and other standard-model backgrounds.  

The focus of this paper is quite different.  We have three goals:  First, to use NRET and its complete set of operators to derive the general form of the inelastic $\mu \rightarrow e$ conversion rate, thereby defining precisely what information is in principle available from studying inelastic $\mu \rightarrow e$ conversion.  The CLFV
physics content is encoded in the coefficients of the 11 response functions and their interference terms that comprise the most general form of the rate.  These coefficients are formed from bilinear combinations of the NRET operator coefficients, the low-energy constants (LECs).  The additional inelastic response functions and the interference terms they generate introduce seven new combinations of the LECs, in principle providing new constraints on CLFV.  Second, we use the example of $^{27}$Al to illustrate how to extract additional information on CLFV by studying the electron spectrum near, but not restricted to, the endpoint region.  A nonzero measurement of elastic $\mu \rightarrow e$ conversion can be attributed to any CLFV operator, but this will not be the case if one adds information from inelastic transitions.  Third, we illustrate how the additional information available from the near-endpoint spectrum, given expected Mu2e and COMET backgrounds and resolutions, can be used to guide the selection of the next target --- what target properties will be important in further distinguishing candidate CLFV sources?  The guidance does not necessarily require a Mu2e or COMET observation of an inelastic signal: the absence of such a signal in the presence of elastic $\mu \rightarrow e$ conversion is also significant.

This paper is organized as follows: In Sec. \ref{sec:leptonic}, following our earlier work on the elastic process, we describe how the Coulomb distortion of leptonic fields can be approximated, without significant loss of accuracy, to dramatically simplify the form of NRET interactions used in nuclear calculations.  The nuclear effective theory is then constructed in two steps.  First, in Sec. \ref{sec:nucleon_level_eft}, we introduce the NRET basis of 16 operators that represents the most general interaction between the leptons and a single nucleon. This single-nucleon effective theory --- originally derived in Refs. \cite{Rule:2021oxe,Haxton:2022piv} --- makes no reference to a particular nuclear target or transition. As the coefficients that govern the theory are target- and transition-independent, the description applies equally well to the elastic and inelastic conversion processes. Second, in Sec. \ref{sec:nuclear_eft}, the single-nucleon theory is embedded into a target nucleus, yielding our central result: the most general expression for the $\mu\rightarrow e$ conversion rate $\Gamma_{\mu e}(gs\rightarrow f)$, valid for transitions to any nuclear final state. The 16 nucleon-level operators embed into 11 nuclear response functions, the coefficients of which encode the CLFV physics. Unlike the elastic process, inelastic $\mu\rightarrow$ e conversion is not constrained by time-reversal symmetry, permitting all 11 response functions to contribute. 

In Sec. \ref{sec:response_props}, we discuss properties of the nuclear response functions, which we view as experimental ``knobs" that can be ``turned," either by changing the final state in a given target or by selecting a new target, to determine the operator(s) responsible for CLFV. In Sec. \ref{sec:exp_sig}, we compute the expected CE signal at Mu2e for a series of simplified scenarios in which only a single nuclear response contributes.  We find that coherent conversion in $^{27}$Al does not yield any significant modification from the inelastic process. On the other hand, spin-dependent conversion mechanisms generate relatively strong transitions to excited states, which modify the shape of the CE spectrum. Thus, if any excess in electron counts is measured relative to the elastic baseline, there must be some other CLFV mechanism besides coherent conversion. 

In Sec. \ref{sec:other_targets}, we discuss inelastic $\mu\rightarrow e$ conversion in targets other than $^{27}$Al, considering in detail the case of a natural titanium target with its multiple isotopes.  In Sec. \ref{sec:conclusion}, we summarize and present our conclusions. 

The main text is supplemented by a series of appendices: Appendix \ref{app:code_desc} describes a publicly available \texttt{Mathematica} script that implements the nuclear effective theory of inelastic $\mu\rightarrow e$ conversion for an $^{27}$Al target. Appendix \ref{app:rate_derivation} provides several intermediate steps in the derivation of the inelastic $\mu\rightarrow e$ conversion rate $\Gamma_{\mu e}(gs\rightarrow f)$ as well as explicit expressions for the requisite nuclear response functions. Finally, in Appendix \ref{app:modeling}, we provide numerical values for each nuclear response function obtained from three different shell-model calculations employing distinct nuclear interactions.

\section{Treatment of Leptonic Fields}
\label{sec:leptonic}
In contrast to the complicated many-body nuclear physics, the leptonic sector of $\mu\rightarrow e$ conversion is relatively straightforward: To begin, the muon is known to occupy the $1s$ orbital of the nuclear Coulomb field. Exploiting the spherical symmetry of the potential, solutions of the Dirac equation can be written as
\begin{equation}
    \psi_\kappa(\vec{r}\,)=\left(\begin{array}{c}
         ig_{\ell j} \left[\begin{array}{c}
              \braket{\frac{1}{2},\frac{1}{2}|(\ell\frac{1}{2})jm}  \\
              \braket{\frac{1}{2},-\frac{1}{2}|(\ell\frac{1}{2})jm}
         \end{array}\right]  \\ 
         \\
         -f_{\ell j} \left[\begin{array}{c}
              \braket{\frac{1}{2},\frac{1}{2}|(\ell\pm 1\frac{1}{2})jm}  \\
              \braket{\frac{1}{2},-\frac{1}{2}|(\ell\pm 1\frac{1}{2})jm}
         \end{array}\right]  
    \end{array}\right),
\end{equation}
where the index $\kappa$ satisfies $j=|\kappa|-\frac{1}{2}$ and
\begin{equation}
    \kappa = \left\{\begin{array}{cc}
        -(\ell + 1), & \kappa < 0  \\
        \ell, & \kappa > 0.
    \end{array}\right.
\end{equation}
The angular-spin eigenfunctions are defined as 
\begin{equation}
    \braket{\theta\phi|(\ell\frac{1}{2})jm}=\sum_{m_\ell m_s}\braket{\ell m_\ell\frac{1}{2}m_s|jm}Y_{\ell m_\ell}(\theta,\phi)\xi_{m_s},
\end{equation}
where $\xi_{m_s}$ is a Pauli spinor. The radial wave functions satisfy the coupled differential equations 
\begin{equation}
    \begin{split}
        \left[E-\mu-V_C(r)\right]g_{\ell j}&=-\frac{d}{dr}f_{\ell j}+\frac{\kappa-1}{r}f_{\ell j}\\
        \left[E+\mu-V_C(r)\right]f_{\ell j}&=\frac{d}{dr}g_{\ell j}+\frac{\kappa+1}{r}g_{\ell j},
    \end{split}
\end{equation}
where $\mu$ is the reduced mass of the lepton and $V_C(r)$ is the Coulomb potential.

 The Coulomb potential is computed by treating the nuclear charge distribution $\rho(r)$ as an extended source, determined from elastic electron scattering data. Specifically, we adopt the 2-parameter Fermi model parameterization,
\begin{equation}
    \rho(r)=\frac{\rho_0}{1+\mathrm{exp}[(r-c)/\beta]},
\end{equation}
where the value of $\rho_0$ is fixed by the normalization condition
\begin{equation}
    \int_0^{\infty} dr~r^2\rho(r)=Z.
\end{equation}
The values of the parameters $c$ and $\beta$ in Table \ref{tab:nuclear_params} are taken from Ref. \cite{DEVRIES1987495}, which performed fits to available electron scattering elastic form factors. Combining this information with the Dirac equation for a spherically symmetric potential, one can precisely determine the binding energy $E_\mu^\mathrm{bind}$ and wave function $\psi^{(\mu)}_{\kappa=-1}$ of the muon. Once the energy of the outgoing electron is determined from the kinematics, as in Eq. \eqref{eq:inelastic_energy}, one can similarly obtain the partial-wave decomposition of the electron wave function.

The vast majority of the $\mu\rightarrow e$ conversion literature focuses on the special case of the charge operator, further simplified by retaining only the two electron partial waves $(\kappa =\pm 1)$ needed for the coherent monopole contribution to the charge response.  It is then computationally feasible to use exact numerical Dirac solutions. Ref. \cite{Kitano:2002mt} is representative of this approach. But in general, significant angular momentum transfer can occur between the leptons and the nucleus, so that a sum over many electron partial waves must be done to compute the $\mu\rightarrow e$ conversion amplitude. The use of exact numerical Dirac solutions then becomes quite complicated, while also obscuring the underlying physics. Instead, we follow Refs. \cite{Rule:2021oxe,Haxton:2022piv} in adopting several simplifying yet still quite accurate approximations. 

The outgoing electron is ultra-relativistic and can be reasonably approximated, well away from the origin, by a Dirac plane wave. Near the nucleus, however, the plane wave is distorted by the nuclear charge. We can account for this Coulomb distortion by evaluating the plane wave not at the physical electron momentum $q$ but at the effective momentum \cite{KNOLL1974462,LENZ1971513}
\begin{equation}
    \vec{q}^{\,2}_\mathrm{eff}=\frac{M_T}{m_\mu + M_T}\left[\left(m_{\mu}-E^\mathrm{bind}_{\mu}-\Delta E_\mathrm{nuc}-\bar{V}_C\right)^2-m_e^2\right],
    \label{eq:q_eff}
\end{equation}
where 
\begin{equation}
    \bar{V}_C\equiv \frac{\int dr~r^2\rho(r)V_C(r)}{\int dr~r^2\rho(r)},
\end{equation}
is the average value of the Coulomb potential over the nuclear charge density. Explicitly, the effective momentum approximation corresponds to the replacement
\begin{equation}
    U(q,s)e^{i\vec{q}\cdot\vec{r}}\rightarrow \frac{q_\mathrm{eff}}{q}\sqrt{\frac{E_\mathrm{CE}}{2m_e}}\left(\begin{array}{c}
         \xi_s \\
         \vec{\sigma}_L\cdot\hat{q}\xi_s
    \end{array}\right)e^{i\vec{q}_\mathrm{eff}\cdot\vec{r}},
    \label{eq:Coulombelectron}
\end{equation}
where $E_\mathrm{CE}=\sqrt{q^2+m_e^2}$ is the physical electron energy, and we have adopted the notation $q=|\vec{q}\,|$, $q_\mathrm{eff}=|\vec{q}_\mathrm{eff}|$. We stress that this form describes the electron wave function in the vicinity of the nucleus, where the integral
over the transition density is performed; it properly captures the effects of Coulomb attraction on the amplitude and local frequency of the electron in this region, both of which are enhanced. This form is not to be used asymptotically, where the wave function heals to its standard plane-wave form with wave number $q$, normalized to unit flux as demanded by consistency with the integration over the density of states.   

In Ref. \cite{Haxton:2022piv}, comparisons of the approximate electron wave function in Eq. \eqref{eq:Coulombelectron} with exact solutions of the Dirac equation are given for various partial waves and for both light and heavy targets.  No fitting was performed in this comparison: the value of $\bar{V}_C$ (and hence $q_\mathrm{eff}$) was computed from the measured nuclear charge distribution, not adjusted to optimize the fit. The agreement between the analytic wave function of Eq. \eqref{eq:Coulombelectron} and the numerically generated Dirac partial waves is excellent.  

\begin{table}
    \centering
    {\renewcommand{\arraystretch}{1.4}
    \begin{tabular}{lcccc}
    \hline
    \hline
    Target & ~~$c$ (fm)~~ & ~~$\beta$ (fm)~~ & ~~$\sqrt{\braket{r^2}}$ (fm)~~ & ~~$E_\mu^\mathrm{bind}$ (MeV)~~ \\
    \hline
    $^{27}$Al  &  3.07 & 0.519 & 3.062 & 0.463 \\
    $^{48}$Ti & 3.843 & 0.588 & 3.693 & 1.262 \\
         \hline
         \hline
    \end{tabular}
    }
    \caption{Nuclear charge distribution parameters $c$ and $\beta$ and the resulting values for the rms charge radius $\sqrt{\braket{r^2}}$ and muon binding energy $E_\mu^\mathrm{bind}$.}
    \label{tab:nuclear_params}
\end{table}

\begin{table*}
    \centering
    {\renewcommand{\arraystretch}{1.4}
    \begin{tabular}{lccc|ccccccc}
    \hline
    \hline
     & & & & \multicolumn{7}{c}{$R=|\phi_{1s}^{Z_\mathrm{eff}}(\vec{0})|^2/ |\phi_{1s}^{Z}(\vec{0})|^2$} \\
    Transition & $\Delta E_\mathrm{nuc}$ (MeV) & $q$ (MeV) & $q_\mathrm{eff}$ (MeV) & ~~$W_{M_0}^{00}$~~ & ~~$W_M^{00}$~~ & ~~$W_M^{11}$~~ & ~~$W^{00}_{\Sigma}+W_{\Sigma'}^{00}$~~ & ~~$W^{11}_{\Sigma}+W_{\Sigma'}^{11}$~~ & ~~$W_{\Sigma''}^{00}$~~ & ~~$W_{\Sigma''}^{11}$~~\\
    \hline
       $^{27}$Al$(5/2+)$ &  0.0 & 104.98 & 110.81 & 0.657 & 0.657 & 0.643 & 0.647 & 0.656 & 0.635 & 0.632 \\
       $^{27}$Al$(1/2^+)$ & 0.844 & 104.13 & 109.97 & --- & 0.603 & 0.588 & 0.593 & 0.591 & 0.592 & 0.592 \\
       $^{27}$Al$(3/2^+)$ & 1.015 & 103.96 & 109.80 & --- & 0.599 & 0.606 & 0.601 & 0.599 & 0.601 & 0.598 \\
       $^{27}$Al$(7/2^+)$ & 2.212 & 102.77 & 108.60 & --- & 0.597 & 0.594 & 0.626 & 0.636 & 0.627 & 0.627 \\[2mm]
       $^{48}$Ti$(0^+)$ & 0.0 & 104.28 & 112.43 & 0.433 & 0.433 & 0.411 & --- & --- & --- & --- \\
       $^{46}$Ti$(2^+)$ & 0.889 & 103.39 & 111.54 & --- & 0.359 & 0.362 & 0.337 & 0.345 & --- & --- \\
       $^{47}$Ti$(3/2^+)$ & 0.159 & 104.11 & 112.27 & --- & 0.359 & 0.362 & 0.390 & 0.374 & 0.384 & 0.367 \\
       $^{48}$Ti$(2^+)$ & 0.984 & 103.29 & 111.45 & --- & 0.359 & 0.357 & 0.341 & 0.346 & --- & --- \\
         \hline
         \hline
    \end{tabular}
    }
    \caption{Left: Transition-dependent excitation energy $\Delta E_\mathrm{nuc}$, electron momentum $q$, and effective momentum $q_\mathrm{eff}$ for $\mu\rightarrow e$ conversions to the ground state and 3 low-lying states of Al and Ti. Right: Comparison of the reduction factor $R$ that describes the effective constant value of the muon wave function obtained by averaging over different nuclear transition densities $W_O^{\tau\tau}$.}
    \label{tab:transition_params}
\end{table*}

In contrast to the electron, the bound muon is highly nonrelativistic, so that its lower component $f(r)$ can be neglected in transition densities, which are all dominated by the upper component $g(r)$.  Nevertheless, for elastic $\mu \rightarrow e$ conversion, the formalism for handling the effects of $f(r)$ has been worked out in detail, extending the NRET to all operators linear in either the nucleon velocity $\vec{v}_N$ or the muon velocity $\vec{v}_\mu$ \cite{Haxton:2022piv}.  These corrections have been incorporated into a numerical script for evaluating elastic $\mu \rightarrow e$ conversion rates \cite{Haxton:2024lyc}, numerically verifying that $f(r)$ produces corrections to nuclear response functions of $O(5\%)$ for $^{27}$Al.  The additional NRET operators associated with $\vec{v}_\mu$ play no role in nuclear selection rules that govern the general form of the rate.

There is some justification for including $f(r)$ in elastic $\mu \rightarrow e$ conversion calculations, as there are certain operators, like the charge, where the associated nuclear response function can be computed to a precision of $\lesssim 5\%$.  There is much less justification for doing so in inelastic $\mu \rightarrow e$ conversion studies, where the nuclear form factors are not tightly constrained by experiment, but instead must be taken from nuclear models.  Consequently, in the present treatment of inelastic conversion, we neglect $f(r)$ and thus corrections associated with $\vec{v}_\mu$, retaining only the dominant upper-component contributions of $g(r)$.

The radial wave function $g(r)$ for the upper component of the muon appears in the nuclear transition density.  In our earlier work on elastic $\mu \rightarrow e$ conversion \cite{Rule:2021oxe,Haxton:2022piv}, we evaluated the integral over the transition density for a specific operator, the isoscalar monopole charge, then employed this result to define an effective average muon density within the nucleus. The latter quantity was then expressed in terms of an equivalent $1s$ muon density for a point nucleus of charge $Z_\mathrm{eff}$, evaluated at $\vec{r}=0$. The value of $Z_\mathrm{eff}$ was determined by matching to the exact result,
\begin{eqnarray}
    |\phi_{1s}^{Z_\mathrm{eff}}(\vec{0})|& = & {1 \over \sqrt{\pi}} \Big[ {Z_\mathrm{eff} \alpha  \mu c \over \hbar} \Big]^{3/2}  \nonumber \\
    & \equiv \Bigg| &\frac{\int d\vec{r} ~\rho_0(r)j_0(q_\mathrm{eff}r)\frac{1}{\sqrt{4\pi}}g(r)}{\int d\vec{r}~\rho_0(r)j_0(q_\mathrm{eff}r)} \Bigg|.
    \label{eq:Zeff}
\end{eqnarray}
Here $\mu$ is the muon's reduced mass and $\rho_0(r)$ is the isoscalar charge density.
The exact result can then be expressed in terms of the undistorted plane-wave result, weighted by the effective Schr\"odinger amplitude,
\begin{equation}
 \int d\vec{r}~\rho_0(r)j_0(q_\mathrm{eff}r) {g(r) \over \sqrt{4 \pi}}= |\phi_{1s}^{Z_\mathrm{eff}}(\vec{0})|{\int d\vec{r}~\rho_0(r)j_0(q_\mathrm{eff}r)}. \nonumber
\end{equation}

The procedure followed in Ref. \cite{Haxton:2022piv} was to assume that the same Schr\"odinger density (same $Z_\mathrm{eff}$) can be used to factor other transition amplitudes.  This of course would not be exact in those cases, but nevertheless can be justified as the first term in a Taylor expansion that assumes the muon wave function varies gently over the region where the nuclear density is significant. Note that the muon's Bohr radius in $^{27}$Al is $\approx$ 20 fm.  This approximation was tested in Ref. \cite{Haxton:2022piv}, where a series of exact calculations were performed for $^{27}$Al in which both the operator and its isospin coupling were varied.  The $Z_\mathrm{eff}$ approximation was found to reproduce the exact transition probability (that is, the exact value of $|\phi_{1s}^{Z_\mathrm{eff}}(\vec{0})|^2$) to about 1\%, and in no case did the discrepancy exceed 4\%.

The earlier work did not explore the validity of this approximation when the final nuclear state is varied, though the underlying Taylor series argument extends to this case as well.  Here we do so, explicitly calculating $Z_\mathrm{eff}$ in the manner of Eq. \eqref{eq:Zeff} for several charge- and spin-dependent nuclear responses. Starting from the isoscalar monopole response $W_{M_0}^{00}$, we can first extend to the full isoscalar charge response $W_M^{00}$, which includes all contributing higher-order even multipoles of the charge operator.  This has effectively no impact on the $Z_\mathrm{eff}$ derived for the elastic case, as the monopole operator is coherent and dominates the elastic response.  But for the inelastic cases, this procedure changes both the operator and the final state.  Thus we need to assess the degree to which a fixed $Z_\mathrm{eff}$ adequately describes the effects of the muon wave function on transition densities, for the transitions of interest in $^{27}$Al.  We do this by performing an exact calculation to determine $Z_\mathrm{eff}$ for each transition, then examining the constancy of the quantity
\begin{equation}
R \equiv {|\phi_{1s}^{Z_\mathrm{eff}}(\vec{0})|^2 \over |\phi_{1s}^{Z}(\vec{0})|^2} = \frac{Z_\mathrm{eff}^3}{Z^3},
\end{equation}
as the transition and the underlying nuclear operator are varied.

The deviation of $R$ from one represents the impact of the finite nuclear size on the convolution integral of the nuclear density and muon wave function. Numerically computed values for $R$ (or equivalently $Z_\mathrm{eff}$) are listed in Table \ref{tab:transition_params} for a variety of responses and elastic and inelastic transitions, following the example of the monopole charge operator described above. If we consider all of the inelastic results of $^{27}$Al given in Table \ref{tab:transition_params}, we find $R=0.604\pm 0.014$. This value is somewhat smaller than that found for the isoscalar elastic response, $R=0.657$, but this is expected because the operators for inelastic transitions act on valence nucleons, with transition densities peaked nearer to the nuclear surface, while the coherent monopole operator acts on all nucleons.  Thus, inelastic transitions should exhibit larger finite-size effects.  Were one to use the coherent elastic value for all transitions, the typical error in the rate would be 8\%.  

If the same exercise is repeated for $^{48}$Ti one finds $R=0.360 \pm 0.015$ for the inelastic transitions considered in Table \ref{tab:transition_params}, compared to the elastic charge response value of $R=0.433$.  There again is good consistency among the $R$'s calculated for inelastic transitions, with the typical variation being 4\%, though the central value is about 20\% smaller than that found in the coherent elastic case.  Finite-size effects are larger in Ti because the Bohr radius is smaller and nuclear radius larger, relative to Al.

The factorization of the muon wave function from the transition density is not necessary: In Ref. \cite{Haxton:2022piv} the complete elastic rate formula is given with both the upper and lower components of the muon wave function explicitly retained.  However, there are advantages to performing the factorization, both in improving the transparency of the final result and in simplifying the evaluation of matrix elements.  Specifically, if harmonic oscillator Slater determinants are employed, the nuclear matrix elements can be evaluated analytically, assuming the finite-size effects associated with the muon are handled as above.  However, the exercise just described shows that results are significantly improved if one uses a different value of $Z_\mathrm{eff}$ for the elastic response than for inelastic responses.  We make this choice here.  In terms of the effective charge, this yields $Z_\mathrm{eff}^\mathrm{el}=11.30$ for the ground-state transition and $Z_\mathrm{eff}^\mathrm{inel}=10.99$ for all inelastic transitions in $^{27}$Al.

In this work, because of Mu2e and COMET plans, the nucleus of primary interest is $^{27}$Al. In Sec. \ref{sec:other_targets}, we also briefly discuss titanium, which was used previously in the SINDRUM II experiment \cite{SINDRUMII:1993gxf} and has been discussed as a future target \cite{Mu2e-II:2022blh}.  The atomic number of Ti is 22, compared to 13 for Al.  The $Z_\mathrm{eff}$ approximation slowly deteriorates with increasing atomic number, as the muon's Bohr radius decreases.  However, in Ref. \cite{Haxton:2022piv} the procedure was tested through Cu (atomic number 29), with satisfactory results.  Unlike $^{27}$Al, Ti has multiple (five) stable isotopes.  We determine the muon binding energy and $Z_\mathrm{eff}$ for natural Ti using $^{48}$Ti, the principal isotope with an abundance of 74\%. As in $^{27}$Al, we adopt different $R$ values for elastic vs inelastic transitions, yielding $Z_\mathrm{eff}^\mathrm{el}=16.64$ for the ground-state transition and $Z_\mathrm{eff}^\mathrm{inel}=15.65$ for all inelastic transitions in Ti.

\section{Nucleon-level Effective Theory}
\label{sec:nucleon_level_eft}
Nucleons that are bound in a nucleus are only mildly relativistic, with typical velocities $v_\mathrm{avg}\approx 0.1$ (in units of $c$). We can therefore perform a nonrelativistic expansion of the nuclear charges and currents.  In the NRET, the nucleon velocity operator $\vec{v}_N$ stands
for the set of $A-1$ independent nucleon Jacobi velocities, e.g., the Galilean-invariant velocities $\vec{v_N} \equiv \{  (\vec{v}_2-\vec{v}_1)/\sqrt{2}, ~(2\vec{v}_3-(\vec{v}_1+\vec{v}_2))/\sqrt{6},~\ldots \}$, where $\vec{v}_i$ is the velocity operator for the $i$-th nucleon and $A$ is the nucleon number.  See Refs. \cite{Fitzpatrick:2012ix,Haxton:2022piv} for details. 

The single-nucleon NRET operators can then be constructed from the available Hermitian vector operators: $i\hat{q}$ where $\hat{q}$ is the velocity of the outgoing ultra-relativistic electron, the nucleon velocity operator $\vec{v}_N$, and the respective lepton and nucleon spin operators, $\vec{\sigma}_L$ and $\vec{\sigma}_N$. To first order in the nucleon velocity $\vec{v}_N$, there are 16 NRET operators that can mediate $\mu\rightarrow e$ conversion \cite{Rule:2021oxe,Haxton:2022piv} 
{\allowdisplaybreaks
\begin{align}
\mathcal{O}_1&=1_L\;1_N, \nonumber\\
\mathcal{O}_2'&=1_L\;i\hat{q}\cdot\vec{v}_N, \nonumber\\
\mathcal{O}_3&=1_L\;i\hat{q}\cdot\left[\vec{v}_N\times\vec{\sigma}_N\right], \nonumber\\
\mathcal{O}_4&=\vec{\sigma}_L\cdot\vec{\sigma}_N, \nonumber\\
\mathcal{O}_5&=\vec{\sigma}_L\cdot\left(i\hat{q}\times\vec{v}_N\right), \nonumber\\
\mathcal{O}_6&=i\hat{q}\cdot\vec{\sigma}_L\;i\hat{q}\cdot\vec{\sigma}_N, \nonumber\\
\mathcal{O}_7&=1_L\;\vec{v}_N\cdot\vec{\sigma}_N, \nonumber\\
\mathcal{O}_8&=\vec{\sigma}_L\cdot\vec{v}_N, \label{eq:basis_NRET}\\
\mathcal{O}_9&=\vec{\sigma}_L\cdot\left(i\hat{q}\times\vec{\sigma}_N\right), \nonumber\\
\mathcal{O}_{10}&=1_L\;i\hat{q}\cdot\vec{\sigma}_N, \nonumber\\
\mathcal{O}_{11}&=i\hat{q}\cdot\vec{\sigma}_L\;1_N, \nonumber\\
\mathcal{O}_{12}&=\vec{\sigma}_L\cdot\left[\vec{v}_N\times\vec{\sigma}_N\right], \nonumber\\
\mathcal{O}'_{13}&=\vec{\sigma}_L\cdot\left(i\hat{q}\times\left[\vec{v}_N\times\vec{\sigma}_N\right]\right), \nonumber\\
\mathcal{O}_{14}&=i\hat{q}\cdot\vec{\sigma}_L\;\vec{v}_N\cdot\vec{\sigma}_N, \nonumber\\
\mathcal{O}_{15}&=i\hat{q}\cdot\vec{\sigma}_L\;i\hat{q}\cdot\left[\vec{v}_N\times\vec{\sigma}_N\right], \nonumber\\
\mathcal{O}_{16}'&=i\hat{q}\cdot\vec{\sigma}_L\;i\hat{q}\cdot\vec{v}_N. \nonumber
\end{align}
}
This operator basis matches closely the one previously derived for dark matter direct detection \cite{Fitzpatrick:2012ix}; we distinguish with a prime the operators for which there are significant differences. The NRET operators $\mathcal{O}_i$ are understood to act between Pauli spinors $\xi_s$ for muon, electron, and nucleons. The NRET can be extended with the leading relativistic corrections to the muon by including the muon velocity operator $\vec{v}_\mu$ to first order \cite{Haxton:2022piv}. As discussed above, these relativistic corrections are always subleading, and we will not consider them further in this work.

We allow for different couplings to protons vs neutrons, or equivalently isoscalar and isovector couplings
\begin{equation}
    \mathcal{L}_\mathrm{NRET}= \sum_{i=1}^{16}\sum_{\tau=0,1}c_i^{\tau}\mathcal{O}_it^\tau,
\end{equation}
where $t^0=1$ is the identity operator and $t^1=\tau_3$ is the 3rd Pauli matrix. The single-nucleon effective theory is specified by the unknown $c_i$ coefficients, which we refer to as low-energy constants (LECs). Conventionally, we define a set of dimensionless NRET coefficients $\tilde{c}_i$ normalized to the weak scale by writing
\begin{equation}
    c_i\equiv \sqrt{2}G_F\tilde{c}_i,
\end{equation}
where $G_F=1.116\times 10^{-5}$ GeV$^{-2}$ is the Fermi constant. 

The nucleon-level effective theory that we have described so far is identical to that developed for the elastic process in Refs. \cite{Rule:2021oxe,Haxton:2022piv}. We have not yet made reference to any particular nuclear target, let alone a specific nuclear transition. As such, the $c_i$ coefficients are target-independent, and the same single-nucleon effective theory governs both the elastic and inelastic conversion processes.

\begin{table*}
\centering
       {\renewcommand{\arraystretch}{1.4}
\begin{tabular}{lcccccl}
\hline
\hline
Projection~~~ & ~~~Charge/Current~~~ & ~~~Operator~~~ & ~~~Range~~~ & ~~~Even J~~~ & ~~~Odd J~~~ & ~~~LECs Probed\\
\hline
Charge & $1_N$ & $M_{JM}$ & $J\geq 0$ & \textbf{E-E} & O-O & ~~~$c_1,c_{11}$\\
Charge & $\vec{v}_N\cdot\vec{\sigma}_N$ & $\tilde{\Omega}_{JM}$ & $J\geq 0$ & O-E & E-O & ~~~$c_7,c_{14}$\\
Longitudinal & $\vec{\sigma}_N$ & $\Sigma''_{JM}$ & $J\geq 0$ & O-O & \textbf{E-E} & ~~~$c_4,c_6,c_{10}$\\
Transverse magnetic & " & $\Sigma_{JM}$ & $J\geq 1$ & E-O & O-E & ~~~$c_4,c_9$\\
Transverse electric & " & $\Sigma'_{JM}$ & $J\geq 1$ & O-O & \textbf{E-E} & ~~~$c_4,c_9$\\
Longitudinal & $\vec{v}_N$ & $\tilde{\Delta}''_{JM}$ & $J\geq 0$ & E-O & O-E & ~~~$c_2,c_8,c_{16}$\\
Transverse magnetic & " & $\Delta_{JM}$ & $J\geq 1$ & O-O & \textbf{E-E} & ~~~$c_5,c_8$\\
Transverse electric & " & $\Delta'_{JM}$ & $J\geq 1$ & E-O & O-E & ~~~$c_5,c_8$\\
Longitudinal & $\vec{v}_N\times\vec{\sigma}_N$ & $\Phi''_{JM}$ & $J\geq 0$ & \textbf{E-E} & O-O & ~~~$c_3,c_{12},c_{15}$ \\
Transverse magnetic & " & $\tilde{\Phi}_{JM}$ & $J\geq 1$ & O-E & E-O & ~~~$c_{12},c_{13}$\\
Transverse electric & " & $\tilde{\Phi}'_{JM}$ & $J\geq 1$ & \textbf{E-E} & O-O & ~~~$c_{12},c_{13}$ \\
\hline
\hline
\end{tabular}
}
\caption{Characteristics of the 11 single-nucleon response functions including the charge/current projection from which they arise, their transformation properties under P-T (even E or odd O), and the LECs of the nucleon-level effective theory that are associated with each response. Based on these results, the elastic $\mu\rightarrow e$ conversion amplitude can depend only on even multipoles of $M$, $\tilde{\Phi}'$, and $\Phi''$, and odd multipoles of $\Delta$, $\Sigma'$, and $\Sigma^{''}$.}
\label{tab:multipole_symmetries}
\end{table*}

\section{Nuclear Effective Theory}
\label{sec:nuclear_eft}
Now we are ready to embed the single-nucleon effective theory described in the previous section into a particular nuclear target. The single-nucleon operators $\{1_N,\vec{v}_N\cdot\vec{\sigma}_N\}$ and $\{\vec{v}_N,\vec{\sigma}_N,\vec{v}_N\times\vec{\sigma}_N\}$ appearing in the NRET basis of Eq. \eqref{eq:basis_NRET} are treated in the impulse approximation, with $\vec{v}_N$ interpreted as the local velocity operator for bound nucleons.
\begin{widetext}
The effective Hamiltonian density can then be expressed as
\begin{equation}
\begin{split}
\mathcal{H}_\mathrm{eff}(\vec{x})&=\sqrt{\frac{E_\mathrm{CE}}{2m_e}}|\phi_{1s}^{Z_\mathrm{eff}}(\vec{0})|\frac{q_\mathrm{eff}}{q}e^{-i\vec{q}_\mathrm{eff}\cdot\vec{x}}\sum_{\tau=0,1}\left[l_0^{\tau}\sum_{i=1}^A\delta(\vec{x}-\vec{x}_i)]\right.\\
&\left.+l_0^{A\;\tau}\sum_{i=1}^A\frac{1}{2m_N}\left(-\frac{1}{i}\overleftarrow{\nabla}_i\cdot\vec{\sigma}_N(i)\delta(\vec{x}-\vec{x}_i)+\delta(\vec{x}-\vec{x}_i)\vec{\sigma}_N(i)\cdot\frac{1}{i}\overrightarrow{\nabla}\right)\right.\\
&+\vec{l}_5^{\tau}\cdot\sum_{i=1}^A\vec{\sigma}_N(i)\delta(\vec{x}-\vec{x}_i)+\vec{l}_M^{\tau}\cdot\sum_{i=1}^A\frac{1}{2m_N}\left(-\frac{1}{i}\overleftarrow{\nabla}_i\delta(\vec{x}-\vec{x}_i)+\delta(\vec{x}-\vec{x}_i)\frac{1}{i}\overrightarrow{\nabla}_i\right)\\
&\left.+\vec{l}_E^{\tau}\cdot\sum_{i=1}^A\frac{1}{2m_N}\left(\overleftarrow{\nabla}_i\times\vec{\sigma}_N(i)\delta(\vec{x}-\vec{x}_i)+\delta(\vec{x}-\vec{x}_i)\vec{\sigma}_N(i)\times\overrightarrow{\nabla}_i\right)\right]_{int} t^{\tau}(i),
\end{split}
\label{eq:H_NRET}
\end{equation}
\end{widetext}
where we have introduced the leptonic charges/currents 
\begin{align}
l_0^{\tau}&\equiv c_1^{\tau}1_L+c_{11}^{\tau}i\hat{q}\cdot\vec{\sigma}_L, \nonumber\\
l_0^{A\;\tau}&\equiv c_7^{\tau}1_L+c_{14}^{\tau}i\hat{q}\cdot\vec{\sigma}_L, \nonumber\\
\vec{l}_5^{\tau}&\equiv c_4^{\tau}\vec{\sigma}_L+c_6^{\tau}i\hat{q}\cdot\vec{\sigma}_Li\hat{q}-c_9^{\tau}i\hat{q}\times\vec{\sigma}_L+c_{10}^{\tau}i\hat{q}1_L, \label{eq:leptonic_currents} \\
\vec{l}_M^{\tau}&\equiv c_2^{\tau}i\hat{q}1_L-c_5^{\tau}i\hat{q}\times\vec{\sigma}_L+c_8^{\tau}\vec{\sigma}_L+c_{16}^{\tau}i\hat{q}\cdot\vec{\sigma}_Li\hat{q}, \nonumber\\
\vec{l}_E^{\tau}&\equiv -c_3^{\tau}\hat{q}1_L+c_{12}^{\tau}i\vec{\sigma}_L+c_{13}^{\tau}\hat{q}\times\vec{\sigma}_L-ic_{15}^{\tau}\hat{q}\cdot\vec{\sigma}_L\hat{q}. \nonumber
\end{align}
The subscript $int$ in Eq. \eqref{eq:H_NRET} indicates that all nuclear operators are understood as acting on \textit{intrinsic} nuclear coordinates, i.e. the Galilean-invariant Jacobi coordinates.  As described in the previous section, the muon wave function has been replaced in transitions by its average value, and Coulomb distortions of the electron have been encoded in $q_\mathrm{eff}$, yielding operators that can be expanded in spherical waves.

A multipole expansion of the nuclear charges and currents is performed by expanding the plane-wave $\exp(-i\vec{q}_\mathrm{eff}\cdot\vec{x})$ in Eq. \eqref{eq:H_NRET} in spherical waves, then employing standard spherical harmonic and vector spherical harmonic techniques. Each of the two charge operators generates a family of multipole operators (indexed by multipolarity $J$), and each of the three spatial currents has transverse-magnetic, transverse-electric, and longitudinal projections that each yield an independent set of multipole operators. Thus, we expect to find a total of 11 independent nuclear responses. Adding a label to denote isospin, the single-nucleon multipole operators are
{\allowdisplaybreaks
\begin{widetext}
\begin{align}
M_{JM;\tau}(q)&\equiv \sum_{i=1}^AM_{JM}(q\vec{x}_i)\;t^{\tau}(i), \nonumber\\
\Omega_{JM;\tau}(q)&\equiv \sum_{i=1}^AM_{JM}(q\vec{x}_i)\vec{\sigma}_N(i)\cdot\frac{1}{q}\vec{\nabla}_i\;t^{\tau}(i), \nonumber\\
\Delta_{JM;\tau}(q)&\equiv \sum_{i=1}^A \vec{M}^M_{JJ}(q\vec{x}_i)\cdot\frac{1}{q}\vec{\nabla}_i\;t^{\tau}(i), \nonumber\\
\Delta'_{JM;\tau}(q)&\equiv -i\sum_{i=1}^A \left\{\frac{1}{q}\vec{\nabla}_i\times\vec{M}^M_{JJ}(q\vec{x}_i)\right\}\cdot\frac{1}{q}\vec{\nabla}_i\;t^{\tau}(i) \nonumber\\
&= \sum_{i=1}^A\left[-\sqrt{\frac{J}{2J+1}}\vec{M}_{JJ+1}^M(q\vec{x}_i)+\sqrt{\frac{J+1}{2J+1}}\vec{M}_{JJ-1}^M(q\vec{x}_i)\right]\cdot \frac{1}{q}\vec{\nabla}_i~t^\tau (i), \nonumber\\
\Delta''_{JM;\tau}(q)&\equiv \sum_{i=1}^A\left(\frac{1}{q}\vec{\nabla}_iM_{JM}(q\vec{x}_i)\right)\cdot\frac{1}{q}\vec{\nabla}_i\;t^{\tau}(i) \nonumber\\
&= \sum_{i=1}^A\left[\sqrt{\frac{J+1}{2J+1}}\vec{M}_{JJ+1}^M(q\vec{x}_i)+\sqrt{\frac{J}{2J+1}}\vec{M}_{JJ-1}^M(q\vec{x}_i)\right]\cdot \frac{1}{q}\vec{\nabla}_i~t^\tau (i), \nonumber\\
\Sigma_{JM;\tau}(q)&\equiv \sum_{i=1}^A\vec{M}^M_{JJ}(q\vec{x}_i)\cdot\vec{\sigma}_N(i)\;t^{\tau}(i), \label{eq:single_nucleon_responses}\\
\Sigma'_{JM;\tau}(q)&\equiv -i\sum_{i=1}^A\left\{\frac{1}{q}\vec{\nabla}_i\times\vec{M}^M_{JJ}(q\vec{x}_i)\right\}\cdot\vec{\sigma}_N(i)\;t^{\tau}(i) \nonumber\\
&= \sum_{i=1}^A\left[-\sqrt{\frac{J}{2J+1}}\vec{M}_{JJ+1}^M(q\vec{x}_i)+\sqrt{\frac{J+1}{2J+1}}\vec{M}_{JJ-1}^M(q\vec{x}_i)\right]\cdot \vec{\sigma}_N(i)~t^\tau (i), \nonumber\\
\Sigma''_{JM;\tau}(q)&\equiv \sum_{i=1}^A\left\{\frac{1}{q}\vec{\nabla}_iM_{JM}(q\vec{x}_i)\right\}\cdot\vec{\sigma}_N(i)\;t^{\tau}(i) \nonumber\\
&= \sum_{i=1}^A\left[\sqrt{\frac{J+1}{2J+1}}\vec{M}_{JJ+1}^M(q\vec{x}_i)+\sqrt{\frac{J}{2J+1}}\vec{M}_{JJ-1}^M(q\vec{x}_i)\right]\cdot \vec{\sigma}_N(i)~t^\tau (i), \nonumber\\
\Phi_{JM;\tau}(q)&\equiv i\sum_{i=1}^A\vec{M}^M_{JJ}(q\vec{x}_i)\cdot\left(\vec{\sigma}_N(i)\times\frac{1}{q}\vec{\nabla}_i\right)\;t^{\tau}(i), \nonumber\\
\Phi'_{JM;\tau}(q)&\equiv\sum_{i=1}^A\left(\frac{1}{q}\vec{\nabla}_i\times\vec{M}^M_{JJ}(q\vec{x}_i)\right)\cdot\left(\vec{\sigma}_N(i)\times\frac{1}{q}\vec{\nabla}_i\right)\;t^{\tau}(i) \nonumber\\
&= \sum_{i=1}^A\left[-\sqrt{\frac{J}{2J+1}}\vec{M}_{JJ+1}^M(q\vec{x}_i)+\sqrt{\frac{J+1}{2J+1}}\vec{M}_{JJ-1}^M(q\vec{x}_i)\right]\cdot \left(\vec{\sigma}_N(i)\times \frac{1}{q}\vec{\nabla}_i\right)~t^\tau (i), \nonumber\\
\Phi''_{JM;\tau}(q)&\equiv i\sum_{i=1}^A\left(\frac{1}{q}\vec{\nabla}_iM_{JM}(q\vec{x}_i)\right)\cdot\left(\vec{\sigma}_N(i)\times\frac{1}{q}\vec{\nabla}_i\right)\;t^{\tau}(i) \nonumber\\
&= \sum_{i=1}^A\left[\sqrt{\frac{J+1}{2J+1}}\vec{M}_{JJ+1}^M(q\vec{x}_i)+\sqrt{\frac{J}{2J+1}}\vec{M}_{JJ-1}^M(q\vec{x}_i)\right]\cdot \left(\vec{\sigma}_N(i)\times \frac{1}{q}\vec{\nabla}_i\right)~t^\tau (i), \nonumber
\end{align}
\end{widetext}
}
where we have used the multipole projections
\begin{equation}
\begin{split}
M_{JM}(q\vec{x})&\equiv j_J(qx)Y_{JM}(\hat{x}),\\
\vec{M}^M_{JL}(q\vec{x})&\equiv j_L(qx)\vec{Y}_{JLM}(\hat{x}),
\end{split}
\end{equation}
and $\vec{Y}_{JLM}$ is a vector spherical harmonic.

In addition to carrying angular momentum $(J,M)$, each multipole operator has a well-defined transformation under parity $\vec{x}\rightarrow -\vec{x}$. The operators $M$, $\Delta'$, $\Delta''$, $\Sigma$, $\Phi'$, and $\Phi''$ are \textit{normal-parity} operators that transform with a phase $(-1)^J$ under parity whereas $\Omega$, $\Delta$, $\Sigma'$, $\Sigma''$, and $\Phi$ are \textit{abnormal-parity} operators that transform with a phase $(-1)^{J+1}$. 

Matrix elements of the above multipole operators evaluated between single-particle harmonic oscillator states can be expressed analytically in terms of the dimensionless quantity $y=(qb/2)^2$, where $b$ is the oscillator length scale. In particular, letting $T_{JM}(q\vec{r}\,)$ represent any of the 11 single-particle operators, we have
\begin{equation}
\begin{split}
&\braket{n'\left(\ell'\;1/2\right)j'||T_J(q\vec{r}\,)||n\left(\ell\;1/2\right)j}\\
&~~~~~~~~~~~~~~~~~~~~~~~=\frac{1}{\sqrt{4\pi}}y^{(J-K)/2}e^{-y}p(y),
\end{split}
\label{eq:multipole_polynomial}
\end{equation}
where $K=(1)2$ for (ab)normal parity operators and $p(y)$ is a finite-degree polynomial in $y$. For the choices of phase conventions in our definitions, all of the matrix elements are real. We also redefine certain operators,
relative to their conventional forms \cite{Donnelly:1979ezn}, so that all of the operators transform simply under time-reversal,
\begin{align}
\tilde{\Omega}_{JM}(q)&\equiv\Omega_{JM}(q)+\frac{1}{2}\Sigma''_{JM}(q), \nonumber\\
\tilde{\Delta}''_{JM}(q)&\equiv\Delta''_{JM}(q)-\frac{1}{2}M_{JM}(q), \nonumber\\
\tilde{\Phi}_{JM}(q)&\equiv \Phi_{JM}(q)-\frac{1}{2}\Sigma'_{JM}(q),\\
\tilde{\Phi}'_{JM}(q)&\equiv \Phi'_{JM}(q)+\frac{1}{2}\Sigma_{JM}(q). \nonumber
\end{align}
With these changes, the single-particle matrix elements of all operators transform simply under the exchange of initial and final states
\begin{equation}
\begin{split}
&\braket{n\left(\ell\;1/2\right)j||T_J(q\vec{r}\,)||n'\left(\ell'\;1/2\right)j'}\\
&~~~~~~=(-1)^{\lambda}\braket{n'\left(\ell'\;1/2\right)j'||T_J(q\vec{r}\,)||n\left(\ell\;1/2\right)j},
\end{split}
\end{equation}
with $\lambda=j'-j$ for the operators $M$, $\Delta$, $\Sigma'$, $\Sigma''$, $\tilde{\Phi}'$, and $\Phi''$, and $\lambda=j'+j$ for the operators $\tilde{\Omega}$, $\Delta'$, $\tilde{\Delta}''$, $\Sigma$, and $\tilde{\Phi}$. The properties of the nuclear response functions are summarized in Table \ref{tab:multipole_symmetries}.

Having introduced the single-nucleon response functions, we can now proceed to write the most general form of the rate for inelastic $\mu\rightarrow e$ conversion $\Gamma_{\mu e}(gs\rightarrow f)$. Details of the derivation are given in Appendix \ref{app:rate_derivation}. The resulting factorized form is 
\begin{widetext}
\begin{equation}
\begin{split}
\Gamma_{\mu e}&(gs\rightarrow f)=\frac{G_F^2}{\pi}\frac{q_\mathrm{eff}^2}{1+\frac{q}{M_T}}|\phi_{1s}^{Z_\mathrm{eff}}(\vec{0})|^2\sum_{\tau=0,1}\sum_{\tau'=0,1} \\
&\Bigg\{\tilde{R}_M^{\tau\tau'}W_M^{\tau\tau'}(q_\mathrm{eff})
+\tilde{R}_{\Sigma''}^{\tau\tau'}W_{\Sigma''}^{\tau\tau'}(q_\mathrm{eff})+\tilde{R}_{\Sigma'}^{\tau\tau'}\left(W_{\Sigma'}^{\tau\tau'}(q_\mathrm{eff})+W_{\Sigma}^{\tau\tau'}(q_\mathrm{eff})\right)\\
&+\frac{q_\mathrm{eff}^2}{m_N^2}\Bigg[\tilde{R}_{\tilde{\Delta}''}^{\tau\tau'}W_{\tilde{\Delta}''}^{\tau\tau'}(q_\mathrm{eff})+\tilde{R}_{\tilde{\Omega}}^{\tau\tau'}W_{\tilde{\Omega}}^{\tau\tau'}(q_\mathrm{eff})+\tilde{R}_{\Phi''}^{\tau\tau'}W_{\Phi''}^{\tau\tau'}(q_\mathrm{eff})+2\tilde{R}_{\tilde{\Delta}'' \Phi'' }^{\tau\tau'}W^{\tau\tau'}_{\tilde{\Delta}''\Phi''}(q_\mathrm{eff})\\
&~~~~~~~+\tilde{R}_{\Delta'}^{\tau\tau'}\left(W_{\Delta'}^{\tau\tau'}(q_\mathrm{eff})+W_{\Delta}^{\tau\tau'}(q_\mathrm{eff})\right)+ \tilde{R}_{\tilde{\Phi}'}^{\tau\tau'}\left(W_{\tilde{\Phi}'}^{\tau\tau'}(q_\mathrm{eff})+W_{\tilde{\Phi}}^{\tau\tau'}(q_\mathrm{eff})\right)+2\tilde{R}^{\tau\tau'}_{\Delta\tilde{\Phi}}\left(W^{\tau\tau'}_{\Delta\tilde{\Phi}}(q_\mathrm{eff})+W^{\tau\tau'}_{\Delta'\tilde{\Phi}'}(q_\mathrm{eff})\right)\Bigg]\\
&-\frac{2q_\mathrm{eff}}{m_N}\Bigg[\tilde{R}^{\tau\tau'}_{\tilde{\Delta}'' M}W^{\tau\tau'}_{\tilde{\Delta}'' M}(q_\mathrm{eff})+\tilde{R}_{ \Phi'' M}^{\tau\tau'}W_{ \Phi'' M}^{\tau\tau'}(q_\mathrm{eff})+\tilde{R}^{\tau\tau'}_{\tilde{\Omega} \Sigma''}W^{\tau\tau'}_{\tilde{\Omega} \Sigma''}(q_\mathrm{eff}) \\
&~~~~~~~~+\tilde{R}_{\Delta \Sigma' }^{\tau\tau'}\left(W_{\Delta \Sigma' }^{\tau\tau'}(q_\mathrm{eff})-W_{\Delta' \Sigma }^{\tau\tau'}(q_\mathrm{eff})\right)+\tilde{R}^{\tau\tau'}_{\tilde{\Phi}' \Sigma}\left(W^{\tau\tau'}_{\tilde{\Phi}' \Sigma}(q_\mathrm{eff})-W^{\tau\tau'}_{\tilde{\Phi} \Sigma'}(q_\mathrm{eff})\right)\Bigg] \Bigg\}.
\end{split}
\label{eq:inelastic_rate}
\end{equation}
\end{widetext}

All of the information about CLFV is encoded in the leptonic response functions $\tilde{R}^{\tau\tau'}_i$. There are 8 independent direct terms
{\allowdisplaybreaks
\begin{align}
\tilde{R}_M^{\tau\tau'}&\equiv \tilde{c}_1^{\tau}\tilde{c}_1^{\tau'*}+\tilde{c}_{11}^{\tau}\tilde{c}_{11}^{\tau'*}, \nonumber\\
\tilde{R}_{\tilde{\Omega}}^{\tau\tau'}&\equiv \tilde{c}_7^{\tau}\tilde{c}_7^{\tau'*}+\tilde{c}_{14}^{\tau}\tilde{c}_{14}^{\tau'*}, \nonumber\\
\tilde{R}_{\Sigma}^{\tau\tau'}=\tilde{R}_{\Sigma'}^{\tau\tau'}&\equiv \tilde{c}_4^{\tau}\tilde{c}_4^{\tau'*}+\tilde{c}_9^{\tau}\tilde{c}_9^{\tau'*}, \nonumber\\
\tilde{R}_{\Sigma''}^{\tau\tau'}&\equiv(\tilde{c}_4^{\tau}-\tilde{c}_6^{\tau})(\tilde{c}_4^{\tau'*}-\tilde{c}_6^{\tau'*})+\tilde{c}_{10}^{\tau}\tilde{c}_{10}^{\tau'*}, \label{eq:R_CLFV_direct}\\
\tilde{R}_{\Phi''}^{\tau\tau'}&\equiv \tilde{c}_3^{\tau}\tilde{c}_3^{\tau'*}+(\tilde{c}_{12}^{\tau}-\tilde{c}_{15}^{\tau})(\tilde{c}_{12}^{\tau'*}-\tilde{c}_{15}^{\tau'*}), \nonumber\\
\tilde{R}_{\tilde{\Phi}}^{\tau\tau'}=\tilde{R}_{\tilde{\Phi}'}^{\tau\tau'}&\equiv \tilde{c}_{12}^{\tau}\tilde{c}_{12}^{\tau'*}+\tilde{c}_{13}^{\tau}\tilde{c}_{13}^{\tau'*}, \nonumber\\
\tilde{R}_{\tilde{\Delta}''}^{\tau\tau'}&\equiv \tilde{c}_2^{\tau}\tilde{c}_2^{\tau'*}+(\tilde{c}_8^{\tau}-\tilde{c}_{16}^{\tau})(\tilde{c}_8^{\tau'*}-\tilde{c}_{16}^{\tau'*}), \nonumber\\
\tilde{R}_{\Delta}^{\tau\tau'}=\tilde{R}_{\Delta'}^{\tau\tau'}&\equiv \tilde{c}_5^{\tau}\tilde{c}_5^{\tau'*}+\tilde{c}_8^{\tau}\tilde{c}_8^{\tau'*},\nonumber
\end{align}
and 7 independent interference terms 
\begin{align}
    \tilde{R}_{ \Phi'' M}^{\tau\tau'}&\equiv \mathrm{Re}\left[\tilde{c}_3^{\tau}\tilde{c}_1^{\tau'*}-(\tilde{c}_{12}^{\tau}-\tilde{c}_{15}^{\tau})\tilde{c}_{11}^{\tau'*}\right], \nonumber\\
\tilde{R}_{ \tilde{\Delta}'' M}^{\tau\tau'}&\equiv\mathrm{Im}\left[\tilde{c}_2^{\tau}\tilde{c}_1^{\tau'*}-(\tilde{c}_8^{\tau}-    \tilde{c}_{16}^{\tau})\tilde{c}_{11}^{\tau'*}\right], \nonumber\\
\tilde{R}_{ \Delta' \Sigma}^{\tau\tau'}=R_{ \Delta \Sigma'}^{\tau\tau'}&\equiv \mathrm{Re}\left[\tilde{c}_5^{\tau}\tilde{c}_4^{\tau'*}+\tilde{c}_8^{\tau}\tilde{c}_9^{\tau'*}\right], \nonumber\\
\tilde{R}_{\tilde{\Phi} \Sigma'}^{\tau\tau'}=\tilde{R}_{\tilde{\Phi}' \Sigma}^{\tau\tau'}&\equiv\mathrm{Im}\left[\tilde{c}_{13}^{\tau}\tilde{c}_{4}^{\tau'*}+\tilde{c}_{12}^{\tau}\tilde{c}_{9}^{\tau'*}\right], \label{eq:R_CLFV_int}\\
\tilde{R}_{\tilde{\Omega} \Sigma''}^{\tau\tau'}&\equiv\mathrm{Im}\left[\tilde{c}_{7}^{\tau}\tilde{c}_{10}^{\tau'*}-\tilde{c}_{14}^{\tau}(\tilde{c}_4^{\tau'*}-\tilde{c}_6^{\tau'*})\right],\nonumber \\
\tilde{R}_{\tilde{\Delta}'' \Phi''}^{\tau\tau'}&\equiv\mathrm{Im}\left[\tilde{c}_2^{\tau}\tilde{c}_3^{\tau'*}+(\tilde{c}_{8}^{\tau}-\tilde{c}_{16}^{\tau})(\tilde{c}_{12}^{\tau'*}-\tilde{c}_{15}^{\tau'*})\right], \nonumber\\
\tilde{R}_{\Delta\tilde{\Phi}}^{\tau\tau'}=\tilde{R}_{\Delta'\tilde{\Phi}'}^{\tau\tau'}&\equiv\mathrm{Im}\left[\tilde{c}_5^{\tau}\tilde{c}_{13}^{\tau'*}+\tilde{c}_8^{\tau}\tilde{c}_{12}^{\tau'*}\right].  \nonumber
\end{align}
}
The notation $\tilde{R}$ indicates that the leptonic responses have been expressed in terms of the dimensionless NRET LECs $\tilde{c}_i$. The leptonic response functions, as bilinear combinations of the single-nucleon NRET LECs, are target-independent. Thus, by studying elastic $\mu \rightarrow e$ conversion in a variety of targets with complementary properties, one can study the LECs while changing the response-function nuclear physics, obtaining new information.  Similarly, as they are also independent of the nuclear transition being studied, inelastic transitions also provide new information, complementing the constraints obtained from elastic conversion.

All of the nuclear physics relevant to $\mu\rightarrow e$ conversion has been factored into the nuclear response functions $W_i^{\tau\tau'}(q_\mathrm{eff})$, which are dimensionless, squared matrix elements of the various operators introduced in Eq. \eqref{eq:single_nucleon_responses}. As indicated, each nuclear response is evaluated at the effective momentum $q_\mathrm{eff}$, determined by Eq. \eqref{eq:q_eff}. The precise definitions of the $W_i^{\tau\tau'}$ are provided in Appendix \ref{app:rate_derivation}. The nuclear responses are entirely determined by standard-model physics. Their numerical values, which can be calculated using the nuclear shell model, are strongly dependent on the nuclear target and transition under consideration.

For a fixed transition, each nuclear response function is composed of a summation over either even or odd multipoles, depending on the parity of the underlying operator and the nuclear transition under consideration. For example, the three low-lying excited states of $^{27}$Al that we consider in this work are positive-parity states, as is the ground state. So, only multipoles that conserve parity can contribute to these transitions. Higher up in the spectrum of Al, there are odd-parity states that would connect to the ground state only through parity-violating multipoles. The constraint of parity also determines the interference terms that can contribute to the conversion rate. For example, $\Delta$ and $\Sigma$ are both transverse-magnetic projections, but they do not interfere because for fixed $J$ they connect to states of opposite parity.

Eq. \eqref{eq:inelastic_rate} is a generalization of the elastic $\mu\rightarrow e$ conversion rate originally derived in Refs. \cite{Rule:2021oxe,Haxton:2022piv}. To recover the elastic rate, Eq. (59) of Ref. \cite{Haxton:2022piv}, one takes $f$ to be the nuclear ground state. In this case, the surviving nuclear responses are those that simultaneously conserve parity and time-reversal --- $M$, $\Sigma'$, $\Sigma''$, $\Delta$, $\tilde{\Phi}'$, and $\Phi''$. Six of the direct responses and two of the interference terms involve only these operators, and thus are nonzero.  The NRET coefficients $c_2$, $c_7$, $c_{14}$, and $c_{16}$ do not appear in the surviving response functions $\tilde{R}_i^{\tau\tau'}$ and thus can only be probed in inelastic conversion.

The responses $\Sigma$, $\Delta'$, and $\tilde{\Phi}$ --- which, by virtue of their P and T transformation properties, contribute only to excited-state transitions --- share the same leptonic response functions as their elastic counterparts $\Sigma'$, $\Delta$, and $\tilde{\Phi}'$. In other words, transverse-magnetic and transverse-electric projections of the same nuclear current share the same leptonic response functions, $\tilde{R}_{O}^{\tau\tau'}=\tilde{R}_{O'}^{\tau\tau'}$ for $O=\Sigma,\Delta,\tilde{\Phi}$. Therefore, although $\Sigma$, $\Delta'$, and $\tilde{\Phi}$ contribute only to excited-state transitions, they are always accompanied by operators that can, in principle, contribute to the ground-state process. In contrast, the operators $\tilde{\Omega}$ and $\tilde{\Delta}''$ are purely inelastic, and through their respective response functions $\tilde{R}_{\tilde{\Omega}}^{\tau\tau'}$ and $\tilde{R}_{\tilde{\Delta}''}^{\tau\tau'}$ they provide sensitivity to the NRET LECs $(c_7,c_{14})$ and $(c_2,c_{16})$, which are not probed by elastic $\mu\rightarrow e$ conversion. 

The previous paragraph could be interpreted to mean that the inelastic responses $\Sigma$, $\Delta'$, and $\tilde{\Phi}$ do not provide any CLFV information that is not already available from the elastic process, since they share the same CLFV response functions as their elastic counterparts. In fact, this is a target-dependent statement that is only true if the nuclear ground state permits the elastic response in question. In each electric/magnetic pair, one response is generated by a normal parity operator, the other by an abnormal parity one. This can lead to situations where nuclear selection rules allow one response but not the other to contribute to a particular transition. 

For example, if the nuclear ground state carries angular momentum $J^\pi=0^+$, then the only response functions that can contribute to the ground-state process are $M$ and $\Phi''$. If the same nucleus were to transition to an excited state with $J^\pi =2^+$, then one would gain sensitivity to additional CLFV coefficients that, although they could be measured in ground-state transitions in other nuclear targets, do not contribute to the elastic process in the nucleus under consideration. In particular, the coefficient $\tilde{R}^{\tau\tau'}_{\Sigma}=\tilde{R}^{\tau\tau'}_{\Sigma'}$ could be measured in the $0^+\rightarrow 2^+$ transition due to contributions from the normal-parity transverse-magnetic multipole operator $\Sigma_2$. The abnormal parity transverse-electric response $\Sigma'$ would not contribute to either the $0^+\rightarrow 0^+$ elastic process or the $0^+\rightarrow 2^+$ inelastic transition.

Five of the interference terms in Eq. \eqref{eq:R_CLFV_int} are unique to the inelastic case and depend on the NRET LECs containing an imaginary component. In general, the $c_i$ coefficients are complex (see Sec 4.3 of Ref. \cite{Haxton:2024lyc} for a detailed discussion of this point). Like the direct contributions, the interference terms are symmetric under the exchange of electric and magnetic components, leading to several redundancies in the CLFV response functions (e.g., $\tilde{R}_{\Sigma\Delta'}^{\tau\tau'}=\tilde{R}_{\Sigma'\Delta}^{\tau\tau'}$).\\
~\\
{\it Current conservation constraints:}  We have not assumed in our multipole operator development that the vector current is conserved.  However, applications of the formalism could include those utilizing the standard model's conserved vector current, e.g., 
when the lepton-nucleus interaction is mediated by exchange of a photon.  In such cases, it can be advantageous to use alternative forms of certain operators.

First consider the longitudinal multipoles $\tilde{\Delta}^{\prime \prime}$, $\Sigma^{\prime \prime}$, and $\Phi^{\prime \prime}$ associated, respectively, with the impulse-approximation velocity (or convection), spin, and spin-velocity currents.  The spin is an
axial vector, while the spin-velocity current is not generated by the standard model (to linear order in velocity). Consequently, the relevant operator is $\tilde{\Delta}^{\prime \prime}$, which by partial integration can be rewritten as
\begin{eqnarray}
    \tilde{\Delta}_{JM}^{\prime \prime}(q) &=& {i m_N \over q^2} \int d \vec{x} \left[ \vec{\nabla} M_{JM}(q \vec{x})\right] \cdot \vec{j}_c(\vec{x}) \nonumber \\
    &\rightarrow&  {i m_N \over q^2} \int d \vec{x} \left[ \vec{\nabla} M_{JM}(q \vec{x})\right] \cdot \vec{j}(\vec{x}) \nonumber \\
     &=&  -{i m_N \over q^2} \int d \vec{x} \, M_{JM}(q \vec{x}) \, \vec{\nabla}  \cdot \vec{j}(\vec{x}).
     \label{eq:cc}
\end{eqnarray}
In the first line above, $\vec{j}_c(\vec{x})$ is the impulse-approximation form of the convection current,
\begin{equation}
  \vec{j}_c(\vec{x}) =  \sum_{i=1}^A {1 \over2 i\, m_N}\left[ - \overleftarrow{\nabla}_i \delta(\vec{x}-\vec{x}_i) +\delta(\vec{x}-\vec{x}_i) \overrightarrow{\nabla}_i \right] ,
\end{equation}
while in the second line it has been replaced by the full nuclear vector current  $\vec{j}(\vec{x})$.
The full current would include additional one-body contributions, for example, magnetization current $i \, \vec{q}/m_N \times \vec{\sigma}_N$, which is transverse and thus
makes no contribution. 
But it would also include all of the corrections beyond the impulse approximation, such as two-nucleon currents.  The last line in Eq. (\ref{eq:cc}) is the general form of $\tilde{\Delta}_{JM}^{\prime \prime}$,
prior to specializing to the impulse approximation.

For a conserved current, the continuity equation yields
\[ \vec{\nabla} \cdot \vec{j}(\vec{x}) = -i \left[H, \rho(\vec{x}) \right], \]
where $H$ is the nuclear Hamiltonian and $\rho$ the vector charge.  Making this replacement yields
\begin{eqnarray} 
 \langle j_f | \tilde{\Delta}''_{JM}(q) | j_i  \rangle &=& -{m_N \over q^2} (E_f-E_i) \langle j_f |M_{JM} (q)| j_i \rangle  \nonumber \\
 &=& \frac{m_N q_0}{q^2} ~\langle j_f | M_{JM}(q) |j_i \rangle .
 \label{eq:siegert}
 \end{eqnarray}
 This result can be substituted into the expressions for the nuclear response functions 
 given in Eqs. \eqref{eq:W_responses_PC} and \eqref{eq:W_responses_PV}, eliminating $\tilde{\Delta}''_J$.

 As we are working in the impulse approximation, what is the advantage of this rewriting?  Both the
 one-body contribution and two-body corrections to the vector current $\vec{j}(\vec{x})$ are $O(\vec{v}_N)$ (although the former
 is enhanced by the large isovector magnetic moment $ \mu_{T=1}\approx 4.7$).  In contrast, the one-body contribution to the charge operator is $O(1)$ while the two-body corrections are $O(\vec{v}_N^{\,2})$.  By using $M_{JM}$, results become
 much less sensitive to such corrections.

 Similar steps can be taken with the transverse electric response, where the convection current 
 contribution enters through $\Delta^\prime_{JM}$.  All contributions of the convection current that are constrained by current
 conservation can be identified and eliminated in terms of the charge operator, through the generalized Siegert's theorem \cite{PhysRevC.29.1645}.  There are several possible re-writings of the resulting operator that are all equivalent if used in many-body
 calculations that respect current conservation: Explicit forms can be found in \cite{PhysRevC.31.2027}.  However, forms that remain well behaved at high $q$, while generating important corrections
 at low $q$, produce quite small corrections once the momentum transfer reaches $q \approx 100$ MeV \cite{PhysRevC.98.065505}.  Consequently, for the present application, there is no motivation for using more complicated expressions, in preference to the simpler transverse electric operator $\Delta^\prime_{JM}$ employed here.

The importance of using the full current operator (for a conserved current) is connected with the
choice we have made to describe the $^{27}$Al states in the $2s$-$1d$ shell model.  By partially
integrating the first line of Eq. (\ref{eq:cc}) one can show
\begin{eqnarray}
    &&\braket{n'\left(\ell'\;1/2\right)j'||\tilde{\Delta}''_J(y)||n\left(\ell\;1/2\right)j} \nonumber \\
    &&={m_N \over q^2} (N-N^\prime) \omega \braket{n'\left(\ell'\;1/2\right)j'||M_J(y)|| n\left(\ell\;1/2\right)j} \nonumber \\
    &&= {1 \over 4 y} (N-N^\prime) \braket{n'\left(\ell'\;1/2\right)j'||M_J(y)|| n\left(\ell\;1/2\right)j},
    \label{eq:delta_HO_mat}
\end{eqnarray}
where $\omega=1/m_Nb^2$ is the oscillator frequency, and $N \equiv 2(n-1)+\ell$ and $N^\prime$ are the  principal oscillator quantum numbers associated with oscillator shell energies $(N+3/2) \hbar \omega$. 
Consequently all impulse-approximation matrix elements of $\tilde{\Delta}^{\prime \prime}$ vanish 
for our shell-model description of $^{27}$Al as $N=N^\prime$ (and similarly for any other target described in a basis of $0\hbar \omega$ shell-model configurations).

We conclude that for such shell-model calculations, transition matrix elements of $\tilde{\Delta}_J^{\prime \prime}$ 
are generated entirely by corrections to the impulse approximation.  By assuming a conserved 
vector current and exploiting the continuity equation, we can take account of these corrections ---
which we have previously noted are of the same order as the one-body contribution, in an expansion
in the nucleon velocity.  In our numerical results below, we will always assume that the nuclear convective current is conserved and apply Eq. \eqref{eq:siegert}, thus avoiding the direct evaluation of $\tilde{\Delta}''$. We again caution the reader that the extent to which current conservation applies is dependent on the underlying CLFV interaction.

\section{Response Function Properties}
\label{sec:response_props}
Having derived the most general expression for the inelastic $\mu\rightarrow e$ conversion rate, Eq. \eqref{eq:inelastic_rate}, we can now calculate the ratio in Eq. \eqref{eq:inelastic_branching}, which determines the relative importance of excited-state contributions. Before undertaking this task, we will discuss a few properties of the nuclear response functions that will aid in our interpretation and understanding of the inelastic process. The explicit
forms of these response functions --- for the phase conventions used here, which produce real nuclear matrix elements --- can be found in Appendix \ref{app:rate_derivation}.

The operators $M$ and $\{\Sigma,\Sigma',\Sigma''\}$ arise, respectively, from couplings to nuclear charge and spin --- macroscopic quantities that are non-vanishing for a point-like nucleus. The remaining responses --- involving the operators $\tilde{\Omega}$, $\Delta$, $\Delta'$, $\tilde{\Delta}''$, $\tilde{\Phi}$, $\tilde{\Phi}'$, and $\Phi''$ --- depend explicitly on $\vec{v}_N$, defined as the set of Galilean-invariant inter-nucleon velocities of the bound nucleons, and thus these operators arise from the composite structure of the nucleus. In the rate formula of Eq. \eqref{eq:inelastic_rate}, their nuclear matrix elements are accompanied by a factor of $q_\mathrm{eff}/m_N$, which vanishes in the long-wavelength (point-like) limit, $q_\mathrm{eff}\rightarrow 0$.

By considering the long-wavelength form of the various multipole operators, we can make several observations that remain true at finite momentum transfer. We begin by discussing only those operators that contribute to elastic $\mu\rightarrow e$ conversion. The small-$q$ forms of the three velocity-independent multipole operators are
\begin{equation}
    \begin{split}
        M_{00}(0)&=\frac{1}{\sqrt{4\pi}}\sum_{i=1}^A1_N(i),\\
        \Sigma'_{1M}(0)&=\frac{1}{\sqrt{6\pi}}\sum_{i=1}^A\sigma_{1M}(i),\\
        \Sigma''_{1M}(0)&=\frac{1}{\sqrt{12\pi}}\sum_{i=1}^A\sigma_{1M}(i).
    \end{split}
\end{equation}
These three operators survive in the point-nucleus limit and are parity even, and thus contribute to elastic scattering. The two spin operators can also drive inelastic $\mu\rightarrow e$ conversion. For isoscalar coupling, $M_{00}(0)$ is coherent, with a matrix element proportional to the total nucleon number $A$.  In contrast, the isovector charge monopole and the two spin operators act only on the single unpaired valence nucleon in $^{27}$Al.  Of course, for the momentum transfers $\approx m_\mu$ relevant for $\mu \rightarrow e$ conversion, the $A^2$ rate coherence from isoscalar $M_{00}(0)$ is reduced somewhat by the elastic form factor, an effect whose fractional impact grows with $A$.

The threshold forms of the velocity-dependent operators that contribute to elastic scattering are
\begin{equation}
    \begin{split}
        \Delta_{1M}(0)&=-\frac{1}{\sqrt{24\pi}}\sum_{i=1}^A\ell_{1M}(i),\\
        \tilde{\Phi}'_{2M}(0)&= -\frac{1}{\sqrt{20\pi}}\sum_{i=1}^A\left[\vec{x}_i\otimes\left(\vec{\sigma}_N(i)\times\frac{1}{i}\vec{\nabla}_i\right)_1\right]_{2M} ,\\
       \Phi''_{JM}(0)&=- \frac{\delta_{J0}}{6\sqrt{\pi}}\sum_{i=1}^A \vec{\sigma}_N(i) \cdot \vec{\ell}(i) \\
       &~~~-{\delta_{J2} \over \sqrt{30 \pi}} \sum_{i=1}^A\left[\vec{x}_i\otimes\left(\vec{\sigma}_N(i)\times\frac{1}{i}\vec{\nabla}_i\right)_1\right]_{2M} ,
    \end{split}
    \label{eq:nucl_resp_v1}
\end{equation}
so that $\Delta_{1M}(0)$ and $\Phi''_{00}(0)$ measure, respectively, the total orbital angular momentum and total spin-orbit response of the nucleus. As we discuss in detail below, the operator $\Phi''_{00}$ is coherently enhanced in certain nuclei \cite{Fitzpatrick:2012ix}, including $^{27}$Al.  All of these operators also contribute to inelastic $\mu \rightarrow e$ conversion. 

There are five additional operators that only contribute to inelastic scattering.  For parity-conserving nuclear transitions, their leading threshold forms are
\begin{equation}
    \begin{split}
    \Sigma_{2M}(q) &={q^2 r^2 \over 15} \sum_{i=1}^A \Big[ Y_2(\Omega_i) \otimes \vec{\sigma}_N(i) \Big]_{2M}, \\ 
    \Delta'_{2M}(0) &={1 \over \sqrt{20 \pi}}\sum_{i=1}^A \Big[ \vec{x}_i \otimes \vec{\nabla}_i \Big]_{2M}, \\
         \tilde{\Delta}''_{JM}(0)&= \frac{\delta_{J0}}{12\sqrt{\pi}}\sum_{i=1}^A \left( \overleftarrow{\nabla}_i \cdot \vec{x}_i-\vec{x}_i \cdot \overrightarrow{\nabla}_i \right) \\
       &~~~+{\delta_{J2} \over \sqrt{30 \pi}} \sum_{i=1}^A\left[\vec{x}_i\otimes \vec{\nabla}_i\right]_{2M}, \\
           \tilde{\Omega}_{1M}(0) &={1 \over 2 \sqrt{12 \pi}}\sum_{i=1}^A \Big[ \vec{x}_i \, \vec{\sigma}_N(i)\cdot  \overrightarrow{\nabla}_i -\overleftarrow{\nabla}_i \cdot \vec{\sigma}_N(i) \, \vec{x}_i \Big]_{1M} ,\\
               \Phi_{1M}(0)&= -\frac{1}{\sqrt{12\pi}}\sum_{i=1}^A\left[\vec{x}_i\otimes\left(\vec{\sigma}_N(i)\times\frac{1}{i}\vec{\nabla}_i\right)_1\right]_{1M} .
    \end{split}
    \label{eq:nucl_resp_v2}
\end{equation}

For elastic scattering, despite the fact that $\mu \rightarrow e$ conversion operators are evaluated for $q_\mathrm{eff} \approx m_\mu$, the threshold forms often provide useful information on the strength of matrix elements, as we know the charge, spin, and angular momentum of the valence nucleons. For example, if we have an operator proportional to $\vec{\ell}(i)$ but a
nucleus with a single unpaired nucleon in an $s$-wave shell-model orbital, we would expect to find a weak matrix element. However, for inelastic transitions, the interplay between the nuclear physics of the transition and operator 
properties is more subtle.  Inelastic transition matrix elements generally must be addressed through detailed nuclear structure calculations; we have done this for $^{27}$Al and report the results below. 

The threshold operators do play another role relevant to transitions, connecting $\mu \rightarrow e$ conversion matrix elements to those mediating standard-model processes like $\gamma$-decay and $\beta$-decay.  While the connections are qualitative because of the momentum-transfer differences, nevertheless they can serve as a crosscheck on the nuclear modeling.

\begin{figure}
    \centering
    \includegraphics[scale=0.65]{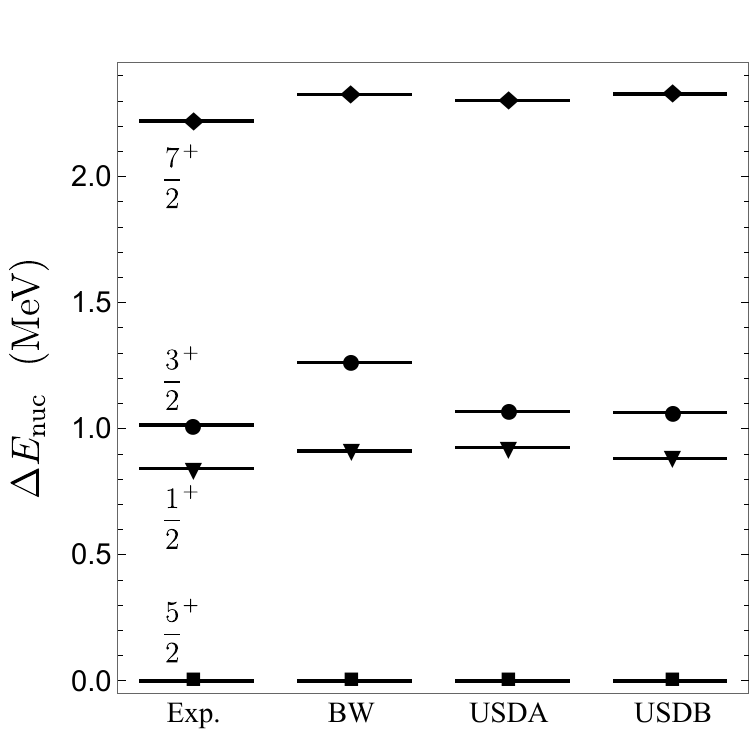}
    \caption{Experimental and theoretical (shell-model) values of the nuclear excitation energy $\Delta E_\mathrm{nuc}$ for low-lying eigenstates of $^{27}$Al.}
    \label{fig:energy_spectrum}
\end{figure}

$^{27}$Al is an attractive target choice for studies of both elastic and inelastic $\mu \rightarrow e$ conversion.  The ground state has $J^\pi=5/2^+$, making it sensitive to all six of the nuclear response functions associated with elastic $\mu\rightarrow e$ conversion \cite{Rule:2021oxe,Haxton:2022piv}. The low-energy spectrum of $^{27}$Al contains three reasonably well spaced excited states, $J^{\pi}=1/2^+$ (0.844 MeV), $3/2^+$ (1.015 MeV), and $7/2^+$ (2.212 MeV), in the energy window where inelastic $\mu \rightarrow e$ conversion might be detectable above background. 

Figure \ref{fig:energy_spectrum} compares the $^{27}$Al spectrum to theoretical estimates obtained from $2s$-$1d$ nuclear shell-model calculations.  We employed three widely used effective interactions: USDA and USDB \cite{PhysRevC.74.034315} and Brown-Wildenthal (BW) \cite{bw}. The agreement among these calculations and with experiment is generally quite good. The one-body density matrices needed in nuclear response function evaluations were calculated from the associated wave functions.  A single-particle harmonic oscillator basis was adopted for Slater determinants, a choice that preserves translational invariance in the computed transition densities.

\begin{figure*}
    \centering
    \includegraphics[scale=0.7]{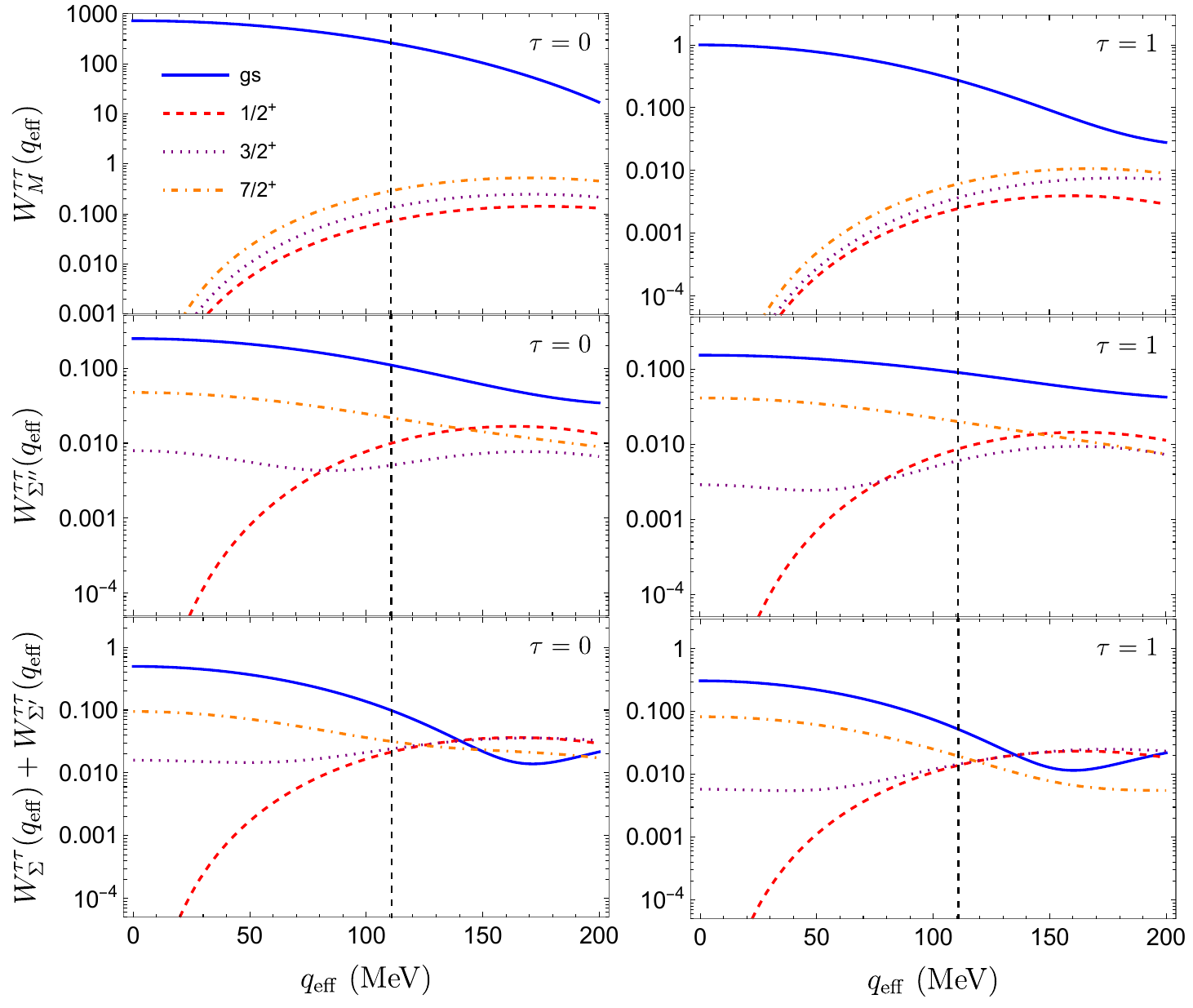}
    \caption{The shell-model charge and spin nuclear response functions for $^{27}$Al as a function of the effective momentum $q_\mathrm{eff}$ evaluated for transitions to the ground-state (solid blue) and first three excited states: $1/2^+$ (dashed red), $3/2^+$ (dotted purple), and $7/2^+$ (dot-dashed orange). Responses that are nonzero at $q_\mathrm{eff}=0$ include allowed contributions.  The isoscalar (isovector) responses are on left (right). The vertical dashed line marks the value $q_\mathrm{eff}=110.81$ MeV of the effective momentum transfer for the ground-state transition.}
    \label{fig:al27_q_response_allowed}
\end{figure*}

\begin{figure*}
    \centering
    \includegraphics[scale=0.7]{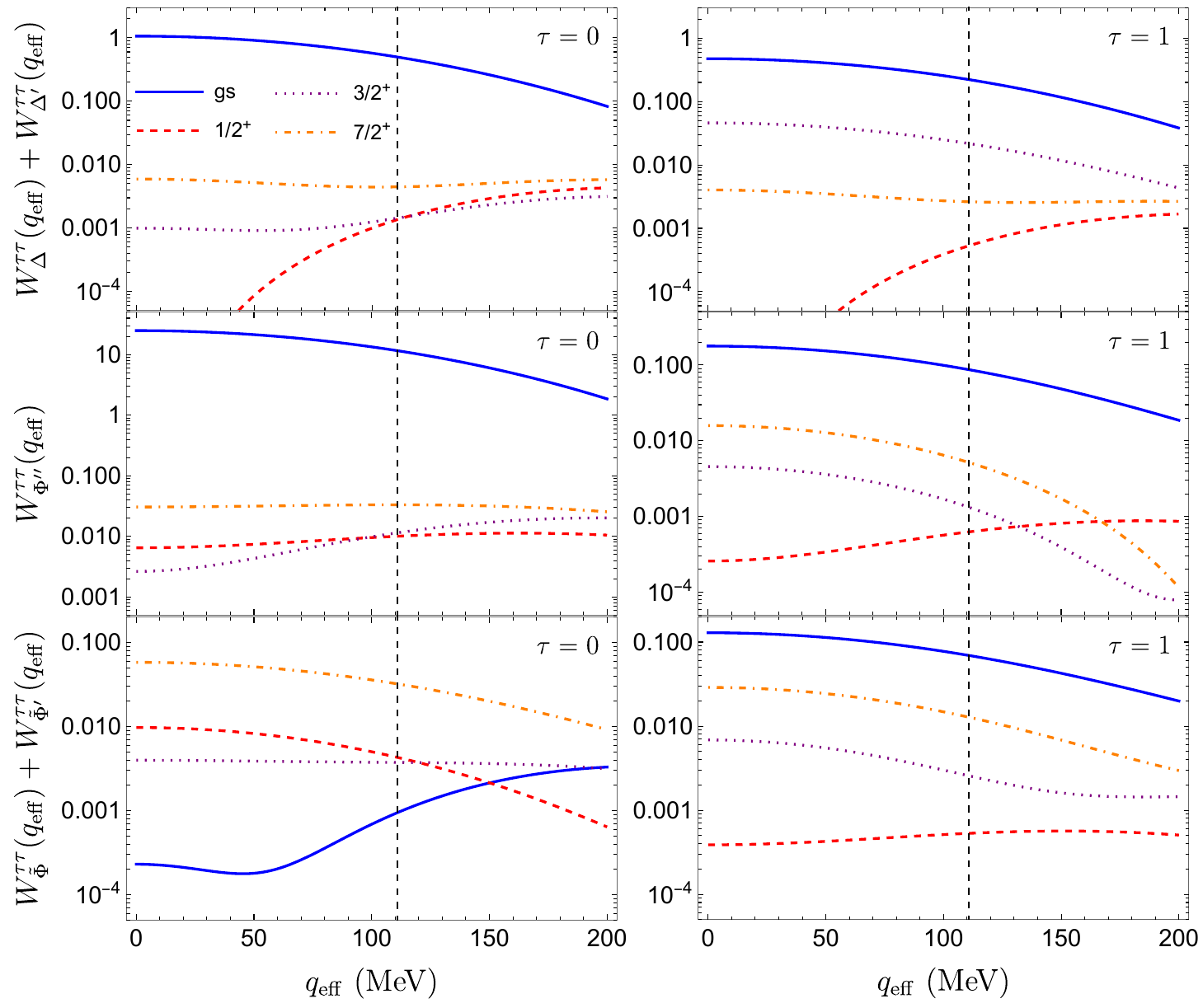}
    \caption{As in Fig. \ref{fig:al27_q_response_allowed} but for velocity-dependent nuclear responses that contribute to the elastic process. In the $\mu\rightarrow e$ conversion rate, these responses enter with an additional factor $q_\mathrm{eff}^2/m_N^2\approx 0.014$ in $^{27}$Al.}
    \label{fig:al27_q_response_v_dep}
\end{figure*}

\begin{figure*}
    \centering
    \includegraphics[scale=0.7]{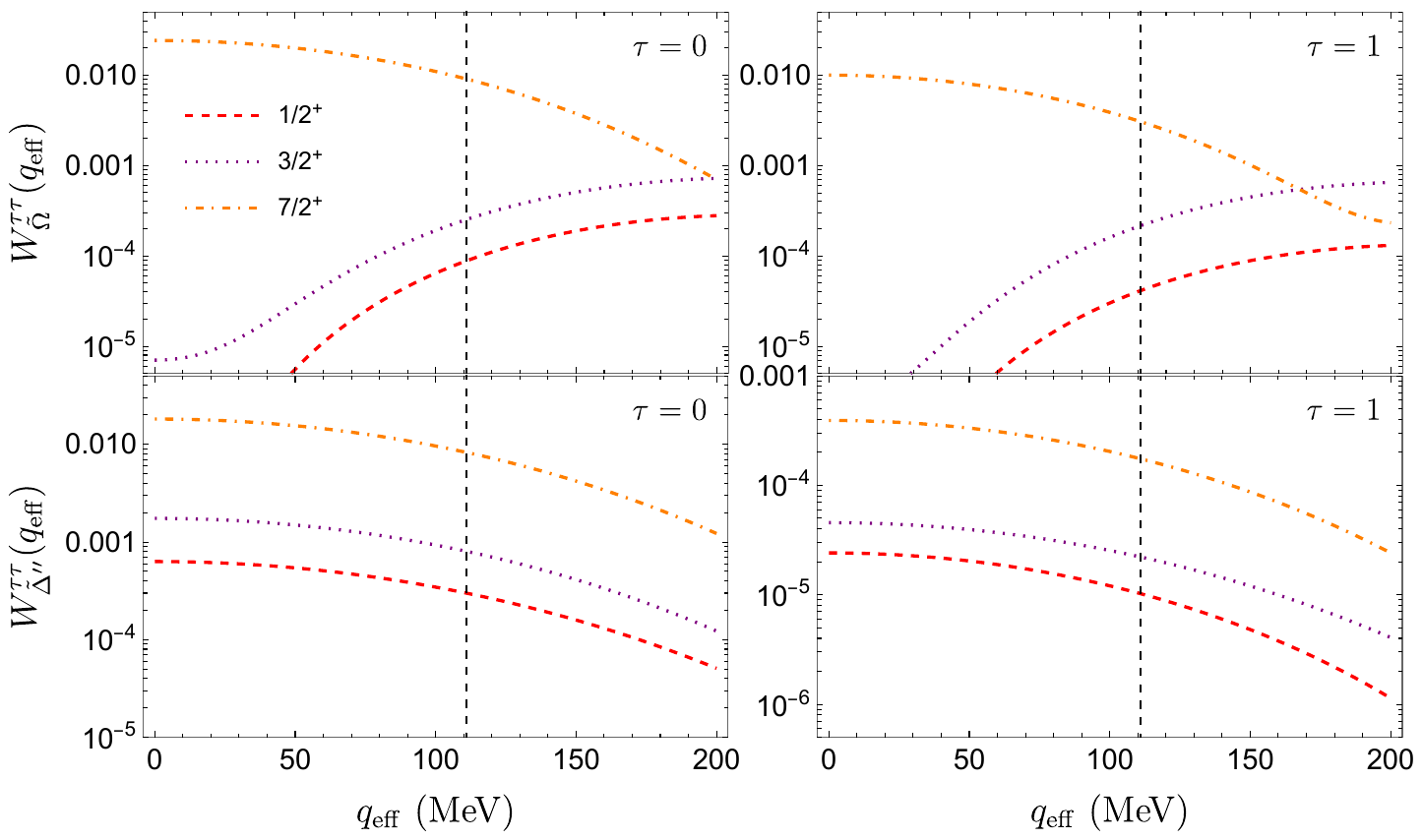}
    \caption{As in Fig. \ref{fig:al27_q_response_allowed} but for purely inelastic responses where the ground-state transition is forbidden. In the $\mu\rightarrow e$ conversion rate, these responses enter with an additional factor $q_\mathrm{eff}^2/m_N^2\approx 0.014$ in $^{27}$Al.}
    \label{fig:al27_q_response_inelastic}
\end{figure*}

Transitions from the $5/2^+$ ground state to the $3/2^+$ and $7/2^+$ second and third excited states of $^{27}$Al are ``allowed'', as they can be driven by the point-nucleus operator $\vec{\sigma}_N$. For small $q_\mathrm{eff}$, these transitions would dominate over ``forbidden'' transitions, such as the $5/2^+\rightarrow 1/2^+$ quadrupole transition to the first excited state. However, the effective momentum transfer in $\mu\rightarrow e$ conversion is not small: the relevant dimensionless quantity governing the nuclear multipole expansion is $y=(q_\mathrm{eff}b/2)^2\approx 0.27$ in $^{27}$Al, where $b$ is the oscillator parameter. The momentum transfer can alter the relative importance of the different transitions compared to the na\"ive long-wavelength estimate. The evolution of the $^{27}$Al response functions with momentum transfer  $q_\mathrm{eff}$ is displayed for each of the 16 nuclear responses in Figs. \ref{fig:al27_q_response_allowed}, \ref{fig:al27_q_response_v_dep}, and \ref{fig:al27_q_response_inelastic}.

The top panels of Fig. \ref{fig:al27_q_response_allowed} show the isoscalar and isovector nuclear charge responses, $W_M^{00}(q)$ and $W_M^{11}(q)$, for transitions to the ground and first three excited states of $^{27}$Al, summed over all contributing multipoles.  At $q_\mathrm{eff}=0$ only the monopole charge operator survives: its threshold form is diagonal and thus only contributes to the elastic transition. The relevant excited-state transitions are generated by the $q_\mathrm{eff}$-dependent higher charge multipoles $M_2$, $M_4$, etc., allowed by parity and angular momentum triangulation.  The $M_2$ contributions are dominant: the isoscalar form factors grow rapidly with $q_\mathrm{eff}$ and by $q_\mathrm{eff} \approx m_\mu$ are approaching their peak values.  The shell model predicts strong $M_2$ responses, and we will see below that this is consistent with experiment.

For isovector coupling, the coherence of the monopole response is lost with only the single unpaired neutron contributing, greatly diminishing the elastic response.  But a similar reduction to that observed in the isoscalar case, though somewhat less in magnitude, is also apparent in the inelastic responses.
The natural explanation for this is that the strong $M_2$ amplitudes noted above are a consequence of a collective quadrupole deformation of the nucleus, an isoscalar effect.

The middle (bottom) panels of Fig. \ref{fig:al27_q_response_allowed} show the longitudinal (transverse) spin response function for the same transitions. The response function for the non-allowed transition to the $1/2^+$ state increases rapidly, reaching a broad peak not far beyond the momentum transfer of interest, $q_\mathrm{eff} \approx m_\mu$.  The transverse magnetic operator $\Sigma_{2M}$ would contribute to this transition, which we noted previously has the leading behavior $q^2[r^2Y_2(\Omega) \otimes \vec{\sigma}_N]_{2M}$. Similarly, the allowed transverse electric $\Sigma_{1M}'$ and longitudinal $\Sigma_{1M}''$ spin operators include, in addition to their leading dependence on $\vec{\sigma}_N$, corrections of the form $q^2[r^2Y_2(\Omega) \otimes \vec{\sigma}_N]_{1M}$. The allowed responses are relatively weak at threshold and their profiles in $q_\mathrm{eff}$ are generally flat or slightly declining, without a diffraction minimum. Consequently, at the physically relevant value of $q_\mathrm{eff}$, the strengths of the allowed ($\Sigma'$ and $\Sigma''$) and momentum-suppressed ($\Sigma$) responses are comparable, for both isoscalar and isovector coupling. These results are consistent with relatively weak allowed responses in two cases and with comparatively large momentum-dependent corrections originating from an enhanced quadrupole interaction --- a phenomenon frequently found in mid-shell nuclei like $^{27}$Al. 

One concludes that nuclear structure effects must be suppressing transitions to the $3/2^+$ and $7/2^+$ states relative to transitions to the $1/2^+$ state. While this conclusion is based on the shell model, it is consistent with experiment: The measured (magnetic) B(M1) values for the gamma decays of the $3/2^+$ and $7/2^+$ states to the ground state are a small fraction of a single-particle unit, 0.0122 W.u. and 0.0627 W.u., respectively, while the (electric) B(E2) 
values for the decays of the $1/2^+$, $3/2^+$, and $7/2^+$ states are all large, 7.86 W.u., 7.8 W.u., and 15.0 W.u., respectively \cite{SHAMSUZZOHABASUNIA20111875}. Recall that the electric quadrupole $\gamma$-decay amplitude can be expressed in terms of the charge quadrupole $M_2$ operator via Siegert's theorem, while the $\gamma$-decay magnetic dipole amplitude involves a combination of $\vec{\sigma}_N$ and $\vec{\ell}$.  Gamma decay probes these operators very near threshold.

A simple exercise shows that indeed some interesting collective physics is at play in the response functions. The na\"ive single-particle shell model of~~$^{27}$Al describes transitions from the ground state to the first 2 excited-states as $2d_{5/2}\rightarrow 2s_{1/2}$ and $2d_{5/2}\rightarrow 2d_{3/2}$, respectively. Evaluating the response function ratio in this simple picture yields
\begin{equation}
\begin{split}
  & \frac{W_{\Sigma''}(gs\rightarrow 1/2^+)}{W_{\Sigma''}(gs\rightarrow 3/2^+)}\rightarrow \frac{|\braket{2s_{1/2}||\Sigma^{''}_3(q_\mathrm{eff}r)||2d_{5/2}}|^2}{|\braket{2d_{3/2}||\Sigma^{''}_1(q_\mathrm{eff}r)||2d_{5/2}}|^2}\approx 0.23, 
  \end{split}
\end{equation}
where we have suppressed isospin labels for simplicity (the response couples only to the odd proton). This calculation properly accounts for the significant momentum transfer and leads one to conclude that the allowed transition will be the stronger.

However, repeating the calculation with our fully correlated $2s$-$1d$ shell-model wave function, we obtain a very different answer
\begin{equation}
    \frac{W_{\Sigma''}(gs\rightarrow 1/2^+)}{W_{\Sigma''}(gs\rightarrow 3/2^+)}\approx 1.9,
\end{equation}
with the forbidden transition being the stronger.  Similar results are obtained for all three of the effective interactions
we employ. We have noted the consistency of this result with $\gamma$-decay.  It is also consistent with the long lifetime of the isospin-analog electron-capture reaction $^{27}$Si($5/2^+$, $gs$)$\rightarrow \,^{27}$Al($3/2^+$, $1.015$ MeV), $\log(ft)=7.23(6)$ \cite{SHAMSUZZOHABASUNIA20111875}.

The nucleus can also alter the na\"ive hierarchy of responses by enhancing certain operators through coherence. The operator $\Phi''_{00}$ is coherently enhanced in nuclei where one of two spin-orbit partners $j=\ell\pm \frac{1}{2}$ is occupied [c.f. Eq. \eqref{eq:nucl_resp_v1}]. $^{27}$Al is an ideal nucleus for exploiting this coherence, as the na\"ive ground-state configuration has 5 protons and 6 neutrons in the $2d_{5/2}$ shell and no nucleons in the $2d_{3/2}$ shell. As a result, the $\Phi''$ response, which is suppressed by $q_\mathrm{eff}^2/m_N^2$ in the conversion rate, can be elevated by coherence to contribute at the same level as the elastic spin responses $\Sigma',\Sigma''$, even including the additional momentum suppression factor.

\section{Experimental signatures}
The electrons produced in $\mu\rightarrow e$ conversion must be distinguished from background electrons originating in standard-model $\mu\rightarrow e+2\nu$ decays. Because the neutrinos always carry away some amount of energy, the continuous spectrum of DIO electrons falls off steeply towards the endpoint energy. Conversion electrons, on the other hand, are emitted exactly at the endpoint energy, assuming that the nucleus remains in the ground state. Any energy that is absorbed by the nucleus to produce an internal excitation must be subtracted from the conversion electron, as in Eq. \eqref{eq:inelastic_energy}, moving the signal into the DIO background. For the low-lying states of $^{27}$Al with excitation energies $\Delta E_\mathrm{nuc}\lesssim 2$ MeV that we consider, the DIO contributions in the relevant spectral region are modest. Fortunately, this background is well understood theoretically \cite{PhysRevD.84.013006,PhysRevD.94.051301,SZAFRON201661} and can be subtracted when performing a shape analysis of the measured electron spectrum.

\label{sec:exp_sig}
The less well understood electron background from radiative muon capture (RMC)
\begin{equation}
\mu^-+A(Z,N)\rightarrow \nu_{\mu}+\gamma+A(Z-1,N+1),
\end{equation}
where the emitted photon subsequently undergoes pair production $\gamma\rightarrow e^+e^-$, does not contribute in this energy window, as the endpoint for RMC electrons in $^{27}$Al is 3.62 MeV below the maximum CE energy.

\begin{table}[]
    \centering
    {\renewcommand{\arraystretch}{1.4}
    \begin{tabular}{l|c|c|c}
    \hline
    \hline 
     & \multicolumn{3}{c}{$R_{\mu e}(gs\rightarrow f)/R_{\mu e}(gs\rightarrow gs)$}\\
     Response & $f=1/2^+$ & $3/2^+$ & $7/2^+$\\
     \hline
    $W_M^{00}$ & $2.40(3)\times 10^{-4}$ & $4.4(2)\times 10^{-4}$ & $9.4(2)\times 10^{-4}$ \\
    $W_M^{11}$ & $8.2(5)\times 10^{-3}$ & $0.0113(5)$ & $0.019(1)$ \\
    $W_{\Sigma''}^{00}$     &  0.084(4) & 0.042(3) & 0.185(7)\\
    $W_{\Sigma''}^{11}$     &  0.081(4) & 0.055(4) & 0.194(10)\\
    $W_{\Sigma}^{00} + W_{\Sigma'}^{00}$ & 0.20(2) & 0.22(2) & 0.30(3) \\
    $W_{\Sigma}^{11} + W_{\Sigma'}^{11}$ & 0.22(3) & 0.22(3) & 0.33(4) \\
    $W_{\Phi''}^{00}$   &  $7(1)\times 10^{-4}$ & $8(1)\times 10^{-4}$ & $2.8(4)\times 10^{-3}$\\
    $W_{\Phi''}^{11}$   &  $6(3)\times 10^{-3}$ & $0.015(2)$ & $0.048(6)$\\
    $W_{\tilde{\Phi}}^{00} + W_{\tilde{\Phi}'}^{00}$ & $3.6(8)$ & $2.7(7)$ & $34(6)$\\
    $W_{\tilde{\Phi}}^{11}+ W_{\tilde{\Phi}'}^{11}$ & $7(3)\times 10^{-3}$ & $0.037(4)$ & $0.163(4)$\\
    $W_{\Delta}^{00} + W_{\Delta'}^{00}$ & $2.41(4)\times 10^{-3}$ & $2.6(2)\times 10^{-3}$ & $7.7(2)\times 10^{-3}$\\
    $W_{\Delta}^{11} + W_{\Delta'}^{11}$ & $2.1(2)\times 10^{-3}$ & $0.084(8)$ & $0.010(2)$\\
    \hline
    $W_{\tilde{\Delta}''}^{00}$ & $0.0361(8)$ & $0.097(4)$ & $1$\\
    $W_{\tilde{\Delta}''}^{11}$ & $0.062(6)$ & $0.12(1)$ & $1$\\
    $W_{\tilde{\Omega}}^{00}$  & $7(3)\times 10^{-3}$ & $0.016(6)$ & $1$\\
    $W_{\tilde{\Omega}}^{11}$ & $0.010(3)$ & $0.04(2)$ & $1$\\
    \hline
    \hline
    \end{tabular}}
    \caption{Relative $\mu\rightarrow e$ conversion strengths for transitions to the first 3 excited states of $^{27}$Al, normalized by either the ground-state (upper) or $7/2^+$ state (lower). Reported errors correspond to 1 standard deviation about the mean of the 3 shell-model calculations performed.}
    \label{tab:inelastic_responses}
\end{table}

\begin{figure*}
    \centering
    \includegraphics[scale=0.7]{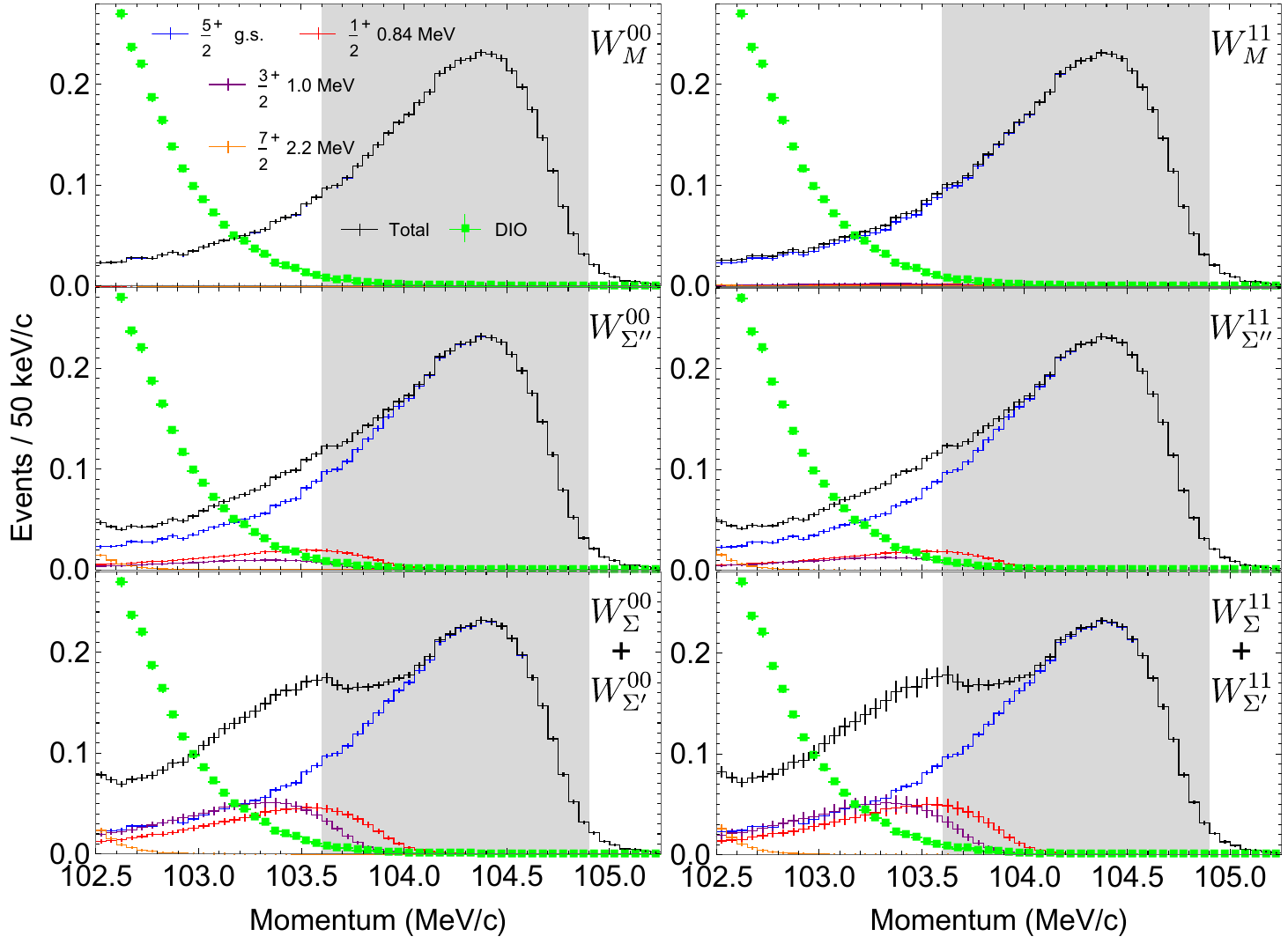}
    \caption{Expected electron counts in reconstructed momentum bins of width $50$ keV/c in Mu2e Run I for 6 different CLFV scenarios: coherent conversion (top panels), longitudinal spin-dependent (middle), and transverse spin-dependent (bottom). The left (right) column corresponds to purely isoscalar (isovector) responses. The total CE signal (black) is separated into contributions from the nuclear ground state (blue) and first 3 excited states at 0.84 MeV (red), 1.0 MeV (purple), and 2.2 MeV (orange). Green squares denote the DIO background, which dominates all other sources \cite{Mu2e:2022ggl}. The gray shading indicates the region $103.60<q_\mathrm{rec}<104.90$ MeV/c where the Mu2e sensitivity has been optimized. Each panel is normalized to produce a ground-state branching ratio $R_{\mu e}(gs\rightarrow gs)=10^{-15}$. The theory uncertainty associated with variations among the shell-model predictions of nuclear response function ratios and the statistical uncertainty in simulations of the Mu2e detector response were combined in computing error bars.}
    \label{fig:mu2e_spec_allowed}
\end{figure*}

\begin{figure*}
    \centering
    \includegraphics[scale=0.7]{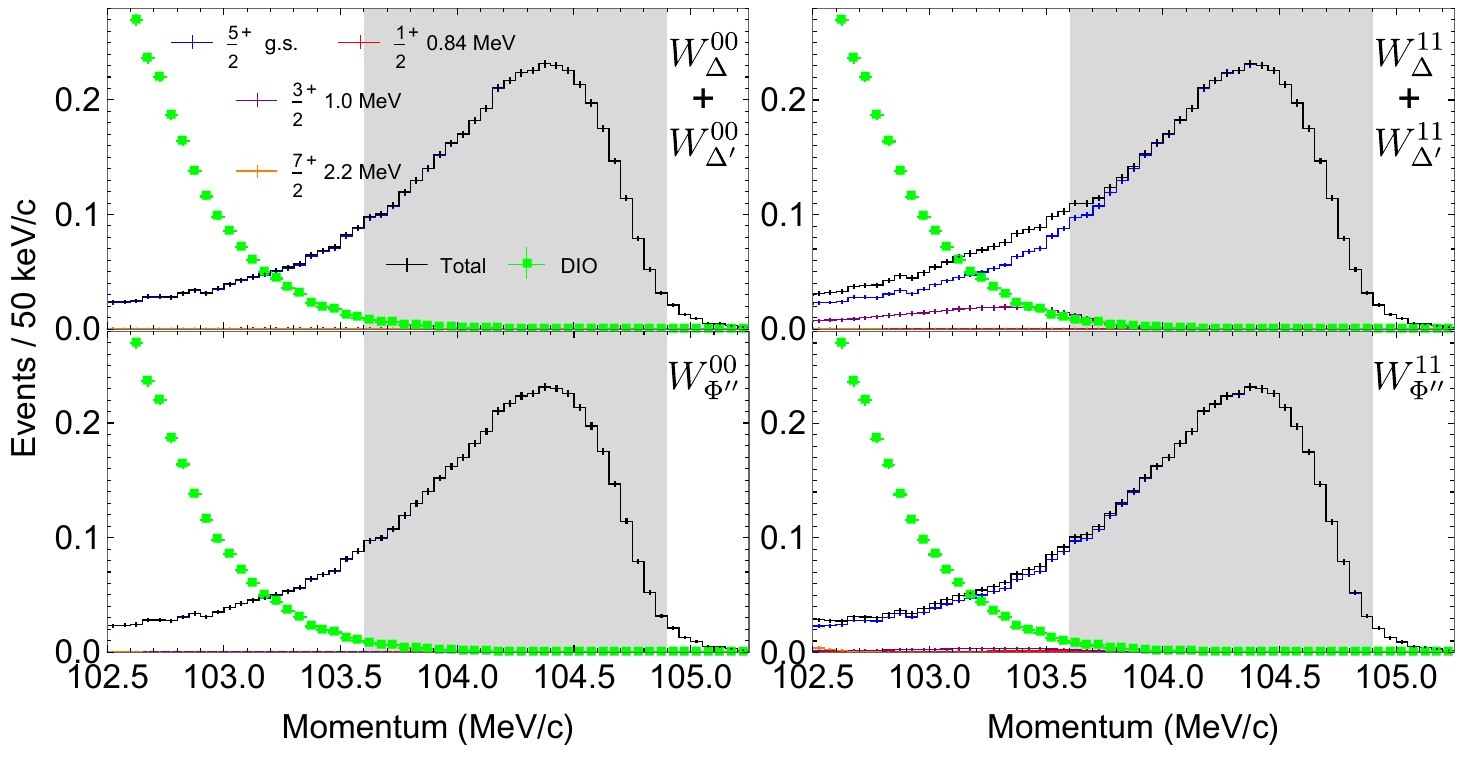}
    \caption{As in Fig. \ref{fig:mu2e_spec_allowed} but for velocity-dependent nuclear response functions.}
    \label{fig:mu2e_spec_v_dep_1}
\end{figure*}

\begin{figure*}
    \centering
    \includegraphics[scale=0.7]{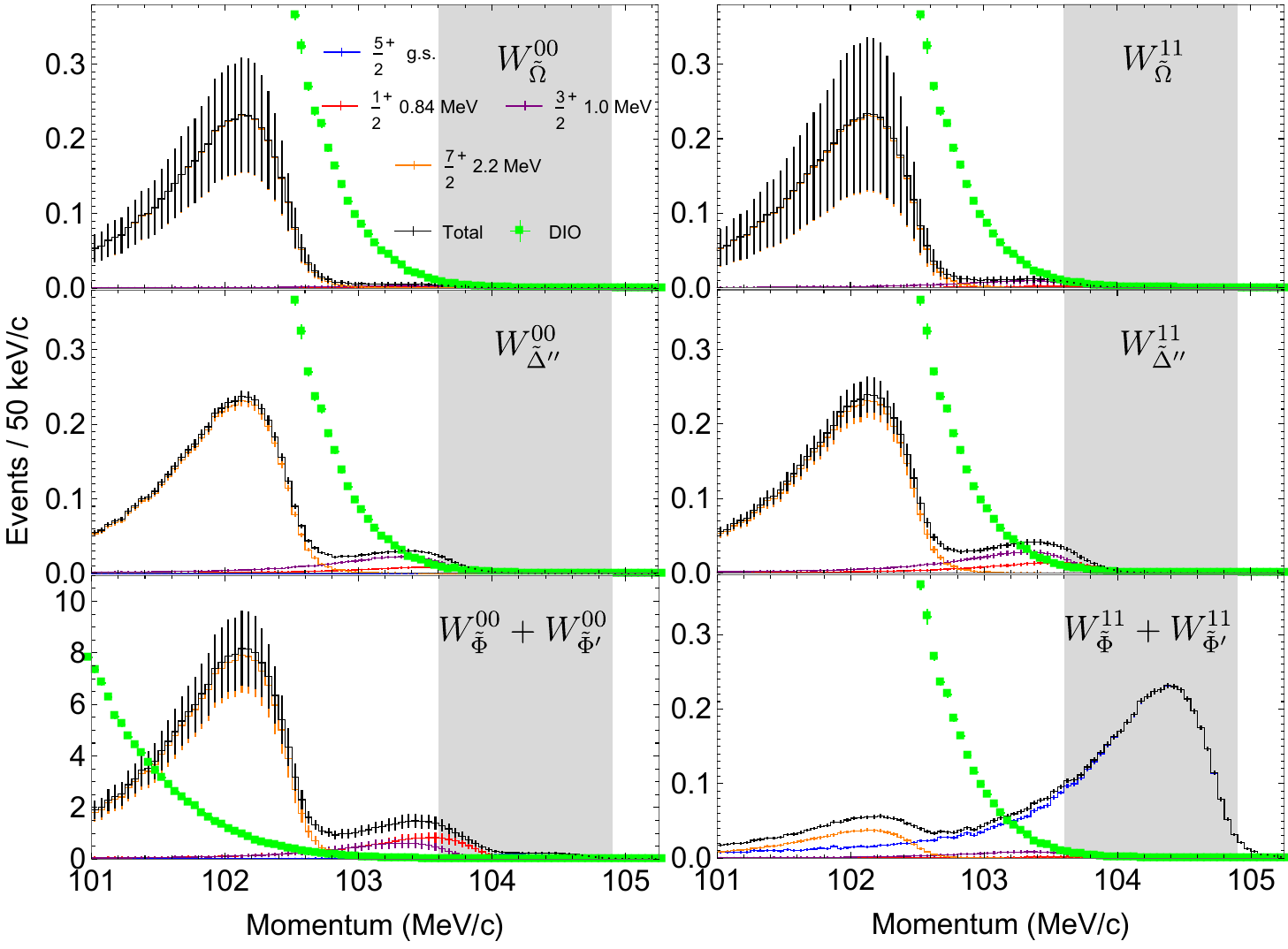}
    \caption{As in Fig. \ref{fig:mu2e_spec_allowed} but for velocity-dependent nuclear response functions. The purely inelastic transitions driven by $\tilde{\Omega}$ and $\tilde{\Delta}''$ are normalized so that the dominant transition (to the $7/2+$ state) has a branching ratio $R_{\mu e}(gs\rightarrow f)=10^{-15}$.}
    \label{fig:mu2e_spec_v_dep_2}
\end{figure*}

The mono-energetic CE will be registered in the Mu2e detector with a reconstructed momentum $q_\mathrm{rec}$ that differs from the initial momentum $q$ due to energy losses incurred as the electron moves through additional layers of the aluminium target and through the proton absorber before reaching the calorimeter. The momentum resolution of the Mu2e detector is significantly more precise than the smearing due to energy losses. Recently, the Mu2e collaboration performed detailed simulation studies in order to characterize the detector response and overall efficiency \cite{Mu2e:2022ggl}. Based on this study, the energy loss experienced by electrons in the signal window ($q\gtrsim 100$ MeV) results in a typical shift in the reconstructed momentum $\delta q_0 = q_\mathrm{rec}-q= -0.5497(26)$ MeV. The smearing is highly asymmetric, with a long low-energy tail. 

The shape of the CE signal generated by individual nuclear transitions is independent of the underlying CLFV mechanism; generically, one expects a mono-energetic electron with momentum $q$ determined by Eq. \eqref{eq:inelastic_energy}, which is then smeared out by energy losses in the tracker. When more than one nuclear final state contributes, the smeared mono-energetic signals from each transition are combined in a unique way depending on the respective excitation energies $\Delta E_\mathrm{nuc}$ and the relative branching ratios [see Eq. \eqref{eq:inelastic_branching}]. As a result, the combined signal contains information about underlying CLFV operators that is not available from the elastic process alone.  

The phenomenology of inelastic $\mu\rightarrow e$ conversion can be quite complicated, due to the large number of terms in Eq. \eqref{eq:inelastic_rate} that can simultaneously contribute. Restricting our attention to the direct terms and considering either purely isoscalar or purely isovector couplings, there are 16 independent ``directions'' in CLFV parameter space that are accessible to experiments. In this work, we perform a sensitivity study by considering each direction independently. That is, we calculate the relative branching ratio\footnote{In our numerical results, we include small phase-space corrections to Eq. \eqref{eq:ratio_single_response} from the variation in $q_\mathrm{eff}^2/(1+q/M_T)$ as the nuclear transition is varied. The same applies below in Eqs. \eqref{eq:inelastic_ratio_Ti_a} and \eqref{eq:inelastic_ratio_Ti_b}.}
\begin{equation}
\begin{split}
    \frac{R_{\mu e}(gs\rightarrow f)}{R_{\mu e}(gs\rightarrow gs)}&=\frac{\Gamma_{\mu e}(gs\rightarrow f)}{\Gamma_{\mu e}(gs\rightarrow gs)}\\
    &\rightarrow \left(\frac{Z_\mathrm{eff}^\mathrm{inel}}{Z_\mathrm{eff}^\mathrm{el}}\right)^3 \frac{W^{\tau\tau}_O(gs\rightarrow f)}{W^{\tau\tau'}_O(gs\rightarrow gs)},
    \end{split}
    \label{eq:ratio_single_response}
\end{equation}
where on the second line we take the limit of a single response function (or minimal pair of responses, e.g., $W_{\Sigma}^{\tau\tau}+W_{\Sigma'}^{\tau\tau}$). In this limit, the transition-independent CLFV response coefficient $\tilde{R}_O^{\tau\tau}$ drops out of the ratio. Aside from the fixed rescaling due to the effective charge, the quantity that we explore is a nuclear response function ratio, depending only on standard-model physics. However, if the spectral endpoint signature of the chosen $W_O^{\tau\tau}$ is distinctive, the presence or absence of this signature would constrain $\tilde{R}_O^{\tau\tau}$ and thus the source of CLFV physics.

We assess the phenomenological importance of inelastic $\mu\rightarrow e$ conversion by estimating the expected electron spectrum in different CFLV scenarios using the Monte Carlo results of the recent Mu2e simulation study. Starting from their result for the elastic electron spectrum, we generate an excited-state signature by shifting the signal by the excitation energy $\Delta E_\mathrm{nuc}$ and rescaling the amplitude by the relative response function ratio in Eq. \eqref{eq:ratio_single_response}. The various excited-state signals are then added together to the elastic signal to generate the total inelastic CE spectrum. 

The needed $^{27}$Al nuclear response functions are taken from the shell model: we use the three $2s$-$1d$ interactions described previously \cite{PhysRevC.74.034315,bw} in order to assess the nuclear modeling uncertainties. The shell-model diagonalizations were performed with the configuration-interaction code BIGSTICK \cite{2013CoPhC.184.2761J,2018arXiv180108432J}.

As the primary quantity derived from nuclear shell-model calculations is a ratio, many systematic uncertainties associated with nuclear response function evaluations should cancel. For example, the well-known ``quenching'' of the Gamow-Teller operator $\sigma \tau$ requires one to employ a renormalized value of the axial coupling constant $g_A$ that is roughly $80\%$ of the free-nucleon value in order to faithfully reproduce the measured $\beta$ decay lifetimes of $sd$-shell nuclei \cite{PhysRevLett.40.1631}. This correction --- and others like it --- would drop out of the required response function ratio in Eq. \eqref{eq:ratio_single_response}. In the general case where multiple nuclear response functions contribute simultaneously, effects of operator mixing and renormalization would need to be more carefully treated.

If the elastic process is not forbidden, then we normalize the amplitude of the expected CE signal by fixing the ground-state CLFV branching ratio to the fiducial value $R_{\mu e}(gs\rightarrow gs)=10^{-15}$. In cases where the elastic process is forbidden ($\tilde{\Delta}''$ or $\tilde{\Omega}$), we normalize by fixing the value of the largest individual branching ratio (which, in all cases considered, is to the $7/2^+$ state at 2.2 MeV) to the fiducial value $R_{\mu e}(gs\rightarrow f)=10^{-15}$. The predicted electron spectra correspond to the expected counts in Mu2e Run I, where $6\times 10^{16}$ muons will be stopped. Table \ref{tab:inelastic_responses} gives the relative strengths of the inelastic transitions for each nuclear response in $^{27}$Al, as well as the theoretical uncertainties inferred from the spread of the three shell-model calculations. The expected electron spectra corresponding to each of the 16 simplified CLFV scenarios are shown in Figs. \ref{fig:mu2e_spec_allowed}, \ref{fig:mu2e_spec_v_dep_1}, and \ref{fig:mu2e_spec_v_dep_2}. The uncertainties in the computed spectra are obtained by combining the estimated theory error on the relative branching ratios with the reported statistical error from the Mu2e Monte Carlo simulations of the elastic spectrum. We now discuss each CLFV scenario in detail:

\textit{Coherent conversion.}
Many previous studies \cite{Kitano:2002mt,Cirigliano:2009bz,Crivellin:2017rmk,Bartolotta:2017mff,DAVIDSON2019380,HEECK2022115833,Cirigliano:2022ekw} have focused on coherent $\mu\rightarrow e$ conversion, which can arise, for example, from a scalar or vector coupling of the leptons to quarks or a dipole coupling to photons. We first take the CLFV coupling to be isoscalar, thereby maximizing the coherent enhancement. As no such coherence arises for inelastic transitions, the elastic contribution dominates the rate. From Table \ref{tab:inelastic_responses}, we see that transitions to excited states constitute $\approx 0.1\%$ of the total response, consistent with the expected elastic enhancement of  $\approx 0.43 A^2\approx 300$ (including effects of the elastic form factor). The top left panel of Fig. \ref{fig:mu2e_spec_allowed} demonstrates that the inclusion of excited-state contributions has no discernible effect on the simulated elastic spectrum of Ref. \cite{Mu2e:2022ggl}.

If the underlying CLFV charge coupling is isovector, then the coherence is lost: the scattering takes place on the unpaired nucleon. The leading multipole operator for the elastic process is monopole, $M_0$, while that for inelastic conversion is quadrupole, $M_2$. 
The resulting $q$-dependent dimensional suppression of inelastic responses is $\approx$ a factor of seven. The computed inelastic contribution is smaller, $\approx 3\%$ of the total, reflecting the specific 
nuclear structure of $^{27}$Al. Consequently, if experiment finds even a modest
inelastic contribution, the CLFV could not be entirely attributed to a charge interaction,
regardless of the charge's isospin couplings. The resulting electron spectrum is shown in the top right panel of Fig. \ref{fig:mu2e_spec_allowed}.

\textit{Spin-dependent conversion.} Spin-dependent operators, which couple primarily to the unpaired proton in $^{27}$Al, have been studied in the form $\vec{\sigma}_L \cdot \vec{\sigma}_N$ \cite{Cirigliano:2017azj,Davidson:2017nrp}.  In the NRET formalism two spin responses arise without velocity suppression, and three more arise when $\vec{v}_N$ is included to first order.  We discuss the former here. Spin-dependent operators primarily couple to the odd proton in $^{27}$Al; as a result, there is typically little difference between isoscalar and isovector variants of the same response.

Longitudinal coupling to spin, associated with the abnormal-parity operator $\Sigma_J^{\prime \prime}$, arises for pseudoscalar or axion-like-particle (ALP) CLFV exchanges \cite{Fuyuto:2024skf,Haxton:2024lyc}.  The contribution of the first excited state is roughly 9\% of the elastic response.  As the peak of the inelastic electron spectrum is displaced from the elastic peak, this produces a 40\% enhancement in the spectrum for $q_\mathrm{rec}\approx 103.5$ MeV/c, distinguishing this case from the charge responses discussed above. (See the middle panels of Fig. \ref{fig:mu2e_spec_allowed}.) The operator dominating this transition is $\Sigma_3^{\prime \prime}$, which has the threshold behavior $q^2 [r^2 Y_2(\Omega) \otimes \vec{\sigma}_N]_{3M}$, underscoring again the importance of the strong quadrupole in enhancing momentum-dependent contributions to $^{27}$Al response functions. 

A transverse coupling to spin generates the electric and magnetic operators $\Sigma_J^\prime$ and $\Sigma_J$,  with parities $(-1)^{J+1}$ and $(-1)^J$, respectively.  The inelastic contributions are quite substantial, with each of the first three excited states contributing with strengths $\approx 20$--$35\%$ that of the ground state. This creates a distinctive second peak in the CE spectrum near $103.5$ MeV/c, shown in the bottom panels of Fig. \ref{fig:mu2e_spec_allowed}.

\textit{Exotic responses.} 
Lastly, we consider several responses that depend explicitly on nuclear compositeness through their dependence on the inter-nucleon velocity operator $\vec{v}_N$ \cite{Rule:2021oxe,Haxton:2022piv}. When matched to a Lorentz-invariant, quark-level effective theory, some fine tuning is required to make $\vec{v}_N$-dependent operators leading \cite{Haxton:2024lyc}. From a bottom-up perspective, however, the associated response functions contain new CLFV information accessible to experiment. 

If CLFV couples to the transverse components of the nuclear convective current, then the response $W_{\Delta}^{\tau\tau}+W_{\Delta'}^{\tau\tau}$ is generated. In the isoscalar case, the excited-state contributions represent $\approx 1\%$ of the total response. Correspondingly, the top left panel of Fig. \ref{fig:al27_q_response_v_dep} shows almost no discernible inelastic contribution. On the other hand, if the coupling is isovector, then the transition to the $3/2^+$ state provides a modest $(\approx 10\%)$ contribution, leading to a roughly $40\%$ enhancement in the number of electron counts around $103.5$ MeV/c.

Next, we consider responses that arise from projections of the nuclear spin-velocity current $\vec{v}_N\times\vec{\sigma}_N$, which generically appears when CLFV is mediated by tensor exchanges \cite{Haxton:2022piv,Haxton:2024lyc}. Taking the longitudinal component yields the response $\Phi''$. As discussed in Sec. \ref{sec:response_props}, this response is coherently enhanced in nuclei, like $^{27}$Al, where one of two spin-orbit partner orbitals is occupied. Similar to the case of the isoscalar charge response $W_{M}^{00}$, the coherent enhancement of the isoscalar tensor response $W_{\Phi''}^{00}$ applies only to the ground-state transition; excited states contribute $\lesssim 0.5\%$ of the total response. The resulting CE spectrum, shown in the bottom left panel of Fig. \ref{fig:mu2e_spec_v_dep_1}, is indistinguishable from the purely elastic signal. The isovector response $W_{\Phi''}^{11}$ shows a very slight excess.

Therefore, while it is true that if any significant excess is measured over the elastic signal, then there must be some CLFV mechanism besides the simple coupling to nuclear charge $W_M$, the converse does not hold: The absence of an appreciable inelastic CE signal does not imply that $\mu\rightarrow e$ conversion is mediated the coherent operator $W_M$. If a signal consistent with purely elastic $\mu\rightarrow e$ conversion is observed, subsequent measurements with additional target isotopes would be required in order to distinguish between $W_M$, $W_{\Phi''}$, and $W_\Delta^{00}+W_{\Delta'}^{00}$ as the CLFV source.

Another interesting case is $W_{\tilde{\Phi}}^{00}+W_{\tilde{\Phi}'}^{00}$, generated from the transverse projection of $\vec{v}_N\times \vec{\sigma}_N$. The transition to the $7/2^+$ state at 2.2 MeV is $\approx 40$ times stronger than the suppressed elastic transition, dominating the response, as shown in the bottom left panel of Fig. \ref{fig:mu2e_spec_v_dep_2}. The expected number of counts is high because of our adopted normalization to the ground-state rate. If the coupling is isovector, then the transition to the $7/2^+$ state provides only a modest contribution, yielding a small second peak in the electron spectrum, deep in the DIO background, which can be seen in the bottom right panel of Fig. \ref{fig:mu2e_spec_v_dep_2}.

The last two responses considered here are associated with CLFV operators that can only be probed in inelastic transitions, as the elastic response vanishes. The response $\tilde{\Omega}$ is generated from interactions that couple to the nuclear axial charge, $\vec{v}_N \cdot \vec{\sigma}_N.$  Our calculations predict that $\gtrsim 95\%$ of the transition strength goes to the $7/2^+$ state at $2.2$ MeV. This places the signal --- shown in the top panels of Fig. \ref{fig:mu2e_spec_v_dep_2} --- in a region where the DIO background is substantial, so that a background subtraction would be needed.  While the DIO shape is known well, ultimately the success of the subtraction will depend on statistical details of future experiments.

Finally, we consider the response $\tilde{\Delta}_J''$, defined as the longitudinal projection of
the full nuclear current $\vec{j}(\vec{x})$, which we assume is conserved in order to rewrite the operator
in terms of $M_J$ and the state energies, as in Eq. (\ref{eq:siegert}).  The resulting response is dominated by the transition to the $7/2^+$ state, although the first two excited states also make modest contributions. This is shown in the middle panels of Fig. \ref{fig:mu2e_spec_v_dep_2}, where the total electron spectrum is double peaked, with the primary response obscured by the DIO background.

While the predicted counts in each 50 keV/c bin are typically small, our estimates are based on Mu2e Run I, where $6\times 10^{16}$ muons will be captured. Over the total experimental lifetime, Mu2e is expected to stop $10^{18}$ muons in its $^{27}$Al target \cite{Bernstein_2019}, improving the statistics by more than an order of magnitude. Mu2e-II \cite{Mu2e:2018osu,Mu2e-II:2022blh}, a proposed extension leveraging proton beamline upgrades at Fermilab, could yield a further order-of-magnitude improvement. There is also the possibility that experiments can enhance their sensitivity to inelastic conversion by detecting the coincident low-energy photons emitted when the nucleus de-excites.

In our analysis, we employed Mu2e simulation data. Our conclusions should apply equally well to COMET, though the exact shape of the reconstructed electron signal will likely differ to some extent. Our analysis is a sensitivity study, exploring separately several charge, spin, and convection current responses generated in NRET, assuming either isoscalar or isovector couplings. Clearly, multiple responses with arbitrary isospin couplings can contribute to total rates, requiring a more general NRET analysis. Still, the basic conclusions reached here should hold up.  In particular, even a modest inelastic signal would rule out the most frequently explored model, CLFV generated entirely by a coherent charge coupling.

\section{Other nuclear targets}
\label{sec:other_targets}
The electron spectra presented in the previous section were derived from simulated Mu2e detector response data, which is only strictly applicable for an aluminum target in their particular experimental design. Even if all design elements of Mu2e were fixed except for the target isotope, the energy loss experienced by electrons in the tracker would change, yielding a different spectral profile for the smeared CE signal. Nonetheless, we can explore --- to some extent --- the physics of inelastic $\mu\rightarrow e$ conversion in other targets of interest by evaluating nuclear response function ratios for relevant low-lying transitions. Although many different nuclei could be analyzed in this manner, here we will consider only titanium, which is illustrative of the potential challenges and rich phenomenology of inelastic $\mu\rightarrow e$ conversion.

For aluminum, which is isotopically pure and has low-lying excited states that are reasonably well spaced in energy, the analysis is comparatively simple. Considering a titanium target, the situation is complicated by the fact that there are five stable isotopes at natural abundance, some of which have a relatively dense low-energy spectrum. For example, $^{47}$Ti --- which constitutes 7.44\% of a natural titanium target --- has 8 excited states with energies $\Delta E_\mathrm{nuc}<2$ MeV, including some states that are inferred from particular nuclear reactions but not confirmed by others and for which the angular momentum has not been experimentally determined \cite{Burrows:2007jwj}. For this reason, we take a simplified approach to titanium by considering only states with $\Delta E_\mathrm{nuc}<1$ MeV, which are generally well established experimentally. There are 3 such excited states: $^{46}$Ti($2^+$, 0.899 MeV), $^{47}$Ti($7/2^-$, 0.159 MeV), and $^{48}$Ti($2^+$, 0.984 MeV). The second of these states has an excitation energy $\Delta E_\mathrm{nuc}=0.159$ MeV that is likely to be very small compared to the characteristic energy width of the detector response. As a result, it would probably be difficult to distinguish electrons originating in transitions to this state vs the ground state, unless one could also detect the coincident de-excitation photons. 

Table \ref{tab:inelastic_responses_ti} compares the relative strength of the 16 independent nuclear response functions, evaluated for the selected transitions in a natural titanium target. In cases where the elastic process is not forbidden, we report the response function ratio for a transition to final state $f$ in isotope $i$ as
\begin{equation}
\left(\frac{Z_\mathrm{eff}^\mathrm{inel}}{Z_\mathrm{eff}^\mathrm{el}}\right)^3\frac{A_i W_O^{\tau\tau}(gs\rightarrow f)_i}{\sum_j A_j W_O^{\tau\tau}(gs\rightarrow gs)_j},
\label{eq:inelastic_ratio_Ti_a}
\end{equation}
where $A_i$ is the natural abundance of the $i$-th isotope, and the denominator is the total (abundance-weighted) elastic response. The notation $W_O^{\tau\tau}(gs\rightarrow f)_i$ indicates that the response function $O$ should be evaluated for a transition to final state $f$ in isotope $i$. In scenarios where the total elastic response is zero ($\tilde{\Delta}''$ or $\tilde{\Omega}$), we report in Table \ref{tab:inelastic_responses_ti} the abundance-weighted response ratio relative to the dominant transition
\begin{equation}
    \frac{A_i W_O^{\tau\tau}(gs\rightarrow f)_i}{\max\left[A_j W_O^{\tau\tau}(gs\rightarrow f)_j\right]}.
    \label{eq:inelastic_ratio_Ti_b}
\end{equation}

Compared to the corresponding results for $^{27}$Al (Table \ref{tab:inelastic_responses}), the most immediate difference is that some of the response functions in titanium vanish identically due to nuclear selection rules. In particular, $0^+\rightarrow 2^+$ transitions in $^{46}$Ti and $^{48}$Ti can only be mediated by parity-conserving $J=2$ multipole operators. As a result, the abnormal-parity responses $\Sigma''$ and $\tilde{\Omega}$ cannot contribute. By the same token, the abnormal-parity responses $\Delta$, $\Sigma'$, and $\tilde{\Phi}$ also vanish for these transitions, but their normal-parity counterparts $\Delta'$, $\Sigma$, and $\tilde{\Phi}'$ provide a non-zero contribution. 

Similar to $^{27}$Al, when CLFV is mediated by the responses $W_{M}^{\tau\tau}$ and $W_{\Phi''}^{\tau\tau}$, the coherently-enhanced ground-state transition tends to dominate over the excited states. Thus, one does not expect to see a strong modification of the CE spectrum in these scenarios, although in the case of isovector charge coupling, $W_M^{11}$, the $0^+\rightarrow 2^+$ transition in $^{48}$Ti approaches $4\%$ of the elastic response. 

If CLFV is mediated by the isoscalar, transverse-spin responses $W_{\Sigma}^{00}+W_{\Sigma'}^{00}$, the transition to the $^{48}$Ti($2^+$, 0.984 MeV) state contributes with strength $\approx 50\%$ that of the total elastic response. The uncertainty on this quantity --- estimated from the spread in the results from three shell-model calculations employing different effective interactions \cite{kb3g,gxpf1,kbp} --- is significant. Nonetheless, given the relative strength and the considerable energy separation ($\Delta E_\mathrm{nuc}\approx 1$ MeV) of this inelastic transition, it seems probable that an excess of electron counts (compared to the elastic signal) could be measured in a natural titanium target in this transverse-spin-mediated CLFV scenario. Of course, additional transitions with energies $\Delta E_\mathrm{nuc}\gtrsim 1$ MeV could also contribute in this same spectral region. Fortunately, the combined spectral shape is strictly additive: the inclusion of these states would only further increase the electron excess compared to the elastic baseline. 

In the exotic scenario where $\mu\rightarrow e$ conversion couples to nuclear axial charge --- leading to the response $\tilde{\Omega}$ --- the only transition that contributes is $^{47}$Ti($5/2^-,gs)\rightarrow$ $^{47}$Ti($7/2^-$, 0.159 MeV). As discussed above, the relatively small excitation energy will likely make it difficult to distinguish this transition from the elastic process. 

\begin{table}[]
    \centering
    {\renewcommand{\arraystretch}{1.4}
    \begin{tabular}{l|c|c|c}
    \hline
    \hline 
     & \multicolumn{3}{c}{$R_{\mu e}(gs\rightarrow f)/R_{\mu e}(gs\rightarrow gs)$}\\
      Response & $f=^{46}$Ti($2^+$) & $^{47}$Ti($7/2^-$) & $^{48}$Ti($2^+$)\\
     \hline
    $W_M^{00}$ & $2.0(1)\times 10^{-4}$ & $6.6(2)\times 10^{-5}$ & $1.6(1)\times 10^{-3}$ \\
    $W_M^{11}$ & $2.9(2)\times 10^{-3}$ & $1.9(1)\times 10^{-3}$ & $0.039(4)$ \\
        $W_{\Sigma''}^{00}$     &  --- & $8.0(1)\times 10^{-3}$ & ---\\
    $W_{\Sigma''}^{11}$     &  --- & 0.013(1)  & ---\\
     $W_{\Sigma}^{00} + W_{\Sigma'}^{00}$ & 0.034(8) & 0.03(1) & 0.5(1) \\
     $W_{\Sigma}^{11} + W_{\Sigma'}^{11}$ & $1.6(9)\times 10^{-3}$ & 0.033(5) & 0.2(1) \\
    $W_{\Phi''}^{00}$      &  $1.0(3)\times 10^{-3}$ & $3.3(8)\times 10^{-4}$ & $6(2)\times 10^{-3}$\\
    $W_{\Phi''}^{11}$   &  $3(4)\times 10^{-5}$ & $9(1)\times 10^{-5}$ & $3(1)\times 10^{-3}$\\
    $W_{\tilde{\Phi}}^{00} + W_{\tilde{\Phi}'}^{00}$ & $2.6(6)$ & $0.9(2)$ & $20(4)$\\
    $W_{\tilde{\Phi}}^{11}+ W_{\tilde{\Phi}'}^{11}$ & $0.4(4)$ & $1.3(3)$ & $0.7(6)$\\
     $W_{\Delta}^{00} + W_{\Delta'}^{00}$ & $1.79(7)\times 10^{-3}$ & $1.0(1)\times 10^{-3}$ & $0.021(3)$\\
    $W_{\Delta}^{11} + W_{\Delta'}^{11}$ & $4.5(2)\times 10^{-4}$ & $1.8(8)\times 10^{-3}$ & $6.6(7)\times 10^{-3}$\\
    \hline
     $W_{\tilde{\Delta}''}^{00}$ & $0.104(8)$ & $1.08(8)\times 10^{-3}$ & $1$\\
     $W_{\tilde{\Delta}''}^{11}$ & $0.060(7)$ & $1.2(1)\times 10^{-3}$ & $1$\\
     $W_{\tilde{\Omega}}^{00}$ & --- & $1$ & ---\\
     $W_{\tilde{\Omega}}^{11}$ & --- & $1$ & ---\\
    \hline
    \hline
    \end{tabular}}
    \caption{Relative $\mu\rightarrow e$ conversion strengths for transitions to 3 low-lying excited states of stable titanium isotopes, normalized by either the total elastic response (upper) or that of the dominant excited state (lower). Reported errors correspond to 1 standard deviation about the mean of the 3 shell-model calculations performed.}
    \label{tab:inelastic_responses_ti}
\end{table}

\section{Summary and Discussion} 
\label{sec:conclusion}
By the end of the current decade, new experiments at Fermilab and J-PARC will probe $\mu\rightarrow e$ conversion with 10,000 times greater sensitivity than previous efforts. A positive signal would be definitive evidence of new physics. As many plausible extensions of the standard model could account for such a signal, additional information would be required in order to connect the low-energy signal to UV physics and thus determine the source of the CLFV. Despite the power of the new experiments, one drawback is that they will provide only a single number, the elastic $\mu \rightarrow e$ conversion rate.  At least, this is the assumption most commonly made in the literature.

What information is available from $\mu \rightarrow e$ conversion?  This question was addressed recently for elastic
$\mu \rightarrow e$ conversion using the tools of nonrelativistic effective theory (NRET). Originally developed for heavy-quark systems, this Galilean-invariant formalism can be applied to $\mu\rightarrow e$ conversion because the relative motion of the constituents of both the initial
muonic atom and final nucleus is nonrelativistic.  The small parameters governing the process --- the momentum transfer through the combination $(q b/2)^2$, the relative nucleon velocity $\vec{v}_N$, and the bound muon velocity $\vec{v}_\mu$ --- form an operator hierarchy that guides the construction of the most general CLFV lepton-nucleon interaction.  This interaction, expanded through order $\vec{v}_N$, contains 16 operators.  The subsequent embedding of this interaction in the nucleus generates an elastic $\mu \rightarrow e$ conversion rate involving six response functions (and two interference terms).  This form of the rate can in fact be derived from symmetry arguments alone but also emerges organically from the NRET because of the completeness of the effective theory's basis of operators. 

The response functions can be viewed as ``knobs" that can be turned by altering the nuclear physics, while the coefficients of the response functions --- bilinear combinations of the LECs of the NRET operator expansion --- represent the CLFV physics that can be extracted from experiment. Thus, in principle, one can extract six constraints on CLFV from elastic $\mu \rightarrow e$ conversion. In practice, the feasibility of this extraction depends both on the operator source of the CLFV and our skill at turning the nuclear knobs.  The most direct way to do this is by selecting new targets and repeating the experiment, clearly a tedious task. Each new experiment will require a major investment of time and effort. Targets would be carefully selected to enhance or suppress particular responses, with each successive experiment informing the choice of the next target. Among the nuclear properties that could be exploited are the nucleus's angular momentum and isospin, and the spin, orbital, and spin-orbit properties of the valence nucleon or nucleons. 

It would be very attractive to find an alternative and more efficient strategy, one that could yield additional information from the $^{27}$Al experiment and any other efforts completed in the next few years.  We point out, in an earlier Letter \cite{Haxton:2024ecp} and in this paper, that inelastic $\mu \rightarrow e$ conversion could provide such an alternative.  The signal of the inelastic contributions would be a modification of the spectrum of observed electrons in a narrow energy window within about two MeV of the endpoint energy.  This is the window where inelastic contributions can stand out above backgrounds.  We argue that $^{27}$Al is an excellent target from this perspective, with three excited states in this energy window that are reasonably well separated in energy, with distinct nuclear properties.  By extracting and utilizing the information from both elastic and inelastic $\mu \rightarrow e$ conversion, Mu2e and COMET experimentalists can learn more about CLFV from their $^{27}$Al measurements and consequently make a more informed choice of follow-up targets, should a CLFV signal be detected.

The generalization from elastic to inelastic $\mu \rightarrow e$ conversion is a substantial exercise, as nuclear operators are no longer constrained to be CP even.  Consequently, when the NRET nucleon-level interaction is embedded in a
nucleus, five new nuclear operators arise (in addition to the six that contribute to the elastic process).  This paper derives, for the first time, the most general expression for the inelastic $\mu \rightarrow e$ conversion rate.  (The elastic rate then is the limit of this more general rate formula obtained by requiring all nuclear operators to be both parity- and CP-even.)  The coefficients of the five new response functions provide new information on CLFV --- including information on four of the NRET operators that do not contribute to the elastic rate.  Further, the CLFV LEC bilinears that appear in the elastic rate can be tested in new combinations in inelastic $\mu \rightarrow e$ conversion, yielding additional information.

Here we illustrate the power of combining these two signals through a shape analysis of the near-endpoint spectrum. Using the anticipated energy resolution function for the Mu2e experiment, we predict the distortions of the electron spectrum for various CLFV scenarios, identifying several very distinctive signals.  We stress that either the presence or absence of an excited-state signal provides new information and thus new guidance for follow-up experiments.

We also extended the treatment to a second target of interest, Ti. The comparison with $^{27}$Al highlights the attractive properties of the lighter target --- one composed of a single isotope, with three states in the energy window of interest.  Titanium's multiple isotopes, some with higher level densities, complicate both the interpretation of spectrum distortions and the underlying nuclear physics.  Nevertheless, this example brings out additional possibilities for exploiting inelastic $\mu \rightarrow e$ conversion. In particular, because the two most important isotopes, $^{48}$Ti and $^{46}$Ti, have $0^+$ ground states, transitions like $0^+ \rightarrow 2^+$ isolate single multipoles carrying a definite parity, simplifying the task of isolating the operator source of the CLFV physics.

In summary, NRET provides a flexible and fully general framework for describing $\mu \rightarrow e$ conversion in the nuclear field.  While most attention has been focused on the elastic process, with its potential for coherent enhancement, there are in fact many candidate operators that might encode the low-energy consequences of CLFV.  The advantage of NRET is that it allows for all such possibilities, eliminating any chance of error by assumption.  Here we have focused on the additional CLFV information that might be extracted from Mu2e and COMET by considering inelastic transitions, showing how different interactions can produce distinctive distortions of the near-endpoint electron spectrum.  We have illustrated several CLFV scenarios in which such inelastic contributions will be detectable, given the anticipated energy resolution of Mu2e.  A great deal of additional information thus might be extracted from upcoming experiments by considering both elastic and inelastic conversion.  The Mu2e and COMET target, $^{27}$Al, is an attractive choice for such a program.  In addition to properties already noted, the various nuclear transitions differ in their underlying nuclear physics, with some admitting an allowed response and others having strong quadrupole amplitudes typically associated with deformed mid-shell nuclei.  By comparing the relative responses of all four states, much more information can be extracted on the source of the CLFV. Thus, inelastic $\mu \rightarrow e$ conversion opens up attractive opportunities for combining ideas from particle and nuclear physics.

\begin{acknowledgments}
    We are grateful to Pavel Murat, Kaori Fuyuto, and Liam Fitzpatrick for helpful discussions and to the Mu2e collaboration for making their simulation data available to us. WH acknowledges support by the US Department of Energy under grants DE-SC0004658, DE-SC0023663, and DE-AC02-05CH11231, the National Science Foundation under cooperative agreement 2020275, and the Heising-Simons Foundation under award 00F1C7. ER is supported by the National Science Foundation under cooperative agreement 2020275 and by the U.S. Department of Energy through the Los Alamos National Laboratory. Los Alamos National Laboratory is operated by Triad National Security, LLC, for the National Nuclear Security Administration of U.S. Department of Energy (Contract No. 89233218CNA000001).
\end{acknowledgments}

\bibliography{inelastic_mu2e_long}

\begin{thebibliography}{60}%
\makeatletter
\providecommand \@ifxundefined [1]{%
 \@ifx{#1\undefined}
}%
\providecommand \@ifnum [1]{%
 \ifnum #1\expandafter \@firstoftwo
 \else \expandafter \@secondoftwo
 \fi
}%
\providecommand \@ifx [1]{%
 \ifx #1\expandafter \@firstoftwo
 \else \expandafter \@secondoftwo
 \fi
}%
\providecommand \natexlab [1]{#1}%
\providecommand \enquote  [1]{``#1''}%
\providecommand \bibnamefont  [1]{#1}%
\providecommand \bibfnamefont [1]{#1}%
\providecommand \citenamefont [1]{#1}%
\providecommand \href@noop [0]{\@secondoftwo}%
\providecommand \href [0]{\begingroup \@sanitize@url \@href}%
\providecommand \@href[1]{\@@startlink{#1}\@@href}%
\providecommand \@@href[1]{\endgroup#1\@@endlink}%
\providecommand \@sanitize@url [0]{\catcode `\\12\catcode `\$12\catcode `\&12\catcode `\#12\catcode `\^12\catcode `\_12\catcode `\%12\relax}%
\providecommand \@@startlink[1]{}%
\providecommand \@@endlink[0]{}%
\providecommand \url  [0]{\begingroup\@sanitize@url \@url }%
\providecommand \@url [1]{\endgroup\@href {#1}{\urlprefix }}%
\providecommand \urlprefix  [0]{URL }%
\providecommand \Eprint [0]{\href }%
\providecommand \doibase [0]{https://doi.org/}%
\providecommand \selectlanguage [0]{\@gobble}%
\providecommand \bibinfo  [0]{\@secondoftwo}%
\providecommand \bibfield  [0]{\@secondoftwo}%
\providecommand \translation [1]{[#1]}%
\providecommand \BibitemOpen [0]{}%
\providecommand \bibitemStop [0]{}%
\providecommand \bibitemNoStop [0]{.\EOS\space}%
\providecommand \EOS [0]{\spacefactor3000\relax}%
\providecommand \BibitemShut  [1]{\csname bibitem#1\endcsname}%
\let\auto@bib@innerbib\@empty
\bibitem [{\citenamefont {Barbieri}\ and\ \citenamefont {Hall}(1994)}]{BARBIERI1994212}%
  \BibitemOpen
  \bibfield  {author} {\bibinfo {author} {\bibfnamefont {R.}~\bibnamefont {Barbieri}}\ and\ \bibinfo {author} {\bibfnamefont {L.}~\bibnamefont {Hall}},\ }\bibfield  {title} {\bibinfo {title} {Signals for supersymmetric unification},\ }\href {https://doi.org/https://doi.org/10.1016/0370-2693(94)91368-4} {\bibfield  {journal} {\bibinfo  {journal} {Physics Letters B}\ }\textbf {\bibinfo {volume} {338}},\ \bibinfo {pages} {212} (\bibinfo {year} {1994})}\BibitemShut {NoStop}%
\bibitem [{\citenamefont {Bernstein}\ and\ \citenamefont {Cooper}(2013)}]{Bernstein:2013hba}%
  \BibitemOpen
  \bibfield  {author} {\bibinfo {author} {\bibfnamefont {R.~H.}\ \bibnamefont {Bernstein}}\ and\ \bibinfo {author} {\bibfnamefont {P.~S.}\ \bibnamefont {Cooper}},\ }\bibfield  {title} {\bibinfo {title} {{Charged Lepton Flavor Violation: An Experimenter's Guide}},\ }\href {https://doi.org/10.1016/j.physrep.2013.07.002} {\bibfield  {journal} {\bibinfo  {journal} {Phys. Rept.}\ }\textbf {\bibinfo {volume} {532}},\ \bibinfo {pages} {27} (\bibinfo {year} {2013})},\ \Eprint {https://arxiv.org/abs/1307.5787} {arXiv:1307.5787 [hep-ex]} \BibitemShut {NoStop}%
\bibitem [{\citenamefont {Calibbi}\ and\ \citenamefont {Signorelli}(2018)}]{Calibbi:2017uvl}%
  \BibitemOpen
  \bibfield  {author} {\bibinfo {author} {\bibfnamefont {L.}~\bibnamefont {Calibbi}}\ and\ \bibinfo {author} {\bibfnamefont {G.}~\bibnamefont {Signorelli}},\ }\bibfield  {title} {\bibinfo {title} {{Charged Lepton Flavour Violation: An Experimental and Theoretical Introduction}},\ }\href {https://doi.org/10.1393/ncr/i2018-10144-0} {\bibfield  {journal} {\bibinfo  {journal} {Riv. Nuovo Cim.}\ }\textbf {\bibinfo {volume} {41}},\ \bibinfo {pages} {71} (\bibinfo {year} {2018})},\ \Eprint {https://arxiv.org/abs/1709.00294} {arXiv:1709.00294 [hep-ph]} \BibitemShut {NoStop}%
\bibitem [{\citenamefont {Suzuki}\ \emph {et~al.}(1987)\citenamefont {Suzuki}, \citenamefont {Measday},\ and\ \citenamefont {Roalsvig}}]{Suzuki:1987jf}%
  \BibitemOpen
  \bibfield  {author} {\bibinfo {author} {\bibfnamefont {T.}~\bibnamefont {Suzuki}}, \bibinfo {author} {\bibfnamefont {D.~F.}\ \bibnamefont {Measday}},\ and\ \bibinfo {author} {\bibfnamefont {J.~P.}\ \bibnamefont {Roalsvig}},\ }\bibfield  {title} {\bibinfo {title} {{Total Nuclear Capture Rates for Negative Muons}},\ }\href {https://doi.org/10.1103/PhysRevC.35.2212} {\bibfield  {journal} {\bibinfo  {journal} {Phys. Rev. C}\ }\textbf {\bibinfo {volume} {35}},\ \bibinfo {pages} {2212} (\bibinfo {year} {1987})}\BibitemShut {NoStop}%
\bibitem [{\citenamefont {Bertl}\ \emph {et~al.}(2006)\citenamefont {Bertl} \emph {et~al.}}]{SINDRUMII:2006dvw}%
  \BibitemOpen
  \bibfield  {author} {\bibinfo {author} {\bibfnamefont {W.~H.}\ \bibnamefont {Bertl}} \emph {et~al.} (\bibinfo {collaboration} {SINDRUM II}),\ }\bibfield  {title} {\bibinfo {title} {{A Search for muon to electron conversion in muonic gold}},\ }\href {https://doi.org/10.1140/epjc/s2006-02582-x} {\bibfield  {journal} {\bibinfo  {journal} {Eur. Phys. J. C}\ }\textbf {\bibinfo {volume} {47}},\ \bibinfo {pages} {337} (\bibinfo {year} {2006})}\BibitemShut {NoStop}%
\bibitem [{\citenamefont {Bernstein}(2019)}]{Bernstein_2019}%
  \BibitemOpen
  \bibfield  {author} {\bibinfo {author} {\bibfnamefont {R.~H.}\ \bibnamefont {Bernstein}},\ }\bibfield  {title} {\bibinfo {title} {The mu2e experiment},\ }\href {https://doi.org/10.3389/fphy.2019.00001} {\bibfield  {journal} {\bibinfo  {journal} {Frontiers in Physics}\ }\textbf {\bibinfo {volume} {7}},\ \bibinfo {pages} {1} (\bibinfo {year} {2019})}\BibitemShut {NoStop}%
\bibitem [{\citenamefont {Lee}(2018)}]{10.3389/fphy.2018.00133}%
  \BibitemOpen
  \bibfield  {author} {\bibinfo {author} {\bibfnamefont {M.}~\bibnamefont {Lee}},\ }\bibfield  {title} {\bibinfo {title} {Comet muon conversion experiment in j-parc},\ }\href {https://doi.org/10.3389/fphy.2018.00133} {\bibfield  {journal} {\bibinfo  {journal} {Frontiers in Physics}\ }\textbf {\bibinfo {volume} {6}},\ \bibinfo {pages} {133} (\bibinfo {year} {2018})}\BibitemShut {NoStop}%
\bibitem [{\citenamefont {Cirigliano}\ \emph {et~al.}(2009)\citenamefont {Cirigliano}, \citenamefont {Kitano}, \citenamefont {Okada},\ and\ \citenamefont {Tuzon}}]{Cirigliano:2009bz}%
  \BibitemOpen
  \bibfield  {author} {\bibinfo {author} {\bibfnamefont {V.}~\bibnamefont {Cirigliano}}, \bibinfo {author} {\bibfnamefont {R.}~\bibnamefont {Kitano}}, \bibinfo {author} {\bibfnamefont {Y.}~\bibnamefont {Okada}},\ and\ \bibinfo {author} {\bibfnamefont {P.}~\bibnamefont {Tuzon}},\ }\bibfield  {title} {\bibinfo {title} {{On the model discriminating power of mu ---\ensuremath{>} e conversion in nuclei}},\ }\href {https://doi.org/10.1103/PhysRevD.80.013002} {\bibfield  {journal} {\bibinfo  {journal} {Phys. Rev. D}\ }\textbf {\bibinfo {volume} {80}},\ \bibinfo {pages} {013002} (\bibinfo {year} {2009})},\ \Eprint {https://arxiv.org/abs/0904.0957} {arXiv:0904.0957 [hep-ph]} \BibitemShut {NoStop}%
\bibitem [{\citenamefont {Davidson}\ \emph {et~al.}(2019)\citenamefont {Davidson}, \citenamefont {Kuno},\ and\ \citenamefont {Yamanaka}}]{DAVIDSON2019380}%
  \BibitemOpen
  \bibfield  {author} {\bibinfo {author} {\bibfnamefont {S.}~\bibnamefont {Davidson}}, \bibinfo {author} {\bibfnamefont {Y.}~\bibnamefont {Kuno}},\ and\ \bibinfo {author} {\bibfnamefont {M.}~\bibnamefont {Yamanaka}},\ }\bibfield  {title} {\bibinfo {title} {Selecting $\mu\rightarrow e$ conversion targets to distinguish lepton flavour-changing operators},\ }\href {https://doi.org/https://doi.org/10.1016/j.physletb.2019.01.042} {\bibfield  {journal} {\bibinfo  {journal} {Physics Letters B}\ }\textbf {\bibinfo {volume} {790}},\ \bibinfo {pages} {380} (\bibinfo {year} {2019})}\BibitemShut {NoStop}%
\bibitem [{\citenamefont {Cirigliano}\ \emph {et~al.}(2022)\citenamefont {Cirigliano}, \citenamefont {Fuyuto}, \citenamefont {Ramsey-Musolf},\ and\ \citenamefont {Rule}}]{Cirigliano:2022ekw}%
  \BibitemOpen
  \bibfield  {author} {\bibinfo {author} {\bibfnamefont {V.}~\bibnamefont {Cirigliano}}, \bibinfo {author} {\bibfnamefont {K.}~\bibnamefont {Fuyuto}}, \bibinfo {author} {\bibfnamefont {M.~J.}\ \bibnamefont {Ramsey-Musolf}},\ and\ \bibinfo {author} {\bibfnamefont {E.}~\bibnamefont {Rule}},\ }\bibfield  {title} {\bibinfo {title} {{Next-to-leading order scalar contributions to \ensuremath{\mu}\textrightarrow{}e conversion}},\ }\href {https://doi.org/10.1103/PhysRevC.105.055504} {\bibfield  {journal} {\bibinfo  {journal} {Phys. Rev. C}\ }\textbf {\bibinfo {volume} {105}},\ \bibinfo {pages} {055504} (\bibinfo {year} {2022})},\ \Eprint {https://arxiv.org/abs/2203.09547} {arXiv:2203.09547 [hep-ph]} \BibitemShut {NoStop}%
\bibitem [{\citenamefont {Heeck}\ \emph {et~al.}(2022)\citenamefont {Heeck}, \citenamefont {Szafron},\ and\ \citenamefont {Uesaka}}]{HEECK2022115833}%
  \BibitemOpen
  \bibfield  {author} {\bibinfo {author} {\bibfnamefont {J.}~\bibnamefont {Heeck}}, \bibinfo {author} {\bibfnamefont {R.}~\bibnamefont {Szafron}},\ and\ \bibinfo {author} {\bibfnamefont {Y.}~\bibnamefont {Uesaka}},\ }\bibfield  {title} {\bibinfo {title} {Isotope dependence of muon-to-electron conversion},\ }\href {https://doi.org/https://doi.org/10.1016/j.nuclphysb.2022.115833} {\bibfield  {journal} {\bibinfo  {journal} {Nuclear Physics B}\ }\textbf {\bibinfo {volume} {980}},\ \bibinfo {pages} {115833} (\bibinfo {year} {2022})}\BibitemShut {NoStop}%
\bibitem [{\citenamefont {Rule}\ \emph {et~al.}(2023)\citenamefont {Rule}, \citenamefont {Haxton},\ and\ \citenamefont {McElvain}}]{Rule:2021oxe}%
  \BibitemOpen
  \bibfield  {author} {\bibinfo {author} {\bibfnamefont {E.}~\bibnamefont {Rule}}, \bibinfo {author} {\bibfnamefont {W.~C.}\ \bibnamefont {Haxton}},\ and\ \bibinfo {author} {\bibfnamefont {K.}~\bibnamefont {McElvain}},\ }\bibfield  {title} {\bibinfo {title} {{Nuclear-Level Effective Theory of \ensuremath{\mu}\textrightarrow{}e Conversion}},\ }\href {https://doi.org/10.1103/PhysRevLett.130.131901} {\bibfield  {journal} {\bibinfo  {journal} {Phys. Rev. Lett.}\ }\textbf {\bibinfo {volume} {130}},\ \bibinfo {pages} {131901} (\bibinfo {year} {2023})},\ \Eprint {https://arxiv.org/abs/2109.13503} {arXiv:2109.13503 [hep-ph]} \BibitemShut {NoStop}%
\bibitem [{\citenamefont {Haxton}\ \emph {et~al.}(2023)\citenamefont {Haxton}, \citenamefont {Rule}, \citenamefont {McElvain},\ and\ \citenamefont {Ramsey-Musolf}}]{Haxton:2022piv}%
  \BibitemOpen
  \bibfield  {author} {\bibinfo {author} {\bibfnamefont {W.~C.}\ \bibnamefont {Haxton}}, \bibinfo {author} {\bibfnamefont {E.}~\bibnamefont {Rule}}, \bibinfo {author} {\bibfnamefont {K.}~\bibnamefont {McElvain}},\ and\ \bibinfo {author} {\bibfnamefont {M.~J.}\ \bibnamefont {Ramsey-Musolf}},\ }\bibfield  {title} {\bibinfo {title} {{Nuclear-level effective theory of \ensuremath{\mu}\textrightarrow{}e conversion: Formalism and applications}},\ }\href {https://doi.org/10.1103/PhysRevC.107.035504} {\bibfield  {journal} {\bibinfo  {journal} {Phys. Rev. C}\ }\textbf {\bibinfo {volume} {107}},\ \bibinfo {pages} {035504} (\bibinfo {year} {2023})},\ \Eprint {https://arxiv.org/abs/2208.07945} {arXiv:2208.07945 [nucl-th]} \BibitemShut {NoStop}%
\bibitem [{\citenamefont {Haxton}\ \emph {et~al.}(2024)\citenamefont {Haxton}, \citenamefont {McElvain}, \citenamefont {Menzo}, \citenamefont {Rule},\ and\ \citenamefont {Zupan}}]{Haxton:2024lyc}%
  \BibitemOpen
  \bibfield  {author} {\bibinfo {author} {\bibfnamefont {W.}~\bibnamefont {Haxton}}, \bibinfo {author} {\bibfnamefont {K.}~\bibnamefont {McElvain}}, \bibinfo {author} {\bibfnamefont {T.}~\bibnamefont {Menzo}}, \bibinfo {author} {\bibfnamefont {E.}~\bibnamefont {Rule}},\ and\ \bibinfo {author} {\bibfnamefont {J.}~\bibnamefont {Zupan}},\ }\bibfield  {title} {\bibinfo {title} {{Effective theory tower for $\mu\rightarrow e$ conversion}},\ }\href@noop {} {\  (\bibinfo {year} {2024})},\ \Eprint {https://arxiv.org/abs/2406.13818} {arXiv:2406.13818 [hep-ph]} \BibitemShut {NoStop}%
\bibitem [{\citenamefont {Aebischer}\ \emph {et~al.}(2018)\citenamefont {Aebischer}, \citenamefont {Kumar},\ and\ \citenamefont {Straub}}]{Aebischer:2018bkb}%
  \BibitemOpen
  \bibfield  {author} {\bibinfo {author} {\bibfnamefont {J.}~\bibnamefont {Aebischer}}, \bibinfo {author} {\bibfnamefont {J.}~\bibnamefont {Kumar}},\ and\ \bibinfo {author} {\bibfnamefont {D.~M.}\ \bibnamefont {Straub}},\ }\bibfield  {title} {\bibinfo {title} {{Wilson: a Python package for the running and matching of Wilson coefficients above and below the electroweak scale}},\ }\href {https://doi.org/10.1140/epjc/s10052-018-6492-7} {\bibfield  {journal} {\bibinfo  {journal} {Eur. Phys. J. C}\ }\textbf {\bibinfo {volume} {78}},\ \bibinfo {pages} {1026} (\bibinfo {year} {2018})},\ \Eprint {https://arxiv.org/abs/1804.05033} {arXiv:1804.05033 [hep-ph]} \BibitemShut {NoStop}%
\bibitem [{\citenamefont {Fuentes-Martin}\ \emph {et~al.}(2021)\citenamefont {Fuentes-Martin}, \citenamefont {Ruiz-Femenia}, \citenamefont {Vicente},\ and\ \citenamefont {Virto}}]{Fuentes-Martin:2020zaz}%
  \BibitemOpen
  \bibfield  {author} {\bibinfo {author} {\bibfnamefont {J.}~\bibnamefont {Fuentes-Martin}}, \bibinfo {author} {\bibfnamefont {P.}~\bibnamefont {Ruiz-Femenia}}, \bibinfo {author} {\bibfnamefont {A.}~\bibnamefont {Vicente}},\ and\ \bibinfo {author} {\bibfnamefont {J.}~\bibnamefont {Virto}},\ }\bibfield  {title} {\bibinfo {title} {{DsixTools 2.0: The Effective Field Theory Toolkit}},\ }\href {https://doi.org/10.1140/epjc/s10052-020-08778-y} {\bibfield  {journal} {\bibinfo  {journal} {Eur. Phys. J. C}\ }\textbf {\bibinfo {volume} {81}},\ \bibinfo {pages} {167} (\bibinfo {year} {2021})},\ \Eprint {https://arxiv.org/abs/2010.16341} {arXiv:2010.16341 [hep-ph]} \BibitemShut {NoStop}%
\bibitem [{\citenamefont {Primakoff}(1959)}]{Primakoff59}%
  \BibitemOpen
  \bibfield  {author} {\bibinfo {author} {\bibfnamefont {H.}~\bibnamefont {Primakoff}},\ }\bibfield  {title} {\bibinfo {title} {{Theory of Muon Capture}},\ }\href {https://doi.org/10.1103/RevModPhys.31.802} {\bibfield  {journal} {\bibinfo  {journal} {Rev. Mod. Phys.}\ }\textbf {\bibinfo {volume} {31}},\ \bibinfo {pages} {802} (\bibinfo {year} {1959})}\BibitemShut {NoStop}%
\bibitem [{\citenamefont {Byrum}\ \emph {et~al.}(2022)\citenamefont {Byrum} \emph {et~al.}}]{Mu2e-II:2022blh}%
  \BibitemOpen
  \bibfield  {author} {\bibinfo {author} {\bibfnamefont {K.}~\bibnamefont {Byrum}} \emph {et~al.} (\bibinfo {collaboration} {Mu2e-II}),\ }\bibfield  {title} {\bibinfo {title} {{Mu2e-II: Muon to electron conversion with PIP-II}},\ }in\ \href@noop {} {\emph {\bibinfo {booktitle} {{Snowmass 2021}}}}\ (\bibinfo {year} {2022})\ \Eprint {https://arxiv.org/abs/2203.07569} {arXiv:2203.07569 [hep-ex]} \BibitemShut {NoStop}%
\bibitem [{\citenamefont {Kuno}(2005)}]{KUNO2005376}%
  \BibitemOpen
  \bibfield  {author} {\bibinfo {author} {\bibfnamefont {Y.}~\bibnamefont {Kuno}},\ }\bibfield  {title} {\bibinfo {title} {Prism/prime},\ }\href {https://doi.org/https://doi.org/10.1016/j.nuclphysbps.2005.05.073} {\bibfield  {journal} {\bibinfo  {journal} {Nuclear Physics B - Proceedings Supplements}\ }\textbf {\bibinfo {volume} {149}},\ \bibinfo {pages} {376} (\bibinfo {year} {2005})},\ \bibinfo {note} {nuFact04}\BibitemShut {NoStop}%
\bibitem [{\citenamefont {Barlow}(2011)}]{BARLOW201144}%
  \BibitemOpen
  \bibfield  {author} {\bibinfo {author} {\bibfnamefont {R.}~\bibnamefont {Barlow}},\ }\bibfield  {title} {\bibinfo {title} {The prism/prime project},\ }\href {https://doi.org/https://doi.org/10.1016/j.nuclphysbps.2011.06.009} {\bibfield  {journal} {\bibinfo  {journal} {Nuclear Physics B - Proceedings Supplements}\ }\textbf {\bibinfo {volume} {218}},\ \bibinfo {pages} {44} (\bibinfo {year} {2011})},\ \bibinfo {note} {proceedings of the Eleventh International Workshop on Tau Lepton Physics}\BibitemShut {NoStop}%
\bibitem [{\citenamefont {Aoki}\ \emph {et~al.}(2022)\citenamefont {Aoki} \emph {et~al.}}]{CGroup:2022tli}%
  \BibitemOpen
  \bibfield  {author} {\bibinfo {author} {\bibfnamefont {M.}~\bibnamefont {Aoki}} \emph {et~al.} (\bibinfo {collaboration} {C. Group}),\ }\bibfield  {title} {\bibinfo {title} {{A New Charged Lepton Flavor Violation Program at Fermilab}},\ }in\ \href@noop {} {\emph {\bibinfo {booktitle} {{Snowmass 2021}}}}\ (\bibinfo {year} {2022})\ \Eprint {https://arxiv.org/abs/2203.08278} {arXiv:2203.08278 [hep-ex]} \BibitemShut {NoStop}%
\bibitem [{\citenamefont {Haxton}\ and\ \citenamefont {Rule}(2024)}]{Haxton:2024ecp}%
  \BibitemOpen
  \bibfield  {author} {\bibinfo {author} {\bibfnamefont {W.~C.}\ \bibnamefont {Haxton}}\ and\ \bibinfo {author} {\bibfnamefont {E.}~\bibnamefont {Rule}},\ }\bibfield  {title} {\bibinfo {title} {{Distinguishing charged lepton flavor violation scenarios with inelastic $\mu\rightarrow e$ conversion}},\ }\href@noop {} {\  (\bibinfo {year} {2024})},\ \Eprint {https://arxiv.org/abs/2404.17166} {arXiv:2404.17166 [hep-ph]} \BibitemShut {NoStop}%
\bibitem [{\citenamefont {Weinberg}\ and\ \citenamefont {Feinberg}(1959)}]{PhysRevLett.3.111}%
  \BibitemOpen
  \bibfield  {author} {\bibinfo {author} {\bibfnamefont {S.}~\bibnamefont {Weinberg}}\ and\ \bibinfo {author} {\bibfnamefont {G.}~\bibnamefont {Feinberg}},\ }\bibfield  {title} {\bibinfo {title} {Electromagnetic transitions between $\ensuremath{\mu}$ meson and electron},\ }\href {https://doi.org/10.1103/PhysRevLett.3.111} {\bibfield  {journal} {\bibinfo  {journal} {Phys. Rev. Lett.}\ }\textbf {\bibinfo {volume} {3}},\ \bibinfo {pages} {111} (\bibinfo {year} {1959})}\BibitemShut {NoStop}%
\bibitem [{\citenamefont {Kosmas}\ and\ \citenamefont {Vergados}(1990)}]{KOSMAS1990641}%
  \BibitemOpen
  \bibfield  {author} {\bibinfo {author} {\bibfnamefont {T.}~\bibnamefont {Kosmas}}\ and\ \bibinfo {author} {\bibfnamefont {J.}~\bibnamefont {Vergados}},\ }\bibfield  {title} {\bibinfo {title} {Study of the flavour violating ($\mu^-$, $e^-$) conversion in nuclei},\ }\href {https://doi.org/https://doi.org/10.1016/0375-9474(90)90353-N} {\bibfield  {journal} {\bibinfo  {journal} {Nuclear Physics A}\ }\textbf {\bibinfo {volume} {510}},\ \bibinfo {pages} {641} (\bibinfo {year} {1990})}\BibitemShut {NoStop}%
\bibitem [{\citenamefont {Chiang}\ \emph {et~al.}(1993)\citenamefont {Chiang}, \citenamefont {Oset}, \citenamefont {Kosmas}, \citenamefont {Faessler},\ and\ \citenamefont {Vergados}}]{CHIANG1993526}%
  \BibitemOpen
  \bibfield  {author} {\bibinfo {author} {\bibfnamefont {H.}~\bibnamefont {Chiang}}, \bibinfo {author} {\bibfnamefont {E.}~\bibnamefont {Oset}}, \bibinfo {author} {\bibfnamefont {T.}~\bibnamefont {Kosmas}}, \bibinfo {author} {\bibfnamefont {A.}~\bibnamefont {Faessler}},\ and\ \bibinfo {author} {\bibfnamefont {J.}~\bibnamefont {Vergados}},\ }\bibfield  {title} {\bibinfo {title} {Coherent and incoherent $(\mu^-, e^-)$ conversion in nuclei},\ }\href {https://doi.org/https://doi.org/10.1016/0375-9474(93)90259-Z} {\bibfield  {journal} {\bibinfo  {journal} {Nuclear Physics A}\ }\textbf {\bibinfo {volume} {559}},\ \bibinfo {pages} {526} (\bibinfo {year} {1993})}\BibitemShut {NoStop}%
\bibitem [{\citenamefont {Kosmas}\ \emph {et~al.}(1994)\citenamefont {Kosmas}, \citenamefont {Vergados}, \citenamefont {Civitarese},\ and\ \citenamefont {Faessler}}]{KOSMAS1994637}%
  \BibitemOpen
  \bibfield  {author} {\bibinfo {author} {\bibfnamefont {T.}~\bibnamefont {Kosmas}}, \bibinfo {author} {\bibfnamefont {J.}~\bibnamefont {Vergados}}, \bibinfo {author} {\bibfnamefont {O.}~\bibnamefont {Civitarese}},\ and\ \bibinfo {author} {\bibfnamefont {A.}~\bibnamefont {Faessler}},\ }\bibfield  {title} {\bibinfo {title} {Study of the muon number violating ($\mu^-$, $e^-$) conversion in a nucleus by using quasi-particle rpa},\ }\href {https://doi.org/https://doi.org/10.1016/0375-9474(94)90077-9} {\bibfield  {journal} {\bibinfo  {journal} {Nuclear Physics A}\ }\textbf {\bibinfo {volume} {570}},\ \bibinfo {pages} {637} (\bibinfo {year} {1994})}\BibitemShut {NoStop}%
\bibitem [{\citenamefont {Siiskonen}\ \emph {et~al.}(2000)\citenamefont {Siiskonen}, \citenamefont {Suhonen},\ and\ \citenamefont {Kosmas}}]{PhysRevC.62.035502}%
  \BibitemOpen
  \bibfield  {author} {\bibinfo {author} {\bibfnamefont {T.}~\bibnamefont {Siiskonen}}, \bibinfo {author} {\bibfnamefont {J.}~\bibnamefont {Suhonen}},\ and\ \bibinfo {author} {\bibfnamefont {T.~S.}\ \bibnamefont {Kosmas}},\ }\bibfield  {title} {\bibinfo {title} {Realistic nuclear matrix elements for the lepton-flavor violating ${\ensuremath{\mu}}^{\ensuremath{-}}\ensuremath{\rightarrow}{e}^{\ensuremath{-}}$ conversion in ${}^{27}\mathrm{Al}$ and ${}^{48}\mathrm{Ti}$},\ }\href {https://doi.org/10.1103/PhysRevC.62.035502} {\bibfield  {journal} {\bibinfo  {journal} {Phys. Rev. C}\ }\textbf {\bibinfo {volume} {62}},\ \bibinfo {pages} {035502} (\bibinfo {year} {2000})}\BibitemShut {NoStop}%
\bibitem [{\citenamefont {Kosmas}(2001)}]{KOSMAS2001443}%
  \BibitemOpen
  \bibfield  {author} {\bibinfo {author} {\bibfnamefont {T.}~\bibnamefont {Kosmas}},\ }\bibfield  {title} {\bibinfo {title} {Exotic $\mu^-\rightarrow e^-$ conversion in nuclei: energy moments of the transition strength and average energy of the outgoing $e^-$},\ }\href {https://doi.org/https://doi.org/10.1016/S0375-9474(00)00471-1} {\bibfield  {journal} {\bibinfo  {journal} {Nuclear Physics A}\ }\textbf {\bibinfo {volume} {683}},\ \bibinfo {pages} {443} (\bibinfo {year} {2001})}\BibitemShut {NoStop}%
\bibitem [{\citenamefont {Civitarese}\ and\ \citenamefont {Tarutina}(2019)}]{PhysRevC.99.065504}%
  \BibitemOpen
  \bibfield  {author} {\bibinfo {author} {\bibfnamefont {O.}~\bibnamefont {Civitarese}}\ and\ \bibinfo {author} {\bibfnamefont {T.}~\bibnamefont {Tarutina}},\ }\bibfield  {title} {\bibinfo {title} {Multipole decomposition of the rate of muon-to-electron $({\ensuremath{\mu}}^{\ensuremath{-}}\ensuremath{\longrightarrow}{e}^{\ensuremath{-}})$ conversion in $^{208}\mathrm{Pb}$},\ }\href {https://doi.org/10.1103/PhysRevC.99.065504} {\bibfield  {journal} {\bibinfo  {journal} {Phys. Rev. C}\ }\textbf {\bibinfo {volume} {99}},\ \bibinfo {pages} {065504} (\bibinfo {year} {2019})}\BibitemShut {NoStop}%
\bibitem [{\citenamefont {{De Vries}}\ \emph {et~al.}(1987)\citenamefont {{De Vries}}, \citenamefont {{De Jager}},\ and\ \citenamefont {{De Vries}}}]{DEVRIES1987495}%
  \BibitemOpen
  \bibfield  {author} {\bibinfo {author} {\bibfnamefont {H.}~\bibnamefont {{De Vries}}}, \bibinfo {author} {\bibfnamefont {C.}~\bibnamefont {{De Jager}}},\ and\ \bibinfo {author} {\bibfnamefont {C.}~\bibnamefont {{De Vries}}},\ }\bibfield  {title} {\bibinfo {title} {Nuclear charge-density-distribution parameters from elastic electron scattering},\ }\href {https://doi.org/https://doi.org/10.1016/0092-640X(87)90013-1} {\bibfield  {journal} {\bibinfo  {journal} {Atomic Data and Nuclear Data Tables}\ }\textbf {\bibinfo {volume} {36}},\ \bibinfo {pages} {495} (\bibinfo {year} {1987})}\BibitemShut {NoStop}%
\bibitem [{\citenamefont {Kitano}\ \emph {et~al.}(2002)\citenamefont {Kitano}, \citenamefont {Koike},\ and\ \citenamefont {Okada}}]{Kitano:2002mt}%
  \BibitemOpen
  \bibfield  {author} {\bibinfo {author} {\bibfnamefont {R.}~\bibnamefont {Kitano}}, \bibinfo {author} {\bibfnamefont {M.}~\bibnamefont {Koike}},\ and\ \bibinfo {author} {\bibfnamefont {Y.}~\bibnamefont {Okada}},\ }\bibfield  {title} {\bibinfo {title} {{Detailed calculation of lepton flavor violating muon electron conversion rate for various nuclei}},\ }\href {https://doi.org/10.1103/PhysRevD.76.059902} {\bibfield  {journal} {\bibinfo  {journal} {Phys. Rev. D}\ }\textbf {\bibinfo {volume} {66}},\ \bibinfo {pages} {096002} (\bibinfo {year} {2002})},\ \bibinfo {note} {[Erratum: Phys.Rev.D 76, 059902 (2007)]},\ \Eprint {https://arxiv.org/abs/hep-ph/0203110} {arXiv:hep-ph/0203110} \BibitemShut {NoStop}%
\bibitem [{\citenamefont {Knoll}(1974)}]{KNOLL1974462}%
  \BibitemOpen
  \bibfield  {author} {\bibinfo {author} {\bibfnamefont {J.}~\bibnamefont {Knoll}},\ }\bibfield  {title} {\bibinfo {title} {An analytic description of inelastic electron scattering on nuclei},\ }\href {https://doi.org/https://doi.org/10.1016/0375-9474(74)90700-3} {\bibfield  {journal} {\bibinfo  {journal} {Nuclear Physics A}\ }\textbf {\bibinfo {volume} {223}},\ \bibinfo {pages} {462} (\bibinfo {year} {1974})}\BibitemShut {NoStop}%
\bibitem [{\citenamefont {Lenz}\ and\ \citenamefont {Rosenfelder}(1971)}]{LENZ1971513}%
  \BibitemOpen
  \bibfield  {author} {\bibinfo {author} {\bibfnamefont {F.}~\bibnamefont {Lenz}}\ and\ \bibinfo {author} {\bibfnamefont {R.}~\bibnamefont {Rosenfelder}},\ }\bibfield  {title} {\bibinfo {title} {Nuclear radii in the high-energy limit of elastic electron scattering},\ }\href {https://doi.org/https://doi.org/10.1016/0375-9474(71)90933-X} {\bibfield  {journal} {\bibinfo  {journal} {Nuclear Physics A}\ }\textbf {\bibinfo {volume} {176}},\ \bibinfo {pages} {513} (\bibinfo {year} {1971})}\BibitemShut {NoStop}%
\bibitem [{\citenamefont {Dohmen}\ \emph {et~al.}(1993)\citenamefont {Dohmen} \emph {et~al.}}]{SINDRUMII:1993gxf}%
  \BibitemOpen
  \bibfield  {author} {\bibinfo {author} {\bibfnamefont {C.}~\bibnamefont {Dohmen}} \emph {et~al.} (\bibinfo {collaboration} {SINDRUM II}),\ }\bibfield  {title} {\bibinfo {title} {{Test of lepton flavor conservation in mu ---\ensuremath{>} e conversion on titanium}},\ }\href {https://doi.org/10.1016/0370-2693(93)91383-X} {\bibfield  {journal} {\bibinfo  {journal} {Phys. Lett. B}\ }\textbf {\bibinfo {volume} {317}},\ \bibinfo {pages} {631} (\bibinfo {year} {1993})}\BibitemShut {NoStop}%
\bibitem [{\citenamefont {Fitzpatrick}\ \emph {et~al.}(2013)\citenamefont {Fitzpatrick}, \citenamefont {Haxton}, \citenamefont {Katz}, \citenamefont {Lubbers},\ and\ \citenamefont {Xu}}]{Fitzpatrick:2012ix}%
  \BibitemOpen
  \bibfield  {author} {\bibinfo {author} {\bibfnamefont {A.~L.}\ \bibnamefont {Fitzpatrick}}, \bibinfo {author} {\bibfnamefont {W.}~\bibnamefont {Haxton}}, \bibinfo {author} {\bibfnamefont {E.}~\bibnamefont {Katz}}, \bibinfo {author} {\bibfnamefont {N.}~\bibnamefont {Lubbers}},\ and\ \bibinfo {author} {\bibfnamefont {Y.}~\bibnamefont {Xu}},\ }\bibfield  {title} {\bibinfo {title} {{The Effective Field Theory of Dark Matter Direct Detection}},\ }\href {https://doi.org/10.1088/1475-7516/2013/02/004} {\bibfield  {journal} {\bibinfo  {journal} {JCAP}\ }\textbf {\bibinfo {volume} {02}},\ \bibinfo {pages} {004}},\ \Eprint {https://arxiv.org/abs/1203.3542} {arXiv:1203.3542 [hep-ph]} \BibitemShut {NoStop}%
\bibitem [{\citenamefont {Donnelly}\ and\ \citenamefont {Haxton}(1979{\natexlab{a}})}]{Donnelly:1979ezn}%
  \BibitemOpen
  \bibfield  {author} {\bibinfo {author} {\bibfnamefont {T.~W.}\ \bibnamefont {Donnelly}}\ and\ \bibinfo {author} {\bibfnamefont {W.~C.}\ \bibnamefont {Haxton}},\ }\bibfield  {title} {\bibinfo {title} {{Multipole operators in semileptonic weak and electromagnetic interactions with nuclei}},\ }\href {https://doi.org/10.1016/0092-640X(79)90003-2} {\bibfield  {journal} {\bibinfo  {journal} {Atom. Data Nucl. Data Tabl.}\ }\textbf {\bibinfo {volume} {23}},\ \bibinfo {pages} {103} (\bibinfo {year} {1979}{\natexlab{a}})}\BibitemShut {NoStop}%
\bibitem [{\citenamefont {Friar}\ and\ \citenamefont {Fallieros}(1984)}]{PhysRevC.29.1645}%
  \BibitemOpen
  \bibfield  {author} {\bibinfo {author} {\bibfnamefont {J.~L.}\ \bibnamefont {Friar}}\ and\ \bibinfo {author} {\bibfnamefont {S.}~\bibnamefont {Fallieros}},\ }\bibfield  {title} {\bibinfo {title} {Extended siegert theorem},\ }\href {https://doi.org/10.1103/PhysRevC.29.1645} {\bibfield  {journal} {\bibinfo  {journal} {Phys. Rev. C}\ }\textbf {\bibinfo {volume} {29}},\ \bibinfo {pages} {1645} (\bibinfo {year} {1984})}\BibitemShut {NoStop}%
\bibitem [{\citenamefont {Friar}\ and\ \citenamefont {Haxton}(1985)}]{PhysRevC.31.2027}%
  \BibitemOpen
  \bibfield  {author} {\bibinfo {author} {\bibfnamefont {J.~L.}\ \bibnamefont {Friar}}\ and\ \bibinfo {author} {\bibfnamefont {W.~C.}\ \bibnamefont {Haxton}},\ }\bibfield  {title} {\bibinfo {title} {Current conservation and the transverse electric multipole field},\ }\href {https://doi.org/10.1103/PhysRevC.31.2027} {\bibfield  {journal} {\bibinfo  {journal} {Phys. Rev. C}\ }\textbf {\bibinfo {volume} {31}},\ \bibinfo {pages} {2027} (\bibinfo {year} {1985})}\BibitemShut {NoStop}%
\bibitem [{\citenamefont {Hayes}\ and\ \citenamefont {Friar}(2018)}]{PhysRevC.98.065505}%
  \BibitemOpen
  \bibfield  {author} {\bibinfo {author} {\bibfnamefont {A.~C.}\ \bibnamefont {Hayes}}\ and\ \bibinfo {author} {\bibfnamefont {J.~L.}\ \bibnamefont {Friar}},\ }\bibfield  {title} {\bibinfo {title} {Effect of siegert's theorem on low-energy neutrino-nucleus interactions},\ }\href {https://doi.org/10.1103/PhysRevC.98.065505} {\bibfield  {journal} {\bibinfo  {journal} {Phys. Rev. C}\ }\textbf {\bibinfo {volume} {98}},\ \bibinfo {pages} {065505} (\bibinfo {year} {2018})}\BibitemShut {NoStop}%
\bibitem [{\citenamefont {Brown}\ and\ \citenamefont {Richter}(2006)}]{PhysRevC.74.034315}%
  \BibitemOpen
  \bibfield  {author} {\bibinfo {author} {\bibfnamefont {B.~A.}\ \bibnamefont {Brown}}\ and\ \bibinfo {author} {\bibfnamefont {W.~A.}\ \bibnamefont {Richter}},\ }\bibfield  {title} {\bibinfo {title} {New ``usd'' hamiltonians for the $\mathit{sd}$ shell},\ }\href {https://doi.org/10.1103/PhysRevC.74.034315} {\bibfield  {journal} {\bibinfo  {journal} {Phys. Rev. C}\ }\textbf {\bibinfo {volume} {74}},\ \bibinfo {pages} {034315} (\bibinfo {year} {2006})}\BibitemShut {NoStop}%
\bibitem [{\citenamefont {Brown}\ and\ \citenamefont {Wildenthal}(1988)}]{bw}%
  \BibitemOpen
  \bibfield  {author} {\bibinfo {author} {\bibfnamefont {B.~A.}\ \bibnamefont {Brown}}\ and\ \bibinfo {author} {\bibfnamefont {B.~H.}\ \bibnamefont {Wildenthal}},\ }\bibfield  {title} {\bibinfo {title} {Status of the nuclear shell model},\ }\href {https://doi.org/10.1146/annurev.ns.38.120188.000333} {\bibfield  {journal} {\bibinfo  {journal} {Annual Review of Nuclear and Particle Science}\ }\textbf {\bibinfo {volume} {38}},\ \bibinfo {pages} {29} (\bibinfo {year} {1988})},\ \Eprint {https://arxiv.org/abs/https://doi.org/10.1146/annurev.ns.38.120188.000333} {https://doi.org/10.1146/annurev.ns.38.120188.000333} \BibitemShut {NoStop}%
\bibitem [{\citenamefont {{Shamsuzzoha Basunia}}(2011)}]{SHAMSUZZOHABASUNIA20111875}%
  \BibitemOpen
  \bibfield  {author} {\bibinfo {author} {\bibfnamefont {M.}~\bibnamefont {{Shamsuzzoha Basunia}}},\ }\bibfield  {title} {\bibinfo {title} {Nuclear data sheets for a = 27},\ }\href {https://doi.org/https://doi.org/10.1016/j.nds.2011.08.001} {\bibfield  {journal} {\bibinfo  {journal} {Nuclear Data Sheets}\ }\textbf {\bibinfo {volume} {112}},\ \bibinfo {pages} {1875} (\bibinfo {year} {2011})}\BibitemShut {NoStop}%
\bibitem [{\citenamefont {Czarnecki}\ \emph {et~al.}(2011)\citenamefont {Czarnecki}, \citenamefont {Garcia~i Tormo},\ and\ \citenamefont {Marciano}}]{PhysRevD.84.013006}%
  \BibitemOpen
  \bibfield  {author} {\bibinfo {author} {\bibfnamefont {A.}~\bibnamefont {Czarnecki}}, \bibinfo {author} {\bibfnamefont {X.}~\bibnamefont {Garcia~i Tormo}},\ and\ \bibinfo {author} {\bibfnamefont {W.~J.}\ \bibnamefont {Marciano}},\ }\bibfield  {title} {\bibinfo {title} {Muon decay in orbit: Spectrum of high-energy electrons},\ }\href {https://doi.org/10.1103/PhysRevD.84.013006} {\bibfield  {journal} {\bibinfo  {journal} {Phys. Rev. D}\ }\textbf {\bibinfo {volume} {84}},\ \bibinfo {pages} {013006} (\bibinfo {year} {2011})}\BibitemShut {NoStop}%
\bibitem [{\citenamefont {Szafron}\ and\ \citenamefont {Czarnecki}(2016{\natexlab{a}})}]{PhysRevD.94.051301}%
  \BibitemOpen
  \bibfield  {author} {\bibinfo {author} {\bibfnamefont {R.}~\bibnamefont {Szafron}}\ and\ \bibinfo {author} {\bibfnamefont {A.}~\bibnamefont {Czarnecki}},\ }\bibfield  {title} {\bibinfo {title} {Bound muon decay spectrum in the leading logarithmic accuracy},\ }\href {https://doi.org/10.1103/PhysRevD.94.051301} {\bibfield  {journal} {\bibinfo  {journal} {Phys. Rev. D}\ }\textbf {\bibinfo {volume} {94}},\ \bibinfo {pages} {051301} (\bibinfo {year} {2016}{\natexlab{a}})}\BibitemShut {NoStop}%
\bibitem [{\citenamefont {Szafron}\ and\ \citenamefont {Czarnecki}(2016{\natexlab{b}})}]{SZAFRON201661}%
  \BibitemOpen
  \bibfield  {author} {\bibinfo {author} {\bibfnamefont {R.}~\bibnamefont {Szafron}}\ and\ \bibinfo {author} {\bibfnamefont {A.}~\bibnamefont {Czarnecki}},\ }\bibfield  {title} {\bibinfo {title} {High-energy electrons from the muon decay in orbit: Radiative corrections},\ }\href {https://doi.org/https://doi.org/10.1016/j.physletb.2015.12.008} {\bibfield  {journal} {\bibinfo  {journal} {Physics Letters B}\ }\textbf {\bibinfo {volume} {753}},\ \bibinfo {pages} {61} (\bibinfo {year} {2016}{\natexlab{b}})}\BibitemShut {NoStop}%
\bibitem [{\citenamefont {Abdi}\ \emph {et~al.}(2023)\citenamefont {Abdi} \emph {et~al.}}]{Mu2e:2022ggl}%
  \BibitemOpen
  \bibfield  {author} {\bibinfo {author} {\bibfnamefont {F.}~\bibnamefont {Abdi}} \emph {et~al.} (\bibinfo {collaboration} {Mu2e}),\ }\bibfield  {title} {\bibinfo {title} {{Mu2e Run I Sensitivity Projections for the Neutrinoless $\mu^- \to e^-$ Conversion Search in Aluminum}},\ }\href {https://doi.org/10.3390/universe9010054} {\bibfield  {journal} {\bibinfo  {journal} {Universe}\ }\textbf {\bibinfo {volume} {9}},\ \bibinfo {pages} {54} (\bibinfo {year} {2023})},\ \Eprint {https://arxiv.org/abs/2210.11380} {arXiv:2210.11380 [hep-ex]} \BibitemShut {NoStop}%
\bibitem [{\citenamefont {{Johnson}}\ \emph {et~al.}(2013)\citenamefont {{Johnson}}, \citenamefont {{Ormand}},\ and\ \citenamefont {{Krastev}}}]{2013CoPhC.184.2761J}%
  \BibitemOpen
  \bibfield  {author} {\bibinfo {author} {\bibfnamefont {C.~W.}\ \bibnamefont {{Johnson}}}, \bibinfo {author} {\bibfnamefont {W.~E.}\ \bibnamefont {{Ormand}}},\ and\ \bibinfo {author} {\bibfnamefont {P.~G.}\ \bibnamefont {{Krastev}}},\ }\bibfield  {title} {\bibinfo {title} {{Factorization in large-scale many-body calculations}},\ }\href {https://doi.org/10.1016/j.cpc.2013.07.022} {\bibfield  {journal} {\bibinfo  {journal} {Computer Physics Communications}\ }\textbf {\bibinfo {volume} {184}},\ \bibinfo {pages} {2761} (\bibinfo {year} {2013})},\ \Eprint {https://arxiv.org/abs/1303.0905} {arXiv:1303.0905 [nucl-th]} \BibitemShut {NoStop}%
\bibitem [{\citenamefont {{Johnson}}\ \emph {et~al.}(2018)\citenamefont {{Johnson}}, \citenamefont {{Ormand}}, \citenamefont {{McElvain}},\ and\ \citenamefont {{Shan}}}]{2018arXiv180108432J}%
  \BibitemOpen
  \bibfield  {author} {\bibinfo {author} {\bibfnamefont {C.~W.}\ \bibnamefont {{Johnson}}}, \bibinfo {author} {\bibfnamefont {W.~E.}\ \bibnamefont {{Ormand}}}, \bibinfo {author} {\bibfnamefont {K.~S.}\ \bibnamefont {{McElvain}}},\ and\ \bibinfo {author} {\bibfnamefont {H.}~\bibnamefont {{Shan}}},\ }\bibfield  {title} {\bibinfo {title} {{BIGSTICK: A flexible configuration-interaction shell-model code}},\ }\href {https://doi.org/10.48550/ARXIV.1801.08432} {\bibfield  {journal} {\bibinfo  {journal} {arXiv e-prints}\ ,\ \bibinfo {eid} {arXiv:1801.08432}} (\bibinfo {year} {2018})},\ \Eprint {https://arxiv.org/abs/1801.08432} {arXiv:1801.08432 [physics.comp-ph]} \BibitemShut {NoStop}%
\bibitem [{\citenamefont {Brown}\ \emph {et~al.}(1978)\citenamefont {Brown}, \citenamefont {Chung},\ and\ \citenamefont {Wildenthal}}]{PhysRevLett.40.1631}%
  \BibitemOpen
  \bibfield  {author} {\bibinfo {author} {\bibfnamefont {B.~A.}\ \bibnamefont {Brown}}, \bibinfo {author} {\bibfnamefont {W.}~\bibnamefont {Chung}},\ and\ \bibinfo {author} {\bibfnamefont {B.~H.}\ \bibnamefont {Wildenthal}},\ }\bibfield  {title} {\bibinfo {title} {Empirical renormalization of the one-body gamow-teller $\ensuremath{\beta}$-decay matrix elements in the $1s\ensuremath{-}0d$ shell},\ }\href {https://doi.org/10.1103/PhysRevLett.40.1631} {\bibfield  {journal} {\bibinfo  {journal} {Phys. Rev. Lett.}\ }\textbf {\bibinfo {volume} {40}},\ \bibinfo {pages} {1631} (\bibinfo {year} {1978})}\BibitemShut {NoStop}%
\bibitem [{\citenamefont {Crivellin}\ \emph {et~al.}(2017)\citenamefont {Crivellin}, \citenamefont {Davidson}, \citenamefont {Pruna},\ and\ \citenamefont {Signer}}]{Crivellin:2017rmk}%
  \BibitemOpen
  \bibfield  {author} {\bibinfo {author} {\bibfnamefont {A.}~\bibnamefont {Crivellin}}, \bibinfo {author} {\bibfnamefont {S.}~\bibnamefont {Davidson}}, \bibinfo {author} {\bibfnamefont {G.~M.}\ \bibnamefont {Pruna}},\ and\ \bibinfo {author} {\bibfnamefont {A.}~\bibnamefont {Signer}},\ }\bibfield  {title} {\bibinfo {title} {{Renormalisation-group improved analysis of $\mu\to e$ processes in a systematic effective-field-theory approach}},\ }\href {https://doi.org/10.1007/JHEP05(2017)117} {\bibfield  {journal} {\bibinfo  {journal} {JHEP}\ }\textbf {\bibinfo {volume} {05}},\ \bibinfo {pages} {117}},\ \Eprint {https://arxiv.org/abs/1702.03020} {arXiv:1702.03020 [hep-ph]} \BibitemShut {NoStop}%
\bibitem [{\citenamefont {Bartolotta}\ and\ \citenamefont {Ramsey-Musolf}(2018)}]{Bartolotta:2017mff}%
  \BibitemOpen
  \bibfield  {author} {\bibinfo {author} {\bibfnamefont {A.}~\bibnamefont {Bartolotta}}\ and\ \bibinfo {author} {\bibfnamefont {M.~J.}\ \bibnamefont {Ramsey-Musolf}},\ }\bibfield  {title} {\bibinfo {title} {{Coherent $\mu-e$ conversion at next-to-leading order}},\ }\href {https://doi.org/10.1103/PhysRevC.98.015208} {\bibfield  {journal} {\bibinfo  {journal} {Phys. Rev. C}\ }\textbf {\bibinfo {volume} {98}},\ \bibinfo {pages} {015208} (\bibinfo {year} {2018})},\ \Eprint {https://arxiv.org/abs/1710.02129} {arXiv:1710.02129 [hep-ph]} \BibitemShut {NoStop}%
\bibitem [{\citenamefont {Cirigliano}\ \emph {et~al.}(2017)\citenamefont {Cirigliano}, \citenamefont {Davidson},\ and\ \citenamefont {Kuno}}]{Cirigliano:2017azj}%
  \BibitemOpen
  \bibfield  {author} {\bibinfo {author} {\bibfnamefont {V.}~\bibnamefont {Cirigliano}}, \bibinfo {author} {\bibfnamefont {S.}~\bibnamefont {Davidson}},\ and\ \bibinfo {author} {\bibfnamefont {Y.}~\bibnamefont {Kuno}},\ }\bibfield  {title} {\bibinfo {title} {{Spin-dependent $\mu \to e$ conversion}},\ }\href {https://doi.org/10.1016/j.physletb.2017.05.053} {\bibfield  {journal} {\bibinfo  {journal} {Phys. Lett. B}\ }\textbf {\bibinfo {volume} {771}},\ \bibinfo {pages} {242} (\bibinfo {year} {2017})},\ \Eprint {https://arxiv.org/abs/1703.02057} {arXiv:1703.02057 [hep-ph]} \BibitemShut {NoStop}%
\bibitem [{\citenamefont {Davidson}\ \emph {et~al.}(2018)\citenamefont {Davidson}, \citenamefont {Kuno},\ and\ \citenamefont {Saporta}}]{Davidson:2017nrp}%
  \BibitemOpen
  \bibfield  {author} {\bibinfo {author} {\bibfnamefont {S.}~\bibnamefont {Davidson}}, \bibinfo {author} {\bibfnamefont {Y.}~\bibnamefont {Kuno}},\ and\ \bibinfo {author} {\bibfnamefont {A.}~\bibnamefont {Saporta}},\ }\bibfield  {title} {\bibinfo {title} {{\textquotedblleft{}Spin-dependent\textquotedblright{} ${\mu \rightarrow e}$ conversion on light nuclei}},\ }\href {https://doi.org/10.1140/epjc/s10052-018-5584-8} {\bibfield  {journal} {\bibinfo  {journal} {Eur. Phys. J. C}\ }\textbf {\bibinfo {volume} {78}},\ \bibinfo {pages} {109} (\bibinfo {year} {2018})},\ \Eprint {https://arxiv.org/abs/1710.06787} {arXiv:1710.06787 [hep-ph]} \BibitemShut {NoStop}%
\bibitem [{\citenamefont {Fuyuto}\ and\ \citenamefont {Mereghetti}(2024)}]{Fuyuto:2024skf}%
  \BibitemOpen
  \bibfield  {author} {\bibinfo {author} {\bibfnamefont {K.}~\bibnamefont {Fuyuto}}\ and\ \bibinfo {author} {\bibfnamefont {E.}~\bibnamefont {Mereghetti}},\ }\bibfield  {title} {\bibinfo {title} {{Axionlike-particle contributions to the \ensuremath{\mu}\textrightarrow{}e conversion}},\ }\href {https://doi.org/10.1103/PhysRevD.109.075014} {\bibfield  {journal} {\bibinfo  {journal} {Phys. Rev. D}\ }\textbf {\bibinfo {volume} {109}},\ \bibinfo {pages} {075014} (\bibinfo {year} {2024})}\BibitemShut {NoStop}%
\bibitem [{\citenamefont {Abusalma}\ \emph {et~al.}(2018)\citenamefont {Abusalma} \emph {et~al.}}]{Mu2e:2018osu}%
  \BibitemOpen
  \bibfield  {author} {\bibinfo {author} {\bibfnamefont {F.}~\bibnamefont {Abusalma}} \emph {et~al.} (\bibinfo {collaboration} {Mu2e}),\ }\bibfield  {title} {\bibinfo {title} {{Expression of Interest for Evolution of the Mu2e Experiment}},\ }\href@noop {} {\  (\bibinfo {year} {2018})},\ \Eprint {https://arxiv.org/abs/1802.02599} {arXiv:1802.02599 [physics.ins-det]} \BibitemShut {NoStop}%
\bibitem [{\citenamefont {Burrows}(2007)}]{Burrows:2007jwj}%
  \BibitemOpen
  \bibfield  {author} {\bibinfo {author} {\bibfnamefont {T.~W.}\ \bibnamefont {Burrows}},\ }\bibfield  {title} {\bibinfo {title} {{Nuclear Data Sheets for A = 47}},\ }\href {https://doi.org/10.1016/j.nds.2007.04.002} {\bibfield  {journal} {\bibinfo  {journal} {Nucl. Data Sheets}\ }\textbf {\bibinfo {volume} {108}},\ \bibinfo {pages} {923} (\bibinfo {year} {2007})}\BibitemShut {NoStop}%
\bibitem [{\citenamefont {Poves}\ \emph {et~al.}(2001)\citenamefont {Poves}, \citenamefont {Sánchez-Solano}, \citenamefont {Caurier},\ and\ \citenamefont {Nowacki}}]{kb3g}%
  \BibitemOpen
  \bibfield  {author} {\bibinfo {author} {\bibfnamefont {A.}~\bibnamefont {Poves}}, \bibinfo {author} {\bibfnamefont {J.}~\bibnamefont {Sánchez-Solano}}, \bibinfo {author} {\bibfnamefont {E.}~\bibnamefont {Caurier}},\ and\ \bibinfo {author} {\bibfnamefont {F.}~\bibnamefont {Nowacki}},\ }\bibfield  {title} {\bibinfo {title} {Shell model study of the isobaric chains a=50, a=51 and a=52},\ }\href {https://doi.org/10.1016/s0375-9474(01)00967-8} {\bibfield  {journal} {\bibinfo  {journal} {Nuclear Physics A}\ }\textbf {\bibinfo {volume} {694}},\ \bibinfo {pages} {157–198} (\bibinfo {year} {2001})}\BibitemShut {NoStop}%
\bibitem [{\citenamefont {Honma}\ \emph {et~al.}(2004)\citenamefont {Honma}, \citenamefont {Otsuka}, \citenamefont {Brown},\ and\ \citenamefont {Mizusaki}}]{gxpf1}%
  \BibitemOpen
  \bibfield  {author} {\bibinfo {author} {\bibfnamefont {M.}~\bibnamefont {Honma}}, \bibinfo {author} {\bibfnamefont {T.}~\bibnamefont {Otsuka}}, \bibinfo {author} {\bibfnamefont {B.~A.}\ \bibnamefont {Brown}},\ and\ \bibinfo {author} {\bibfnamefont {T.}~\bibnamefont {Mizusaki}},\ }\bibfield  {title} {\bibinfo {title} {New effective interaction for $pf$-shell nuclei and its implications for the stability of the $n=z=28$ closed core},\ }\href {https://doi.org/10.1103/PhysRevC.69.034335} {\bibfield  {journal} {\bibinfo  {journal} {Phys. Rev. C}\ }\textbf {\bibinfo {volume} {69}},\ \bibinfo {pages} {034335} (\bibinfo {year} {2004})}\BibitemShut {NoStop}%
\bibitem [{\citenamefont {McGrory}\ \emph {et~al.}(1970)\citenamefont {McGrory}, \citenamefont {Wildenthal},\ and\ \citenamefont {Halbert}}]{kbp}%
  \BibitemOpen
  \bibfield  {author} {\bibinfo {author} {\bibfnamefont {J.~B.}\ \bibnamefont {McGrory}}, \bibinfo {author} {\bibfnamefont {B.~H.}\ \bibnamefont {Wildenthal}},\ and\ \bibinfo {author} {\bibfnamefont {E.~C.}\ \bibnamefont {Halbert}},\ }\bibfield  {title} {\bibinfo {title} {Shell-model structure of $^{42\ensuremath{-}50}\mathrm{Ca}$},\ }\href {https://doi.org/10.1103/PhysRevC.2.186} {\bibfield  {journal} {\bibinfo  {journal} {Phys. Rev. C}\ }\textbf {\bibinfo {volume} {2}},\ \bibinfo {pages} {186} (\bibinfo {year} {1970})}\BibitemShut {NoStop}%
\bibitem [{\citenamefont {Donnelly}\ and\ \citenamefont {Haxton}(1979{\natexlab{b}})}]{DONNELLY1979103}%
  \BibitemOpen
  \bibfield  {author} {\bibinfo {author} {\bibfnamefont {T.}~\bibnamefont {Donnelly}}\ and\ \bibinfo {author} {\bibfnamefont {W.}~\bibnamefont {Haxton}},\ }\bibfield  {title} {\bibinfo {title} {Multipole operators in semileptonic weak and electromagnetic interactions with nuclei: Harmonic oscillator single-particle matrix elements},\ }\href {https://doi.org/https://doi.org/10.1016/0092-640X(79)90003-2} {\bibfield  {journal} {\bibinfo  {journal} {Atomic Data and Nuclear Data Tables}\ }\textbf {\bibinfo {volume} {23}},\ \bibinfo {pages} {103} (\bibinfo {year} {1979}{\natexlab{b}})}\BibitemShut {NoStop}%
\end{thebibliography}%

\appendix
\begin{widetext}
\section{Code Description}
\label{app:code_desc}
The effective theory of inelastic $\mu\rightarrow e$ conversion has been distilled into a publicly available \texttt{Mathematica} script, which we call \texttt{Mu2e\_Inelastic\_v1}, available at
\begin{center}\label{url:Mu2e_Inelastic}
 \href{https://github.com/Berkeley-Electroweak-Physics/Mu2e_Inelastic}{https://github.com/Berkeley-Electroweak-Physics/Mu2e\_Inelastic}\,.
\end{center}
Details on obtaining and running the code can be found in the corresponding README file. Here we provide only a brief overview of the software's scope and utility: The choice of nuclear target is currently limited to $^{27}$Al, although this can be extended in the future. The user specifies the form of the CLFV interaction by entering the dimensionless NRET coefficients $\tilde{c}^{\tau}_i$. The code then produces several outputs: First, it evaluates the requisite nuclear response functions\footnote{In evaluating the multipole operator $\tilde{\Delta}''$, the code assumes current conservation and evaluates the equivalent charge operator, as in Eq. \eqref{eq:siegert}. Nonetheless, the subroutines to compute matrix elements of $\tilde{\Delta}''$ are included in the \texttt{Mathematica} library and could be straightforwardly utilized by an enterprising reader.} in order to calculate the expected $\mu\rightarrow e$ conversion decay rates and branching ratios for the four low-lying transitions in $^{27}$Al considered in this work. Each calculation is performed with three different shell-model interactions, yielding an estimate of the theory uncertainty in the computed decay rates. However, as additional effects of operator mixing and renormalization have been neglected, we recommend that the reported errors be regarded as minimum uncertainties. 

Combining the decay rates with the simulated elastic $\mu\rightarrow e$ conversion spectrum from Ref. \cite{Mu2e:2022ggl}, \texttt{Mu2e\_Inelastic\_v1} calculates the conversion-electron spectrum expected in Mu2e run I. The results are output in a series of files, containing separate CE spectra for each transition as well as the combined signal. Thus, one can reproduce Figs. \ref{fig:mu2e_spec_allowed}, \ref{fig:mu2e_spec_v_dep_1}, and \ref{fig:mu2e_spec_v_dep_2} or explore more general scenarios where multiple nuclear response functions contribute simultaneously. In principle, one can use \texttt{Mu2e\_Inelastic\_v1} in combination with the existing package \texttt{MuonConverter}{\interfootnotelinepenalty=10000\footnote{\href{https://github.com/Berkeley-Electroweak-Physics/MuonConverter}{https://github.com/Berkeley-Electroweak-Physics/MuonConverter}}} in order to connect to quark-level WET and SMEFT. Using \texttt{MuonConverter}, one can match a set of WET Wilson coefficients at $\mu=2$ GeV to the set of 16 NRET coefficients, which can then be input into \texttt{Mu2e\_Inelastic\_v1}.

\section{Derivation of rate formula}
\label{app:rate_derivation}

Starting from Eq. \eqref{eq:H_NRET}, we multipole expand the nuclear charges/currents. We specialize to the case of even-parity nuclear transitions, take the nuclear states to be eigenstates of parity and CP, and adopt a sign convention where the single-particle and nuclear matrix elements of the various one-body operators are real \cite{DONNELLY1979103}.  Averaging over initial nuclear spins and summing of final nuclear spins yields the $\mu\rightarrow e$ transition probability, 
{\allowdisplaybreaks
\begin{align*}
&\frac{1}{2j_i+1}\sum_{m_f,m_i}|\braket{\frac{1}{2}s_f;j_fm_f|\mathcal{M}|\frac{1}{2}s_i;j_im_i}|^2=\frac{E_\mathrm{CE}}{2m_e}|\phi_{1s}^{Z_\mathrm{eff}}(\vec{0})|^2\frac{q_\mathrm{eff}^2}{q^2}\frac{4\pi}{2j_i+1}\sum_{\tau=0,1}\sum_{\tau'=0,1}\\
&\Bigg\{\sum_{J=0,2,...}^{\infty}\Bigg[\braket{l_0^{\tau}}\braket{l_0^{\tau'}}^*\braket{j_f||M_{J,\tau}(q_\mathrm{eff})||j_i}\braket{j_f||M_{J,\tau'}(q_\mathrm{eff})||j_i}\\
&+\frac{\vec{q}_\mathrm{eff}}{m_N}\cdot\braket{\vec{l}_E^{\tau}}\frac{\vec{q}_\mathrm{eff}}{m_N}\cdot\braket{\vec{l}_E^{\tau'}}^*\braket{j_f||\Phi''_{J,\tau}(q_\mathrm{eff})||j_i}\braket{j_f||\Phi''_{J,\tau'}(q_\mathrm{eff})||j_i}\\
&+\frac{\vec{q}_\mathrm{eff}}{m_N}\cdot\braket{\vec{l}_M^{\tau}}\frac{\vec{q}_\mathrm{eff}}{m_N}\cdot\braket{\vec{l}_M^{\tau'}}^*\braket{j_f||\tilde{\Delta}''_{J,\tau}(q_\mathrm{eff})||j_i}\braket{j_f||\tilde{\Delta}''_{J,\tau'}(q_\mathrm{eff})||j_i}\\
&+\frac{2\vec{q}_\mathrm{eff}}{m_N}\cdot\mathrm{Re}\left[\braket{\vec{l}_E^{\tau}}\braket{l_0^{\tau'}}^*\right]\braket{j_f||\Phi''_{J,\tau}(q_\mathrm{eff})||j_i}\braket{j_f||M_{J,\tau'}(q_\mathrm{eff})||j_i}\\
&+\frac{2\vec{q}_\mathrm{eff}}{m_N}\cdot\mathrm{Re}\left[\braket{\vec{l}_M^{\tau}}\braket{l_0^{\tau'}}^*\right]\braket{j_f||\tilde{\Delta}''_{J,\tau}(q_\mathrm{eff})||j_i}\braket{j_f||M_{J,\tau'}(q_\mathrm{eff})||j_i}\\
&+2\mathrm{Re}\left[\frac{\vec{q}_\mathrm{eff}}{m_N}\cdot\braket{\vec{l}_E^{\tau}}\frac{\vec{q}_\mathrm{eff}}{m_N}\cdot\braket{\vec{l}_M^{\tau'}}^*\right]\braket{j_f||\Phi''_{J,\tau}(q_\mathrm{eff})||j_i}\braket{j_f||\tilde{\Delta}''_{J,\tau'}(q_\mathrm{eff})||j_i}\Bigg]\\
&+\sum_{J=2,4,...}^{\infty}\Bigg[\frac{1}{2}\left(\frac{q^2_\mathrm{eff}}{m_N^2}\braket{\vec{l}_E^{\tau}}\cdot\braket{\vec{l}_E^{\tau'}}^*-\frac{\vec{q}_\mathrm{eff}}{m_N}\cdot\braket{\vec{l}_E^{\tau}}\frac{\vec{q}_\mathrm{eff}}{m_N}\cdot\braket{\vec{l}_E^{\tau'}}^*\right)\braket{j_f||\tilde{\Phi}'_{J,\tau}(q_\mathrm{eff})||j_i}\braket{j_f||\tilde{\Phi}'_{J,\tau'}(q_\mathrm{eff})||j_i}\\
&+\frac{1}{2}\left(\frac{q_{\mathrm{eff}}^2}{m_N^2}\braket{\vec{l}_M^{\tau}}\cdot\braket{\vec{l}_M^{\tau'}}^*-\frac{\vec{q}_\mathrm{eff}}{m_N}\cdot\braket{\vec{l}_M^{\tau}}\frac{\vec{q}_\mathrm{eff}}{m_N}\cdot\braket{\vec{l}_M^{\tau'}}^*\right)\braket{j_f||\Delta'_{J,\tau}(q_\mathrm{eff})||j_i}\braket{j_f||\Delta'_{J,\tau'}(q_\mathrm{eff})||j_i}\\
&+\frac{1}{2}\left(\braket{\vec{l}_5^{\tau}}\cdot\braket{\vec{l}_5^{\tau'}}^*-\hat{q}\cdot\braket{\vec{l}_5^{\tau}}\hat{q}\cdot\braket{\vec{l}_5^{\tau'}}^*\right)\braket{j_f||\Sigma_{J,\tau}(q_\mathrm{eff})||j_i}\braket{j_f||\Sigma_{J,\tau'}(q_\mathrm{eff})||j_i}\\
&-\frac{\vec{q}_\mathrm{eff}}{m_N}\cdot\mathrm{Re}\left[i\braket{\vec{l}_5^{\tau}}\times\braket{\vec{l}_M^{\tau'}}^*\right]\braket{j_f||\Sigma_{J,\tau}(q_\mathrm{eff})||j_i}\braket{j_f||\Delta'_{J,\tau'}(q_\mathrm{eff})||j_i}\tag{\refstepcounter{equation}\theequation}\\
&-\frac{\vec{q}_\mathrm{eff}}{m_N}\cdot\mathrm{Re}\left[i\braket{\vec{l}_5^{\tau}}\times\braket{\vec{l}_E^{\tau'}}^*\right]\braket{j_f||\Sigma_{J,\tau}(q_\mathrm{eff})||j_i}\braket{j_f||\tilde{\Phi}'_{J,\tau'}(q_\mathrm{eff})||j_i}\\
&+\mathrm{Re}\left[\frac{q_\mathrm{eff}^2}{m_N^2}\braket{\vec{l}_M^{\tau}}\cdot\braket{\vec{l}_E^{\tau'}}^*-\frac{\vec{q}_\mathrm{eff}}{m_N}\cdot\braket{\vec{l}_M^{\tau}}\frac{\vec{q}_\mathrm{eff}}{m_N}\cdot\braket{\vec{l}_E^{\tau'}}^*\right]\braket{j_f||\Delta'_{J,\tau}(q_\mathrm{eff})||j_i}\braket{j_f||\tilde{\Phi}'_{J,\tau'}(q_\mathrm{eff})||j_i}\Bigg]\\
&+\sum_{J=1,3,...}^{\infty}\Bigg[\frac{q_\mathrm{eff}^2}{m_N^2}\braket{l_0^{A\;\tau}}\braket{l_0^{A\;\tau'}}^*\braket{j_f||\tilde{\Omega}_{J,\tau}(q_\mathrm{eff})||j_i}\braket{j_f||\tilde{\Omega}_{J,\tau'}(q_\mathrm{eff})||j_i}\\
&+\hat{q}\cdot\braket{\vec{l}_5^{\tau}}\hat{q}\cdot\braket{\vec{l}_5^{\tau'}}^*\braket{j_f||\Sigma''_{J,\tau}(q_\mathrm{eff})||j_i}\braket{j_f||\Sigma''_{J,\tau'}(q_\mathrm{eff})||j_i}\\
&+\frac{1}{2}\left(\braket{\vec{l}_5^{\tau}}\cdot\braket{\vec{l}_5^{\tau'}}^*-\hat{q}\cdot\braket{\vec{l}_5^{\tau}}\hat{q}\cdot\braket{\vec{l}_5^{\tau'}}^*\right)\braket{j_f||\Sigma'_{J,\tau}(q_\mathrm{eff})||j_i}\braket{j_f||\Sigma'_{J,\tau'}(q_\mathrm{eff})||j_i}\\
&+\frac{1}{2}\left(\frac{q^2_\mathrm{eff}}{m_N^2}\braket{\vec{l}_M^{\tau}}\cdot\braket{\vec{l}_M^{\tau'}}^*-\frac{\vec{q}_\mathrm{eff}}{m_N}\cdot\braket{\vec{l}_M^{\tau}}\frac{\vec{q}_\mathrm{eff}}{m_N}\cdot\braket{\vec{l}_M^{\tau'}}^*\right)\braket{j_f||\Delta_{J,\tau}(q_\mathrm{eff})||j_i}\braket{j_f||\Delta_{J,\tau'}(q_\mathrm{eff})||j_i}\\
&+\frac{1}{2}\left(\frac{q_\mathrm{eff}^2}{m_N^2}\braket{\vec{l}_E^{\tau}}\cdot\braket{\vec{l}_E^{\tau'}}^*-\frac{\vec{q}_\mathrm{eff}}{m_N}\cdot\braket{\vec{l}_E^{\tau}}\frac{\vec{q}_\mathrm{eff}}{m_N}\cdot\braket{\vec{l}_E^{\tau'}}^*\right)\braket{j_f||\tilde{\Phi}_{J,\tau}(q_\mathrm{eff})||j_i}\braket{j_f||\tilde{\Phi}_{J,\tau'}(q_\mathrm{eff})||j_i}\\
&+\frac{\vec{q}_\mathrm{eff}}{m_N}\cdot\mathrm{Re}\left[i\braket{\vec{l}_M^{\tau}}\times\braket{\vec{l}_5^{\tau'}}^*\right]\braket{j_f||\Delta_{J,\tau}(q_\mathrm{eff})||j_i}\braket{j_f||\Sigma'_{J,\tau'}(q_\mathrm{eff})||j_i}\\
&+\frac{\vec{q}_\mathrm{eff}}{m_N}\cdot\mathrm{Re}\left[i\braket{\vec{l}_E^{\tau}}\times\braket{\vec{l}_5^{\tau'}}^*\right]\braket{j_f||\tilde{\Phi}_{J,\tau}(q_\mathrm{eff})||j_i}\braket{j_f||\Sigma'_{J,\tau'}(q_\mathrm{eff})||j_i}\\
&-\frac{2\vec{q}_\mathrm{eff}}{m_N}\cdot\mathrm{Re}\left[\braket{\vec{l}_5^{\tau}}\braket{l_0^{A\;\tau'}}^*\right]\braket{j_f||\Sigma''_{J,\tau}(q_\mathrm{eff})||j_i}\braket{j_f||\tilde{\Omega}_{J,\tau'}(q_\mathrm{eff})||j_i}\\
&+\mathrm{Re}\left[\frac{q_\mathrm{eff}^2}{m_N^2}\braket{\vec{l}_M^{\tau}}\cdot\braket{\vec{l}_E^{\tau'}}^*-\frac{\vec{q}_\mathrm{eff}}{m_N}\cdot\braket{\vec{l}_M^{\tau}}\frac{\vec{q}_\mathrm{eff}}{m_N}\cdot\braket{\vec{l}_E^{\tau'}}^*\right]\braket{j_f||\Delta_{J,\tau}(q_\mathrm{eff})||j_i}\braket{j_f||\tilde{\Phi}_{J,\tau'}(q_\mathrm{eff})||j_i}\Bigg]\Bigg\}.
\end{align*}
}
Here $||$ indicates a matrix element reduced in angular momentum. We see that all 11 nuclear response functions --- and various interference terms --- contribute to this process. Next, we average over spin states of the bound muon and sum over spin states of the outgoing electron. Evaluating the resulting leptonic traces yields 
\begin{equation}
\begin{split}
&\frac{1}{2}\frac{1}{2j_i+1}\sum_\mathrm{spins}|\braket{\frac{1}{2}s_f;j_fm_f|\mathcal{M}|\frac{1}{2}s_i;j_im_i}|^2=\frac{E_\mathrm{CE}}{2m_e}|\phi_{1s}^{Z_\mathrm{eff}}(0)|^2\frac{q_\mathrm{eff}^2}{q^2}\frac{4\pi}{2j_i+1}\sum_{\tau=0,1}\sum_{\tau'=0,1}\\
&\Bigg\{\sum_{J=0,2,...}^{\infty}\Bigg(R_M^{\tau\tau'}\braket{j_f||M_{J,\tau}(q_\mathrm{eff})||j_i}\braket{j_f||M_{J,\tau'}(q_\mathrm{eff})||j_i}+\frac{q^2_\mathrm{eff}}{m_N^2}R_{\Phi''}^{\tau\tau'}\braket{j_f||\Phi''_{J,\tau}(q_\mathrm{eff})||j_i}\braket{j_f||\Phi''_{J,\tau'}(q_\mathrm{eff})||j_i}\\
&+\frac{q^2_\mathrm{eff}}{m_N^2}R_{\tilde{\Delta}''}^{\tau\tau'}\braket{j_f||\tilde{\Delta}''_{J,\tau}(q_\mathrm{eff})||j_i}\braket{j_f||\tilde{\Delta}''_{J,\tau'}(q_\mathrm{eff})||j_i}-\frac{2q_\mathrm{eff}}{m_N}R_{ \Phi'' M}^{\tau\tau'}\braket{j_f||\Phi''_{J,\tau}(q_\mathrm{eff})||j_i}\braket{j_f||M_{J,\tau'}(q_\mathrm{eff})||j_i}\\
&-\frac{2q_\mathrm{eff}}{m_N}R_{ \tilde{\Delta}'' M}^{\tau\tau'}\braket{j_f||\tilde{\Delta}''_{J,\tau}(q_\mathrm{eff})||j_i}\braket{j_f||M_{J,\tau'}(q_\mathrm{eff})||j_i}+\frac{2q_\mathrm{eff}^2}{m_N^2}R_{\tilde{\Delta}''\Phi''}^{\tau\tau'}\braket{j_f||\tilde{\Delta}''_{J,\tau'}(q_\mathrm{eff})||j_i}\braket{j_f||\Phi''_{J,\tau}(q_\mathrm{eff})||j_i}\Bigg)\\
&+\sum_{J=2,4,...}^{\infty}\Bigg(R_{\Sigma}^{\tau\tau'}\braket{j_f||\Sigma_{J,\tau}(q_\mathrm{eff})||j_i}\braket{j_f||\Sigma_{J,\tau'}(q_\mathrm{eff})||j_i}+\frac{q_{\mathrm{eff}}^2}{m_N^2}R_{\Delta'}^{\tau\tau'}\braket{j_f||\Delta'_{J,\tau}(q_\mathrm{eff})||j_i}\braket{j_f||\Delta'_{J,\tau'}(q_\mathrm{eff})||j_i}\\
&+\frac{q^2_\mathrm{eff}}{m_N^2}R_{\tilde{\Phi}'}^{\tau\tau'}\braket{j_f||\tilde{\Phi}'_{J,\tau}(q_\mathrm{eff})||j_i}\braket{j_f||\tilde{\Phi}'_{J,\tau'}(q_\mathrm{eff})||j_i}+\frac{2q_\mathrm{eff}}{m_N}R_{ \Delta' \Sigma}^{\tau\tau'}\braket{j_f||\Delta'_{J,\tau}(q_\mathrm{eff})||j_i}\braket{j_f||\Sigma_{J,\tau'}(q_\mathrm{eff})||j_i}\\
&-\frac{2q_\mathrm{eff}}{m_N}R_{\tilde{\Phi}' \Sigma}^{\tau\tau'}\braket{j_f||\tilde{\Phi}'_{J,\tau}(q_\mathrm{eff})||j_i}\braket{j_f||\Sigma_{J,\tau'}(q_\mathrm{eff})||j_i}+\frac{2q_\mathrm{eff}^2}{m_N^2}R_{\Delta'\tilde{\Phi}'}^{\tau\tau'}\braket{j_f||\Delta'_{J,\tau}(q_\mathrm{eff})||j_i}\braket{j_f||\tilde{\Phi}'_{J,\tau'}(q_\mathrm{eff})||j_i}\Bigg)\\
&+\sum_{J=1,3,...}^{\infty}\Bigg(R_{\Sigma'}^{\tau\tau'}\braket{j_f||\Sigma'_{J,\tau}(q_\mathrm{eff})||j_i}\braket{j_f||\Sigma'_{J,\tau'}(q_\mathrm{eff})||j_i}+R_{\Sigma''}^{\tau\tau'}\braket{j_f||\Sigma''_{J,\tau}(q_\mathrm{eff})||j_i}\braket{j_f||\Sigma''_{J,\tau'}(q_\mathrm{eff})||j_i}\\
&+\frac{q_\mathrm{eff}^2}{m_N^2}R_{\tilde{\Omega}}^{\tau\tau'}\braket{j_f||\tilde{\Omega}_{J,\tau}(q_\mathrm{eff})||j_i}\braket{j_f||\tilde{\Omega}_{J,\tau'}(q_\mathrm{eff})||j_i}+\frac{q^2_\mathrm{eff}}{m_N^2}R_{\Delta}^{\tau\tau'}\braket{j_f||\Delta_{J,\tau}(q_\mathrm{eff})||j_i}\braket{j_f||\Delta_{J,\tau'}(q_\mathrm{eff})||j_i}\\
&+\frac{q_\mathrm{eff}^2}{m_N^2}R_{\tilde{\Phi}}^{\tau\tau'}\braket{j_f||\tilde{\Phi}_{J,\tau}(q_\mathrm{eff})||j_i}\braket{j_f||\tilde{\Phi}_{J,\tau'}(q_\mathrm{eff})||j_i}-\frac{2q_\mathrm{eff}}{m_N}R_{ \Delta \Sigma'}^{\tau\tau'}\braket{j_f||\Delta_{J,\tau}(q_\mathrm{eff})||j_i}\braket{j_f||\Sigma'_{J,\tau'}(q_\mathrm{eff})||j_i}\\
&+\frac{2q_\mathrm{eff}}{m_N}R_{\tilde{\Phi} \Sigma'}^{\tau\tau'}\braket{j_f||\tilde{\Phi}_{J,\tau}(q_\mathrm{eff})||j_i}\braket{j_f||\Sigma'_{J,\tau'}(q_\mathrm{eff})||j_i}-\frac{2q_\mathrm{eff}}{m_N}R_{\tilde{\Omega} \Sigma''}^{\tau\tau'}\braket{j_f||\tilde{\Omega}_{J,\tau}(q_\mathrm{eff})||j_i}\braket{j_f||\Sigma''_{J,\tau'}(q_\mathrm{eff})||j_i}\\
&+\frac{2q_\mathrm{eff}^2}{m_N^2}R_{\Delta\tilde{\Phi}}^{\tau\tau'}\braket{j_f||\Delta_{J,\tau}(q_\mathrm{eff})||j_i}\braket{j_f||\tilde{\Phi}_{J,\tau'}(q_\mathrm{eff})||j_i}\Bigg)\Bigg\},
\label{eq:full_inelastic_prob}
\end{split}
\end{equation}
where the leptonic response functions are given in Eqs. \eqref{eq:R_CLFV_direct} and \eqref{eq:R_CLFV_int}. For transitions that conserve parity, the nuclear response functions are 
{\allowdisplaybreaks
\begin{align}
W_O^{\tau\tau'}(q)&\equiv \frac{4\pi}{2j_i+1}\sum_{J=0,2,...}^{\infty}\braket{j_f||O_{J;\tau}(q)||j_i}\braket{j_f||O_{J;\tau'}(q)||j_i}\;\mathrm{for}\;O=M,\tilde{\Delta}'',\Phi'', \nonumber\\
W_O^{\tau\tau'}(q)&\equiv \frac{4\pi}{2j_i+1} \sum_{J=1,3,...}^{\infty}\braket{j_f||O_{J;\tau}(q)||j_i}\braket{j_f||O_{J;\tau'}(q)||j_i}\;\mathrm{for}\;O=\Sigma',\Sigma'',\Delta,\tilde{\Omega},\tilde{\Phi}, \nonumber\\
W_{O}^{\tau\tau'}(q)&\equiv \frac{4\pi}{2j_i+1} \sum_{J=2,4,...}^{\infty}\braket{j_f||O_{J;\tau}(q)||j_i}\braket{j_f||O_{J;\tau'}(q)||j_i}\;\mathrm{for}\;O=\Sigma,\Delta',\tilde{\Phi}', \nonumber\\
W_{ \Phi'' M}^{\tau\tau'}(q)&\equiv \frac{4\pi}{2j_i+1} \sum_{J=0,2,...}^{\infty} \braket{j_f||\Phi''_{J;\tau}(q)||j_i}\braket{j_f||M_{J;\tau'}(q)||j_i}, \nonumber\\
{W_{\tilde{\Delta}'' O }^{\tau\tau'}(q)}&\equiv \frac{4\pi}{2j_i+1} \sum_{J=0,2,...}^{\infty} \braket{j_f||\tilde{\Delta}''_{J;\tau'}(q)||j_i}\braket{j_f||O_{J;\tau}(q)||j_i}\;\mathrm{for}\;O=M,\Phi'', \nonumber\\
W_{ \Delta \Sigma'}^{\tau\tau'}(q)&\equiv \frac{4\pi}{2j_i+1} \sum_{J=1,3,...}^{\infty}\braket{j_f||\Delta_{J;\tau}(q)||j_i}\braket{j_f||\Sigma'_{J;\tau'}(q)||j_i}, \label{eq:W_responses_PC}\\
W^{\tau\tau'}_{\tilde{\Phi} \Sigma'}(q)&\equiv\frac{4\pi}{2j_i+1}\sum_{J=1,3,...}^{\infty}\braket{j_f||\tilde{\Phi}_{J;\tau}(q)||j_i}\braket{j_f||\Sigma'_{J;\tau'}(q)||j_i}, \nonumber\\
W_{\Delta' \Sigma}^{\tau\tau'}(q)&\equiv \frac{4\pi}{2j_i+1} \sum_{J=2,4,...}^{\infty}\braket{j_f||\Delta'_{J;\tau}(q)||j_i}\braket{j_f||\Sigma_{J;\tau'}(q)||j_i}, \nonumber\\
W^{\tau\tau'}_{\tilde{\Phi}' \Sigma}(q)&\equiv\frac{4\pi}{2j_i+1}\sum_{J=2,4,...}^{\infty}\braket{j_f||\tilde{\Phi}'_{J;\tau}(q)||j_i}\braket{j_f||\Sigma_{J;\tau'}(q)||j_i}, \nonumber\\
W^{\tau\tau'}_{\tilde{\Omega} \Sigma''}(q)&\equiv\frac{4\pi}{2j_i+1}\sum_{J=1,3,...}^{\infty}\braket{j_f||\tilde{\Omega}_{J;\tau}(q)||j_i}\braket{j_f||\Sigma''_{J;\tau'}(q)||j_i}, \nonumber\\
W^{\tau\tau'}_{\Delta\tilde{\Phi}}(q)&\equiv\frac{4\pi}{2j_i+1}\sum_{J=1,3,...}^{\infty}\braket{j_f||\Delta_{J;\tau}(q)||j_i}\braket{j_f||\tilde{\Phi}_{J;\tau'}(q)||j_i}, \nonumber\\
W^{\tau\tau'}_{\Delta'\tilde{\Phi}'}(q)&\equiv\frac{4\pi}{2j_i+1}\sum_{J=2,4,...}^{\infty}\braket{j_f||\Delta'_{J;\tau}(q)||j_i}\braket{j_f||\tilde{\Phi}'_{J;\tau'}(q)||j_i}.\nonumber
\end{align}
}
The summations over nuclear multipoles $J$ are restricted by the triangle inequality $|j_i-j_f|\leq J\leq j_i+j_f$, where $j_i$ and $j_f$ are the total angular momentum of the initial and final nuclear states, respectively. As discussed in Sec. \ref{sec:nuclear_eft}, if the vector current is known to be conserved, the operator $\tilde{\Delta}''$ can be replaced by the charge multipole $M$, as in Eq. \eqref{eq:siegert}.

If the nucleus transitions to a state of opposite parity, then the nuclear response functions must change to opposite parity. In this case, the conversion rate is still described by Eq. \eqref{eq:inelastic_rate}, the leptonic response functions $R$ are unchanged, but the nuclear response functions are given by
{\allowdisplaybreaks
\begin{align}
W_O^{\tau\tau'}(q)&\equiv \frac{4\pi}{2j_i+1}\sum_{J=1,3,...}^{\infty}\braket{j_f||O_{J;\tau}(q)||j_i}\braket{j_f||O_{J;\tau'}(q)||j_i}\;\mathrm{for}\;O=M,\Sigma,\Delta',\tilde{\Delta}'',\tilde{\Phi}',\Phi'', \nonumber\\
W_O^{\tau\tau'}(q)&\equiv \frac{4\pi}{2j_i+1} \sum_{J=0,2,...}^{\infty}\braket{j_f||O_{J;\tau}(q)||j_i}\braket{j_f||O_{J;\tau'}(q)||j_i}\;\mathrm{for}\;O=\Sigma'',\tilde{\Omega}, \nonumber\\
W_{O}^{\tau\tau'}(q)&\equiv \frac{4\pi}{2j_i+1} \sum_{J=2,4,...}^{\infty}\braket{j_f||O_{J;\tau}(q)||j_i}\braket{j_f||O_{J;\tau'}(q)||j_i}\;\mathrm{for}\;O=\Sigma',\Delta,\tilde{\Phi}, \nonumber\\
W_{\Phi'' M}^{\tau\tau'}(q)&\equiv \frac{4\pi}{2j_i+1} \sum_{J=1,3,...}^{\infty}\braket{j_f||\Phi''_{J;\tau}(q)||j_i}\braket{j_f||M_{J;\tau'}(q)||j_i}, \nonumber\\
W_{\tilde{\Delta}'' O }^{\tau\tau'}(q)&\equiv \frac{4\pi}{2j_i+1} \sum_{J=1,3,...}^{\infty} \braket{j_f||\tilde{\Delta}''_{J;\tau'}(q)||j_i}\braket{j_f||O_{J;\tau}(q)||j_i}\;\mathrm{for}\;O=M,\Phi'', \nonumber\\
W_{\Delta \Sigma'}^{\tau\tau'}(q)&\equiv \frac{4\pi}{2j_i+1} \sum_{J=2,4,...}^{\infty}\braket{j_f||\Delta_{J;\tau}(q)||j_i}\braket{j_f||\Sigma'_{J;\tau'}(q)||j_i}, \label{eq:W_responses_PV}\\
W^{\tau\tau'}_{\tilde{\Phi} \Sigma'}(q)&\equiv\frac{4\pi}{2j_i+1}\sum_{J=2,4,...}^{\infty}\braket{j_f||\tilde{\Phi}_{J;\tau}(q)||j_i}\braket{j_f||\Sigma'_{J;\tau'}(q)||j_i}, \nonumber\\
W_{ \Delta' \Sigma}^{\tau\tau'}(q)&\equiv \frac{4\pi}{2j_i+1} \sum_{J=1,3,...}^{\infty}\braket{j_f||\Delta'_{J;\tau}(q)||j_i}\braket{j_f||\Sigma_{J;\tau'}(q)||j_i}, \nonumber\\
W^{\tau\tau'}_{\tilde{\Phi}' \Sigma}(q)&\equiv\frac{4\pi}{2j_i+1}\sum_{J=1,3,...}^{\infty}\braket{j_f||\tilde{\Phi}'_{J;\tau}(q)||j_i} \braket{j_f||\Sigma_{J;\tau'}(q)||j_i}, \nonumber\\
W^{\tau\tau'}_{\tilde{\Omega} \Sigma''}(q)&\equiv\frac{4\pi}{2j_i+1}\sum_{J=0,2,...}^{\infty}\braket{j_f||\tilde{\Omega}_{J;\tau}(q)||j_i}\braket{j_f||\Sigma''_{J;\tau'}(q)||j_i}, \nonumber\\
W^{\tau\tau'}_{\Delta\tilde{\Phi}}(q)&\equiv\frac{4\pi}{2j_i+1}\sum_{J=1,3,...}^{\infty}\braket{j_f||\Delta_{J;\tau}(q)||j_i}\braket{j_f||\tilde{\Phi}_{J;\tau'}(q)||j_i}, \nonumber\\
W^{\tau\tau'}_{\Delta'\tilde{\Phi}'}(q)&\equiv\frac{4\pi}{2j_i+1}\sum_{J=1,3,...}^{\infty}\braket{j_f||\Delta'_{J;\tau}(q)||j_i}\braket{j_f||\tilde{\Phi}'_{J;\tau'}(q)||j_i}. \nonumber
\end{align}
}

\section{Nuclear modeling uncertainties}
\label{app:modeling}
The $^{27}$Al wave functions used in calculations were taken from shell-model diagonalizations that included
all Slater determinants in the $2s1d$ shell.  Three commonly used effective interactions \cite{PhysRevC.74.034315,bw} were employed.  The numerical values are given in Table \ref{tab:nuc_responses}, so that future studies can use these results
as a cross check.  Differences among the three calculations provide a measure of nuclear modeling uncertainties,
though we stress that the calculations all share certain assumptions. Thus, their differences should be taken as minimum uncertainties. As is the case throughout this work, numerical evaluations of $\tilde{\Delta}_J''$ are performed assuming a conserved vector current, so that we can use Eq. \eqref{eq:siegert} to rewrite this operator in terms of $M_J$.

\begin{table*}[]
 {\renewcommand{\arraystretch}{1.4}
    \centering
    \begin{tabular}{c||c|c|c||c|c|c||c|c|c}
    \hline
    \hline  
     & \multicolumn{3}{c||}{$W_M^{00}$}  & \multicolumn{3}{c||}{$W_{\Sigma''}^{00}$} & \multicolumn{3}{c}{$W_{\Sigma}^{00}+W_{\Sigma'}^{00}$}\\
         &  BW & USDA & USDB &  BW & USDA & USDB &  ~~BW~~ & ~~USDA~~ & ~~USDB~~ \\
         \hline
        $^{27}$Al$(5/2^+)$ &  262.85 & 262.86 & 262.86 & 0.113 & 0.107 & 0.110 & 0.105 & 0.091 & 0.099\\
           $^{27}$Al$(1/2^+)$   &  0.0690 & 0.0694 & 0.0708 & 0.011 & 0.010 & 0.010 & 0.0219 & 0.0216 & 0.0211\\
            $^{27}$Al$(3/2^+)$  &  0.124  & 0.133  & 0.131  & $5.0\times 10^{-3}$  & $5.5\times 10^{-3}$ & $5.0\times 10^{-3}$ & 0.023 & 0.025 & 0.024\\
           $^{27}$Al$(7/2^+)$   & 0.281 & 0.285 & 0.276  & 0.023 & 0.024 & 0.022 & 0.034 & 0.032 & 0.033\\
           \hline
           \hline
            & \multicolumn{3}{c||}{$W_M^{11}$} & \multicolumn{3}{c||}{$W_{\Sigma''}^{11}$} & \multicolumn{3}{c}{$W_{\Sigma}^{11}+W_{\Sigma'}^{11}$}\\
         &  BW & USDA & USDB &  BW & USDA & USDB &  BW & USDA & USDB \\
         \hline
        $^{27}$Al$(5/2^+)$ &  0.274 & 0.272 & 0.273 & 0.099 & 0.092 & 0.091 & 0.066 & 0.054 & 0.052\\
           $^{27}$Al$(1/2^+)$ & $2.7\times10^{-3}$ & $2.4\times10^{-3}$ & $2.4\times10^{-3}$ & $8.6\times10^{-3}$ & $8.2\times10^{-3}$ & $8.5\times10^{-3}$ & 0.0137 & 0.0136 & 0.0137 \\
            $^{27}$Al$(3/2^+)$  & $3.3\times10^{-3}$ & $3.4\times10^{-3}$ & $3.6\times10^{-3}$  & $5.4\times10^{-3}$ & $5.8\times10^{-3}$ & $6.0\times10^{-3}$ & 0.0131 & 0.0149 & 0.0143\\
           $^{27}$Al$(7/2^+)$   & $6.3\times10^{-3}$ & $5.5\times10^{-3}$ & $5.8\times10^{-3}$  & 0.0206 & 0.0207 & 0.0206 & 0.0221 & 0.0215 & 0.0206 \\
           \hline
           \hline
            & \multicolumn{3}{c||}{$W_{\Delta}^{00}+W_{\Delta'}^{00}$}  & \multicolumn{3}{c||}{$W_{\Phi''}^{00}$} & \multicolumn{3}{c}{$W_{\tilde{\Phi}}^{00}+W_{\tilde{\Phi}'}^{00}$}\\
         &  BW & USDA & USDB &  BW & USDA & USDB &  ~~BW~~ & ~~USDA~~ & ~~USDB~~ \\
         \hline
        $^{27}$Al$(5/2^+)$ &  0.494 & 0.504 & 0.498 & 11.0 & 10.7 & 11.6 & $1.0\times 10^{-3}$ & $8.8\times 10^{-4}$ & $9.4\times 10^{-4}$\\
           $^{27}$Al$(1/2^+)$   &  $2.0\times 10^{-5}$ & $2.5\times 10^{-5}$ & $2.3\times 10^{-5}$ & $9.3\times 10^{-3}$ & $7.8\times 10^{-3}$ & 0.010 & $4.2\times 10^{-3}$ & $3.0\times 10^{-3}$ & $4.4\times 10^{-3}$\\
            $^{27}$Al$(3/2^+)$  &  $8.9\times 10^{-4}$  & $1.1\times 10^{-3}$  & $9.3\times 10^{-4}$  & $8.6\times 10^{-3}$  & $8.5\times 10^{-3}$ & 0.011 & $2.5\times 10^{-3}$ & $2.4\times 10^{-3}$ & $3.7\times 10^{-3}$\\
           $^{27}$Al$(7/2^+)$   & $3.1\times 10^{-3}$ & $3.1\times 10^{-3}$ & $3.0\times 10^{-3}$  & 0.042 & 0.034 & 0.033 & 0.045 & 0.037 & 0.033\\
           \hline
           \hline
            & \multicolumn{3}{c||}{$W_{\Delta}^{11}+W_{\Delta'}^{11}$}  & \multicolumn{3}{c||}{$W_{\Phi''}^{11}$} & \multicolumn{3}{c}{$W_{\tilde{\Phi}}^{11}+W_{\tilde{\Phi}'}^{11}$}\\
         &  BW & USDA & USDB &  BW & USDA & USDB &  BW & USDA & USDB \\
         \hline
        $^{27}$Al$(5/2^+)$ &  0.225 & 0.206 & 0.222 & 0.087 & 0.083 & 0.087 & 0.068 & 0.066 & 0.069\\
           $^{27}$Al$(1/2^+)$ & $4.0\times10^{-5}$ & $3.8\times10^{-5}$ & $4.9\times10^{-5}$ & $8.0\times10^{-4}$ & $3.0\times10^{-4}$ & $6.2\times10^{-4}$ & $7.5\times10^{-4}$ & $2.5\times10^{-4}$ & $5.3\times10^{-4}$ \\
            $^{27}$Al$(3/2^+)$  & 0.019 & 0.021 & 0.022  & $1.7\times10^{-3}$ & $1.4\times10^{-3}$ & $1.4\times10^{-3}$ & $3.2\times10^{-3}$ & $2.7\times10^{-3}$ & $2.6\times10^{-3}$\\
           $^{27}$Al$(7/2^+)$   & $1.6\times10^{-3}$ & $2.6\times10^{-3}$ & $2.1\times10^{-3}$  & $4.2\times 10^{-3}$ & $5.0\times 10^{-3}$ & $5.4\times 10^{-3}$ & 0.0129 & 0.0129 & 0.0132 \\
           \hline
           \hline
                & \multicolumn{3}{c||}{$W_{\tilde{\Delta}''}^{00}$}  & \multicolumn{3}{c||}{$W_{\tilde{\Omega}}^{00}$} & \multicolumn{3}{c}{}\\
         &  BW & USDA & USDB &  BW & USDA & USDB &  \multicolumn{3}{c}{} \\
          \cline{1-7}
           $^{27}$Al$(1/2^+)$   &  $2.96\times 10^{-4}$ & $2.98\times 10^{-4}$ & $3.04\times 10^{-4}$ & $5.0\times 10^{-5}$ & $1.2\times 10^{-4}$ & $8.6\times 10^{-5}$ & \multicolumn{3}{c}{}\\
            $^{27}$Al$(3/2^+)$  &  $7.76\times 10^{-4}$  & $8.31\times 10^{-4}$  & $8.16\times 10^{-4}$  & $1.5\times 10^{-4}$  & $1.4\times 10^{-4}$ & $2.5\times 10^{-4}$ & \multicolumn{3}{c}{}\\
           $^{27}$Al$(7/2^+)$   & $8.72\times 10^{-3}$ & $8.85\times 10^{-3}$ & $8.55\times 10^{-3}$  & 0.015 & 0.012 & 0.0095 & \multicolumn{3}{c}{}\\
            \cline{1-7}
            \cline{1-7}
            & \multicolumn{3}{c||}{$W_{\tilde{\Delta}''}^{11}$}  & \multicolumn{3}{c||}{$W_{\tilde{\Omega}}^{11}$} & \multicolumn{3}{c}{}\\
         &  BW & USDA & USDB &  BW & USDA & USDB & \multicolumn{3}{c}{} \\
         \cline{1-7}
           $^{27}$Al$(1/2^+)$ & $1.14\times10^{-5}$ & $1.02\times10^{-5}$ & $1.04\times10^{-5}$ & $4.4\times10^{-5}$ & $3.4\times10^{-5}$ & $4.0\times10^{-5}$ & \multicolumn{3}{c}{} \\
            $^{27}$Al$(3/2^+)$  & $2.06\times10^{-5}$ & $2.10\times10^{-5}$ & $2.25\times10^{-5}$ & $1.7\times10^{-4}$ & $1.5\times10^{-4}$ & $2.1\times10^{-4}$ & \multicolumn{3}{c}{}\\
           $^{27}$Al$(7/2^+)$   & $1.96\times10^{-4}$ & $1.71\times10^{-4}$ & $1.81\times10^{-4}$  & $5.9\times 10^{-3}$ & $4.0\times 10^{-3}$ & $3.2\times 10^{-3}$ & \multicolumn{3}{c}{} \\
            \hline
            \hline
    \end{tabular}
    }
    \caption{Comparison of nuclear response functions computed for low-lying transitions in $^{27}$Al for three different $2s1d$-shell effective interactions.}
    \label{tab:nuc_responses}
\end{table*}

\end{widetext}

\end{document}